\documentclass[12pt]{article}
\usepackage{graphicx}
\usepackage{epsfig,amsmath,amssymb}
\allowdisplaybreaks

\title{{\bf Spin Effects in Long Range Electromagnetic Scattering}}
\author{Barry R. Holstein$^a$\footnote{\tt holstein@physics.umass.edu} \hspace*{1pt} and Andreas
Ross$^{a,b}$\footnote{\tt
andreas.ross@yale.edu} \\ \\
$^a$ Department of Physics -- LGRT\\
University of Massachusetts\\
Amherst, MA  01003, USA\\  \\
$^b$ Department of Physics \\
 Yale University \\
New Haven, CT 06520, USA \\ \\}

\begin{document}
\maketitle
\thispagestyle{empty}

\begin{abstract}
We analyze the electromagnetic scattering of massive particles with
and without spin and, using the techniques of effective field
theory, we isolate the leading long distance effects beyond one
photon exchange, both classical and quantum mechanical.
Spin-independent and spin-dependent effects are isolated and shown
to have a universal structure.
\end{abstract}

\vspace{0.2 in}
\setcounter{page}{0}
\newpage

\section{Introduction}
There has been a good deal of recent interest in higher order
corrections to Coulomb scattering.  In particular the
one-photon exchange approximation, which has traditionally been used
to analyze electron scattering has been shown to be inadequate when
applied to the problem of isolating nucleon form factors via a
Rosenbluth separation.  Inclusion of two-photon exchange
contributions have been found to be essential in resolving small
discrepancies with the values of these same form factors as obtained
from spin correlation measurements \cite{ros}. A second arena where
two-photon exchange effects are needed is in the analysis of
transverse polarization asymmetry measurements in electron
scattering.  Such quantities vanish in the one-photon exchange
approximation meaning that the sizable effects found experimentally
must arise from two-photon effects \cite{tpa}.

Much has been written about such higher order photon processes and a
number of groups have undertaken precision calculation of such
effects \cite{the}.  It is not our purpose here to attempt such
detailed calculations or to confront experimental data.  Rather our
goal is to use the methods of effective field theory in order to
analyze the very longest range (smallest momentum transfer)
contributions to the scattering process.  These long range
components are associated with pieces of the scattering amplitude
which are nonanalytic in the momentum transfer, and most of them
are also singular in the limit of a vanishing momentum transfer
(the exception being part of the the correction to the spin-spin
coupling component where an extra factor of $q^2$ arises). Some
of these corrections are classical ($\hbar$-independent) and behave
as $1/\sqrt{-q^2}$ while others are quantum mechanical
($\hbar$-dependent) and behave as $\log -q^2$, where $q^2$ is the
invariant momentum transfer squared \cite{jdh}.  Below we shall
examine both types of structures in the context of the
electromagnetic scattering of two distinguishable massive particles
of unit charge $e$ with and without spin. In this case the lowest order interaction,
which arises from one-photon exchange, is the simple Coulomb
interaction
\begin{equation}
 V(r) = \frac{\alpha}{r}
\end{equation}
where $\alpha=e^2/4\pi$ is the fine structure constant. We find that
two-photon exchange processes at threshold ($v \rightarrow 0$) yield
corrections of the form
\begin{equation}
 V(r) = \frac{\alpha}{r} \left(1 + A_C \frac{\alpha}{mr} + A_Q \, \hbar \frac{\alpha}{(mr)^2} \right)
\end{equation}
where $A_C$ and $A_Q$ are the coefficients of the classical and
quantum corrections respectively and are evaluated below.

To see how such terms arise, in the next section we sketch our
calculational techniques in the context of spin-independent
scattering. This is a problem addressed nearly two decades ago by
Feinberg and Sucher using dispersive methods \cite{fs}. Even earlier
Iwasaki had studied the classical piece of this problem using
standard noncovariant perturbation theory \cite{iwa}. Our quantum
corrections are found to agree completely with those found by
Feinberg and Sucher.  However, our classical potential is at
variance with that found both in \cite{fs} and \cite{iwa, sp}.  The
origin of these differences is found in terms of differing
contributions from the iterated piece of the lowest order potential,
which must be subtracted from the scattering amplitude in order to
produce a properly defined higher order potential \cite{js}.
Our work has also been motivated by more recent calculations
in gravitational scattering where corrections to Newton's law
are obtained \cite{bdh}.

In the following sections we extend these effective field theoretic
methods to the problem of spin-dependent scattering and demonstrate
that the results are {\it universal}, in that they can be written in
terms of forms which are independent of spin. The calculation of
spin-0 -- spin-1/2 scattering reveals new structures of spin-orbit
character whose universal form is also obtained when we consider
spin-0 -- spin-1 scattering. The next extension consists of spin-1/2
-- spin-1/2 scattering where again we find the universal
spin-independent and spin-orbit pieces as well as new (presumably
universal) spin-spin coupling interactions. Our results are
summarized in a short concluding chapter and the calculational details
are found in the appendices. In Appendix \ref{app_general} we give
generalized results for arbitrary charges and g-factors of the scattered
particles and argue for a multipole expansion like scheme which explains
the universalities found.

\section{Spin-Independent Scattering}

\begin{figure}
\begin{center}
\epsfig{file=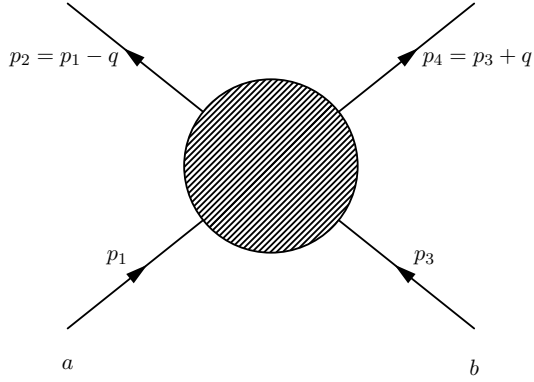,width=7cm} \caption{Basic kinematics of
electromagnetic scattering. } \label{fig_kinem}
\end{center}
\end{figure}

We first set the generic framework for our study.  We examine the
electromagnetic scattering of two charged particles---particle $a$
with mass $m_a$, charge $e$ and incoming four-momentum $p_1$ and
particle $b$ with mass $m_b$, charge $e$ and incoming four-momentum
$p_3$.  After undergoing scattering the final four-momenta of
particle $a$ is $p_2=p_1-q$ and that of particle $b$ is
$p_4=p_3+q$---{\it cf.} Fig. \ref{fig_kinem}.  Now we need to be
more specific.

\subsection{Spin-0 -- Spin-0 Scattering}

We begin by examining the electromagnetic scattering of two spinless
particles.  The electromagnetic interaction follows from making the
minimal substitution in the Klein-Gordon Lagrangian density,
yielding
\begin{equation}
{\cal L}=(iD_\mu\phi)^\dagger iD^\mu\phi-m^2\phi^\dagger\phi
\end{equation}
where $iD_\mu=i\partial_\mu-eA_\mu$ is the covariant derivative, and
leads to the one- and two-photon vertices
\begin{eqnarray}
\tau^{(1)}_\mu(p_2,p_1)&=&-ie(p_2+p_1)_\mu\nonumber\\
\tau^{(2)}_{\mu\nu}(p_2,p_1)&=&2ie^2\eta_{\mu\nu}
\end{eqnarray}
Single photon exchange then leads to the familiar amplitude (in
Feynman gauge)
\begin{eqnarray}
{}^0 \! {\cal M}^{(1)}(q)&=&{-i\over \sqrt{2E_12E_22E_32E_4} } \,
\tau^{(1)}_\mu(p_1,p_2)\hspace*{1pt}{-i\eta^{\mu\nu}\over
q^2} \, \tau^{(1)}_\nu(p_3,p_4)\nonumber\\
&=&{8\pi\alpha\over \sqrt{2E_12E_22E_32E_4}} \,
\frac{s-m_a^2-m_b^2+\frac{1}{2} q^2}{q^2}
\end{eqnarray}
with $s=(p_1+p_3)^2$ the square of the center of mass energy.

One way to define the nonrelativistic potential is as the Fourier
transform of the nonrelativistic amplitude evaluated in the center
of mass frame. We will use a symmetric center of mass frame\footnote{These
symmetric momentum labels of the center of mass frame are chosen so that
the leading order coordinate space potential is real in the calculation
of spin-0 -- spin-1 scattering presented below.}
with incoming momenta $\vec p_1 = \vec p + \vec q / 2$ and $\vec p_3 = -
\vec p_1 = - \vec p - \vec q / 2$ and with outgoing momenta $\vec
p_2 = \vec p - \vec q / 2$ and $\vec p_4 = - \vec p + \vec q / 2$.
Conservation of energy then requires $\vec p \cdot \vec q = 0$ so
that $\vec p_i^{\hspace*{1.4pt} 2} = \vec p^{\hspace*{1.4pt} 2} +
\vec q^{\hspace*{1.4pt} 2} / 4$ for $i = 1, 2, 3, 4$ and $q^2 = -
\vec q^{\hspace*{1.4pt} 2}$. In the nonrelativistic limit--- $\vec
q^{\hspace*{1.4pt} 2}, \vec p^{\hspace*{1.4pt} 2} \ll m^2$ ---the
        amplitude reads
\begin{equation}
{}^0 \! {\cal M}^{(1)}(\vec q) \simeq - \frac{4 \pi \alpha} {\vec
q^{\hspace*{1.4pt} 2}} \hspace*{-1pt} \left(\hspace*{-1pt} 1 \hspace*{-1pt} + \frac{\vec p^{\hspace*{1.4pt}
2}}{m_a m_b} \hspace*{-1pt} + \hspace*{-1pt} \ldots \hspace*{-1pt} \right) + \frac{\pi \alpha}{m_a m_b} \hspace*{-1pt} \left( \hspace*{-1pt} 0 +
\frac{(m_a^2 + m_b^2) \vec p^{\hspace*{1.4pt} 2}}{2 m_a^2 m_b^2} +
\ldots \hspace*{-1pt} \right) + \ldots
\end{equation}
yielding the potential
\begin{eqnarray}
{}^0V^{(1)}_C(\vec{r}) \hspace*{-3pt} & = \hspace*{-3pt} &-\int {d^3q\over (2\pi)^3} \,
{}^0 \!{\cal M}^{(1)}(\vec q) \, e^{-i\vec{q}\cdot\vec{r}} \nonumber\\
&= \hspace*{-3pt} &{\alpha\over r}\left(1+{\vec p^{\hspace*{1.4pt} 2}\over m_am_b
}+\ldots\right) -{\pi\alpha\over m_am_b}\,
\delta^3(\vec{r})\left(0+{(m_a^2+m_b^2)\vec p^{\hspace*{1.4pt}
2}\over
2m_a^2m_b^2}+\ldots\right) \nonumber\\
\label{eq:fs}
\end{eqnarray}
The first component of Eq. (\ref{eq:fs}) is recognized as the usual
Coulomb potential (accompanied by small kinematic effects) while the
second piece is a short range correction.

\begin{figure}
\begin{center}
\epsfig{file=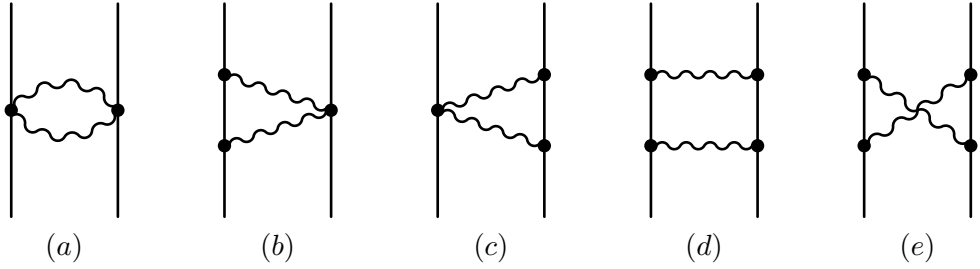,width=0.95\textwidth} \caption{One loop
diagrams in electromagnetic scattering. } \label{fig_diags}
\end{center}
\end{figure}

Our purpose in this paper is to study the long distance corrections
to this form which arise from the two-photon exchange diagrams shown
in Fig. \ref{fig_diags}.  This problem has been previously studied
by Iwasaki using nonrelativistic perturbation theory \cite{iwa} and
by Feinberg and Sucher using dispersive methods \cite{fs}. Our
approach will be to use the methods of effective field theory,
wherein we evaluate these diagrams by keeping only the leading
nonanalytic structure in $q^2$, since it is these pieces which
lead to the long range corrections to the potential.  This nonanalytic behavior is of two
forms---
\begin{itemize}
\item [i)] terms in $1/\sqrt{-q^2}$ which are $\hbar$-independent
and therefore classical
\item [ii)] terms in $\log -q^2$ which are $\hbar$-dependent and
therefore quantum mechanical
\end{itemize}
The former terms, when Fourier transformed lead to corrections to
the nonrelativistic potential of the form $V_{classical}(r)\sim
1/r^2$ while the latter lead to $V_{quantum}(r)\sim \hbar/mr^3$
corrections.  For typical masses and separations the quantum
mechanical forms are themselves numerically insignificant. However,
they are intriguing in that their origin appears to be associated
with zitterbewegung.  That is, classically we can define the
potential by measuring the energy when two objects are separated by
distance $r$.  However, in the quantum mechanical case the distance
between two objects is uncertain by an amount of order the Compton
wavelength due to zero point motion---$\delta r\sim \hbar/ m$. This
leads to the replacement
$$V(r)\sim {1\over r^2}\longrightarrow {1\over (r\pm\delta r)^2}
\sim {1\over r^2}\mp 2{\hbar\over mr^3}$$ which is the form found in
our calculations.

The calculational details are described in Appendix \ref{app_int}.  Here we
present only the results.   Defining
$$S = \frac{\pi^2}{\sqrt{-q^2}}\quad{\rm and}\quad L=\log -q^2$$
we have, from diagrams (a)-(e) of Fig. \ref{fig_diags} respectively
\begin{eqnarray}
{}^0 \! {\cal M}_{\ref{fig_diags}a}^{(2)}(q)&=&{\alpha^2\over
m_am_b}(-2L)\nonumber\\
{}^0 \! {\cal M}_{\ref{fig_diags}b}^{(2)}(q)&=&{\alpha^2\over m_am_b}(2L+m_aS)\nonumber\\
{}^0 \! {\cal M}_{\ref{fig_diags}c}^{(2)}(q)&=&{\alpha^2\over m_am_b}(2L+m_bS)\nonumber\\
{}^0 \! {\cal M}_{\ref{fig_diags}d}^{(2)}(q)&=&{\alpha^2\over
m_am_b}\left[L\left({4m_am_b\over q^2}+{5(m_a^2+m_b^2)\over
4m_am_b}-{1\over 2}\right)+S(m_a+m_b)\right]\nonumber\\
&-&i4\pi\alpha^2 {L\over q^2}\sqrt{m_am_b\over s-s_0}\nonumber\\
{}^0 \! {\cal M}^{(2)}_{\ref{fig_diags}e}(q)&=&{\alpha^2\over m_am_b}\left[
L\left(-{4m_am_b\over q^2}-{5(m_a^2+m_b^2)\over
4m_am_b}-{23\over 6}\right)-S(m_a+m_b)\right]\nonumber\\
\quad
\end{eqnarray}
where $s=(p_1+p_3)^2$ is the square of the center of mass energy and
$s_0=(m_a+m_b)^2$ is its threshold value.  Summing, we find the
final result
\begin{equation}
{}^0 \! {\cal M}_{tot}^{(2)}(q)={\alpha^2\over
m_am_b}\left[(m_a+m_b)S-{7L\over 3}\right]-i4\pi\alpha^2{L\over
q^2}\sqrt{m_am_b\over s-s_0}
\end{equation}
We observe that in addition to the expected terms involving $L$ and
$S$ there arises a piece of the second order amplitude which is
imaginary. The origin of this imaginary piece is, of course, from
the second Born approximation to the Coulomb potential, and reminds
us that in order to define a proper correction to the first order
Coulomb potential we must subtract off such terms. Before performing
the necessary subtraction, we also point out that the imaginary part
arising at this order of the amplitude is very singular in the
nonrelativistic limit, even more than the leading order amplitude.
It stems from an overall phase of the amplitude
\cite{Weinberg:1965nx} and thus cannot contribute to any observable
since observables are proportional to $|{\cal M}|^2$. At our order
we have calculated ${\cal M} = {}^0 \! {\cal M}^{(1)} + {}^0 \!
{\cal M}^{(2)}$ where an observable such as a differential cross
section has a leading piece ${\cal O}(\alpha^2)$ and the corrections
we have calculated contribute to order ${\cal O}(\alpha^3)$, but the
imaginary part does not contribute to ${\cal O}(\alpha^3)$.

In order to subtract the second Born piece, we will work in the
nonrelativistic limit and the center of mass
frame---$\vec{p}_1+\vec{p}_3=0$. We have then
\begin{equation}
s-s_0=2\sqrt{m_a^2+\vec p_1^{\hspace*{1.4pt} 2}}\sqrt{m_b^2+\vec
p_1^{\hspace*{1.4pt} 2}}+2\vec p_1^{\hspace*{1.4pt} 2}-2m_am_b \label{eq_sminuss0_p0}
\end{equation}
and
\begin{equation}
\sqrt{m_am_b\over s-s_0}\simeq {m_r\over p_0}
\end{equation}
where $m_r=m_am_b/(m_a+m_b)$ is the reduced mass and
$p_0 \equiv |\vec{p}_i|, \  i=1,2,3,4$.  The transition amplitude then assumes the form
\begin{equation}
{}^0 \! {\cal M}_{tot}^{(2)}(\vec q) \simeq {\alpha^2\over
m_am_b}\left[(m_a+m_b)S-{7L\over 3}\right]-i4\pi\alpha^2{L\over
q^2}{m_r\over p_0}\label{eq:lj}
\end{equation}
For the iteration we shall use the simple potential
\begin{equation}
{}^0V^{(1)}_C(\vec r)={\alpha\over r}\label{eq:co}
\end{equation}
which reproduces the lowest order amplitude for spin-0 -- spin-0
Coulomb scattering---Eq. (\ref{eq:fs})---in the nonrelativistic
limit and which reads in momentum space
\begin{equation}
 {}^0V^{(1)}_C(\vec q) \equiv \left<\vec p_f \left| {}^0 \hat V^{(1)}_C \right| \vec p_i \right> = \frac{e^2}{\vec q^{\hspace*{1.4pt}2}} = \frac{e^2}{(\vec p_i - \vec p_f)^2}
\end{equation}
where we identify $\vec p_i = \vec p_1$ and $\vec p_f = \vec p_2$.
The second Born term is then
\begin{eqnarray} \label{eq:iteration00a}
{}^0{\rm Amp}_C^{(2)}(\vec q)&=&- \int{d^3\ell\over
(2\pi)^3} \, \frac{\left<\vec p_f \left| {}^0 \hat V^{(1)}_C \right| \vec \ell \, \right> \left<\vec \ell \left| {}^0 \hat V^{(1)}_C \right| \vec p_i \right>}{E(p_0) - E(\ell) + i \epsilon}\nonumber\\
&=&i\int{d^3\ell\over (2\pi)^3}{}^0V_C^{(1)}(\vec \ell - \vec p_f)
\, G^{(0)}(\vec{\ell}) \, {}^0V_C^{(1)}(\vec{p}_i-\vec{\ell} \, )
\end{eqnarray}
where
\begin{equation}
G^{(0)}(\ell)={i\over {p_0^2\over 2m_r}-{\ell^2\over
2m_r}+i\epsilon}
\end{equation}
is the free propagator. Note that in Eq. (\ref{eq:iteration00a}) we
take both the leading order potential as well as the total energies
$E(p_0)$ and $E(\ell)$ in the nonrelativistic limit.  The remaining
integration can be performed exactly, as discussed in Appendix \ref{app_iter}, by
including a "photon mass" term $\lambda^2$ as a regulator, yielding
\begin{eqnarray} \label{eq:iteration00b}
{}^0{\rm Amp}_C^{(2)}(\vec q)&=&i\int {d^3\ell\over (2\pi)^3}
{e^2\over |\vec{p}_2-\vec{\ell}|^2+\lambda^2}{i\over {p_0^2\over
2m_r}-{\ell^2\over 2m_r}+i\epsilon}{e^2\over
|\vec{\ell}-\vec{p}_1|^2+\lambda^2}\nonumber\\
&\stackrel{\lambda\rightarrow 0}{\longrightarrow}&H
=-i4\pi\alpha^2 \hspace*{0.2pt} \frac{L}{q^2} \frac{m_r}{p_0}
\end{eqnarray}
which reproduces the imaginary component of ${}^0 \! {\cal
M}_{tot}^{(2)}(\vec q)$, as expected.\footnote{We have omitted the IR singularity
in the limit $\lambda \rightarrow 0$ since it does not contain nonanalytic
dependence on $q^2$. However, we note that it is present in the iteration as
well as in the amplitude of the box diagram, and it is easily shown to
cancel for the potential.}

In order to produce a
properly defined second order potential ${}^0V^{(2)}_C(\vec r)$, we must
then subtract this second order Born term from the second order
scattering amplitude Eq. (\ref{eq:lj}), yielding the result
\begin{eqnarray}
{}^0V_C^{(2)}(\vec r)&=&-\int{d^3q\over
(2\pi)^3}e^{-i\vec{q}\cdot\vec{r}}\left[{}^0 \! {\cal
M}_{tot}^{(2)}(\vec q)-{}^0{\rm Amp}_C^{(2)}(\vec q)\right]\nonumber\\
&=&\int{d^3q\over (2\pi)^3}e^{-i\vec{q}\cdot\vec{r}}{\alpha^2\over
m_am_b}\left[-S(m_a+m_b)+{7\over 3}L\right]\nonumber\\
&=&-{\alpha^2(m_a+m_b)\over 2m_am_b r^2 }-{7\alpha^2\hbar\over 6\pi
m_am_br^3} \quad\label{eq:so}
\end{eqnarray}
The result given in Eq. (\ref{eq:so}) agrees with that previously
given by Feinberg and Sucher for the quantum mechanical---$\sim
1/r^3$---piece but disagrees for the classical---$1/r^2$---term. The
classical contribution has also been calculated by Iwasaki
\cite{iwa}, who determined zero for this second order potential. The
resolution of this issue was given by Sucher, who pointed out that
the classical term depends upon the precise definition of the first
order potential used in the iteration \cite{js} and on whether one
uses relativistic expressions in the iteration. Use of the simple
lowest order form Eq. (\ref{eq:co}) within a nonrelativistic
iteration yields our result for the iteration amplitude given in Eq. (\ref{eq:iteration00b}) and
is sufficient to remove the offending imaginary piece of the
scattering amplitude. However, if one uses relativistic expressions
for the potential and the relativistic form of the energy in the
iteration then alternate forms result with a different classical
piece of the potential. Moreover, there are ambiguities in the
leading order potential used for the iteration so that a unique
definition of the second order potential does not exist \cite{js}.

For example, Feinberg and Sucher \cite{fs} calculate the iteration
amplitude
\begin{equation} 
 {}^0{\rm Amp}_{FS}^{(2)}(\vec q) = - \int{d^3\ell\over
  (2\pi)^3} \, \frac{\left<\vec p_f \left| {}^0 \hat V^{(1)}_{FS} \right|
  \vec \ell \, \right> \left<\vec \ell \left| {}^0 \hat V^{(1)}_{FS} \right|
   \vec p_i \right>}{E(p_0) - E(\ell) + i \epsilon}\label{eq:ab}
\end{equation}
using fully relativistic expressions for the total energies in the
denominator $E(p_0) = E_a(p_0) + E_b(p_0) = \sqrt{m_a^2 + p_0^2} +
\sqrt{m_b^2 + p_0^2}$ and $E(\ell) = E_a(\ell) + E_b(\ell)$ and a potential
including relativistic corrections
\begin{equation}
{}^0 \hat V^{(1)}_{FS} = \sqrt{1+\frac{\hat p^2}{\hat E_a(\hat p)
\hat E_b(\hat p)}} \ \frac{e^2}{\hat q^2} \ \sqrt{1+\frac{\hat
p^2}{\hat E_a(\hat p) \hat E_b(\hat p)}}.\label{eq:cd}
\end{equation}
where we do not display the short distance part of the potential and
with hats denoting operators whose ordering matters. Sucher calls
this a ``Feynman gauge inspired'' potential whose operator ordering
would be a natural choice when working in Feynman gauge \cite{js}.
One can evaluate the iteration integral by keeping the leading
relativistic modifications of our previous results. Thus the
Feinberg-Sucher potential---Eq. (\ref{eq:cd})---becomes
\begin{equation}
\left<\vec p_f \left| {}^0 \hat V^{(1)}_{FS} \right| \vec p_i \right> \simeq {e^2\over \vec{q}^{\hspace*{1.4pt}2}}\left(1+{\vec p_i^{\hspace*{1.4pt}2} + \vec p_f^{\hspace*{1.4pt}2} \over
2m_am_b}\right)
\end{equation}
Similarly, the energy difference appearing in the propagator becomes
\begin{eqnarray}
E(p_0)-E(\ell)&\simeq& \left(m_a+m_b+{p_0^2\over 2m_a}+{p_0^2\over 2m_b}-{p_0^2\over 8m_a^3}-{p_0^2\over 8m_b^3} \right)\nonumber\\
&-&\left(m_a+m_b+{\ell^2\over 2m_a}+{\ell^2\over 2m_b}-{\ell^2\over 8m_a^3}-{\ell^2\over 8m_b^3} \right)\nonumber\\
&=&\left({p_0^2\over 2m_r}-{\ell^2\over 2m_r}\right)
  \left[1-\left({p_0^2\over 4m_r^2}+{\ell^2\over 4m_r^2}\right)\left(1-3{m_r^2\over m_am_b}\right)\right] \nonumber \\ \quad
\end{eqnarray}
We can now perform the Feinberg-Sucher iteration integral given in
Eq. (\ref{eq:ab})
\begin{eqnarray}
{}^0{\rm Amp}^{(2)}_{FS}(\vec{q})&\simeq&-\int{d^3\ell\over
(2\pi)^3} {e^2\over |\vec{p}_f-\vec{\ell}|^2}{1\over {p_0^2\over
2m_r}-{\ell^2\over 2m_r} + i \epsilon} {e^2\over
|\vec{\ell}-\vec{p}_i|^2}\nonumber\\
&& \hspace*{44pt} \times \left\{1+{(p_0^2+\ell^2)\over m_am_b}\left[1+{m_am_b\over
4m_r^2}\left(1-3{m_r^2\over m_am_b}\right) \right]\right\}\nonumber\\
& \simeq &H+{1\over m_am_b}\left(p_0^2H+\delta_{rs}H_{rs}\right)\left[1+{m_am_b\over
4m_r^2}\left(1-3{m_r^2\over m_am_b}\right)\right]\nonumber\\
&\simeq& - i4\pi\alpha^2{m_r\over p_0}{L\over q^2}+{4\alpha^2\over
m_a+m_b}S\left[1+{(m_a+m_b)^2\over 4m_am_b}-{3\over 4}\right]\nonumber\\
&\simeq&- i4\pi\alpha^2{m_r\over p_0}{L\over q^2}+{\alpha^2\over
m_am_b} \left(m_a+m_b+{m_am_b\over m_a+m_b}\right) S \label{eq:jk}
\end{eqnarray}
which agrees precisely to this order with the exact relativistic
evaluation given in Ref. \cite{fs}.

Subtracting from ${}^0 \! {\cal M}^{(2)}_{tot}(\vec{q})$ we find
then the second order potential
\begin{eqnarray}
{}^0V_{FS}^{(2)}(\vec r)&=&-\int{d^3q\over
(2\pi)^3}e^{-i\vec{q}\cdot\vec{r}}\left[{}^0 \! {\cal
M}^{(2)}_{tot}(\vec q)-{}^0{\rm Amp}_{FS}^{(2)}(\vec q)\right]\nonumber\\
&=&\int{d^3q\over (2\pi)^3}e^{-i\vec{q}\cdot\vec{r}}
\frac{\alpha^2}{m_a m_b}
\left[\frac{m_a m_b}{m_a + m_b} S +{7\over 3}L\right]\nonumber\\
&=& {\alpha^2\over 2(m_a+m_b) r^2 }-{7\alpha^2\hbar\over 6\pi
m_am_br^3} \label{eq:sf}
\end{eqnarray}
which is the form given by Feinberg and Sucher \cite{fs}.

On the other hand Sucher \cite{js} also discusses an alternative
version of the one-photon exchange potential which is a natural
choice when working in Coulomb gauge. It includes relativistic
expressions, and we refer the reader to \cite{js} for the details of
this ``Coulomb gauge inspired'' order $\alpha$ potential ${}^0 \hat
V^{(1)}_{SP}$. We note that {\it on-shell} the two potentials coincide
\begin{equation}
 \left<\vec p_f \left| {}^0 \hat V^{(1)}_{FS} \right| \vec p_i \right>
  = \left<\vec p_f \left| {}^0 \hat V^{(1)}_{SP} \right| \vec p_i \right>
\end{equation}
but when used {\it off-shell} in the iteration then the second order
amplitude becomes
\begin{eqnarray}
{}^0{\rm Amp}_{SP}^{(2)}(\vec{q}) &=& - \int{d^3\ell\over
  (2\pi)^3} \, \frac{\left<\vec p_f \left| {}^0
  \hat V^{(1)}_{SP} \right| \vec \ell \, \right>
  \left<\vec \ell \left| {}^0 \hat V^{(1)}_{SP} \right|
  \vec p_i \right>}{E(p_0) - E(\ell) + i \epsilon} \nonumber \\
&=&\frac{\alpha^2}{m_am_b}\, (m_a+m_b) S - i4\pi\alpha^2{m_r\over
p_0}{L\over q^2}+\ldots
\end{eqnarray}
where fully relativistic expressions for the total energies in the
denominator were used. Subtracting from ${}^0 \! {\cal
M}^{(2)}_{tot}(\vec{q})$ we find then the second order potential
\begin{eqnarray}
{}^0V_{SP}^{(2)}(\vec r)&=&-\int{d^3q\over
(2\pi)^3}e^{-i\vec{q}\cdot\vec{r}}\left[{}^0 \! {\cal
M}^{(2)}(\vec q)-{}^0{\rm Amp}_{SP}^{(2)}(\vec q)\right]\nonumber\\
&=&\int{d^3q\over (2\pi)^3}e^{-i\vec{q}\cdot\vec{r}}{7\over
3}{\alpha^2L\over m_am_b}=-{7\alpha^2\hbar\over 6\pi m_am_br^3}
\label{eq:sp}
\end{eqnarray}
This vanishing result for the classical piece of the $\mathcal
O(\alpha^2)$ potential was also obtained by Iwasaki \cite{iwa} and
by Spruch \cite{sp}. Comparing Eqs. (\ref{eq:so}), (\ref{eq:sf}) and
(\ref{eq:sp}) we note that the quantum mechanical contribution is
{\it invariant}---only the classical term changes. As we have seen,
even two fully relativistic iteration treatments do not yield the
same answer for the classical piece, and the reason is that
ambiguities arise when defining an $\mathcal O(\alpha)$ potential
operator ${}^0 \hat V^{(1)}$. We have used an on-shell matrix
element to find ${}^0 \hat V^{(1)}$ whereas the
``Coulomb gauge inspired'' leading order potential of \cite{js}
is defined from an off-shell matrix element.
When we use these leading order potentials in an iteration
where we integrate over all possible intermediate states,
terms that vanish on-shell can contribute and yield differing
results for the iteration amplitude and thus for the $\mathcal O(\alpha^2)$ potential.
However, since the potential itself is not an observable this is not an
issue.  What {\it is} an observable is the transition amplitude.  In
each case we find the same result
\begin{equation}
{}^0 \!{\cal M}_{tot}(\vec q)=-\int
d^3re^{i\vec{q}\cdot\vec{r}}\left[{}^0V_i^{(1)}(\vec r)+{}^0V_i^{(2)}(\vec r)\right]+{}^0{\rm
Amp}_i(\vec{q})\quad i=C,FS,SP
\end{equation}
In this paper, since we are interested only in the threshold
behavior of the transition amplitude, we shall utilize the simple
Coulomb form for the potential and a nonrelativistic iteration,
since this is sufficient to remove any pieces of the amplitude that
would prevent us from writing down a well defined second order
potential.

Before moving on to details, it is useful also to point out the
parallels between our calculational methods and those of the
effective field theory NRQED \cite{NRQED}
which has been set up to analyze
non-relativistic bound states. The latter involves a systematic
expansion in powers of the relative velocity $v$ of the two
particles. In this picture the one loop corrections to the amplitude
involve terms of order $1/v^3$ and higher. For example, the calculation
of Manohar and Stewart\footnote{Manohar and Stewart (and collaborators)
have performed a number of NRQED and NRQCD calculations of great phenomenological
importance \cite{smmore}, such as the Lamb shift and hyperfine splitting or the
$t \bar t$ production cross section near threshold. Much of their work includes
impressive two loop results and involved applications of renormalization group methods.
When we refer to Manohar and Stewart in our comparison here, we refer specifically to Ref. \cite{sm}
which we found most suitable when comparing with our work.}
evaluates corrections for the scattering of a
particle-antiparticle pair up to order $v^0$ \cite{sm}. In coordinate
space such terms include a combination of both short and long distance
corrections. Our calculation involves a different sort of expansion
looking only at the {\it longest range} terms in coordinate space, and it
is not optimized for bound states since we do not use a power counting
based on the virial theorem.
Correspondingly, in momentum space we look for the nonanalytic
components of the scattering amplitude.

In their work in NRQED, Manohar and Stewart have performed a one loop matching
calculation to $\mathcal O(v^0)$ \cite{sm}. They calculate the
full QED amplitude as well as the amplitude in their formulation of
NRQED---vNRQED---which describes interactions of nonrelativistic
fermions, ultra-soft photons and soft photons. For our discussion, the
essential difference between vNRQED and full QED is that potential photons
have been integrated out and their effects are described by effective
four-fermion operators in vNRQED. The coefficients in front of these
four-fermion operators are the potential of Manohar and Stewart.

In this paper we do not match onto a well-defined theory such as NRQED.
Instead, our matching corresponds solely to the subtraction of the
second Born iteration amplitude from the QED amplitude before Fourier
transforming to a second order potential in coordinate space. Thus we regard
our potential as a nice way to display our resulting scattering amplitudes
in coordinate space, but we emphasize that our main results are the
long distance components of the scattering amplitude.
Since we do not match onto a theory containing photons whereas
the Manohar-Stewart potential arises in the matching to vNRQED which
has soft and ultrasoft photons as degrees of freedom, we expect
that the potentials differ. It is seen that our quantum corrections
proportional to $1/r^3$ stemming from $\log q^2$ pieces of the amplitude
are absent from Manohar and Stewart's potential because the exchange of
soft photons in vNRQED yield the complete QED contribution of
$\log q^2$ terms so that their matching does not yield any quantum pieces
of the potential\footnote{The alternative formulation of potential NRQED (pNRQED)
of Pineda and Soto \cite{Pineda:1997bj}, \cite{Pineda:1998kn} differs from vNRQED in that it only
contains ultra-soft photons as degrees of freedom whereas soft photons are integrated
out. (Further differences between vNRQED and pNRQED are not important for our discussion here.)
Therefore, the potential in pNRQED does exhibit the same quantum pieces of the potential
as ours---cf. Eq. (2.17) in Ref. \cite{Pineda:1998kn} for example. However, we note that Pineda
and Soto use off-shell matching in Coulomb gauge yielding a vanishing classical piece of the potential.}.

Despite these differences, Manohar and Stewart's calculation \cite{sm} of
spin-1/2 -- spin-1/2 scattering gives us the opportunity to compare
our results for the scattering amplitudes and iterations with theirs.
Note that one further difference is that our calculation deals with
non-identical particle scattering, while that of Manohar and Stewart
involves quark-quark pairs or quark-antiquark pairs of equal mass.
Thus we do not have the exchange piece of the amplitude or the
annihilation channel contributions given in \cite{sm}.
We verify that the classical spin-independent iteration amplitude
in vNRQED found by these authors using on-shell matching---given by the sum of iterated terms
involving ${\cal V}_c\times{\cal V}_c$ (including relativistic corrections
to the free propagator) and ${\cal V}_c\times{\cal
V}_r$ in their notation---yields for the classical piece
\begin{equation}
 {\rm Amp}_{MS}^{(2)}(\vec{q}) ={5\over 2m}\alpha^2S
\end{equation}
and agrees with the iterated Feinberg-Sucher amplitude Eq.
(\ref{eq:jk}) when we set $m_a=m_b=m$.  Note that Manohar and Stewart
also emphasize the nonuniqueness of the classical potential, which
they associate in part with the existence of off-shell terms of the
lowest order potentials such as
$${\cal V}_{\Delta 2}(\vec{q})\propto {(\vec p_f^{\hspace*{1.4pt}2} - \vec p_i^{\hspace*{1.4pt}2})^2\over 4m^2\vec{q}^{\hspace*{1.4pt}4}}$$
which vanish on-shell. Since we perform the iteration using on-shell
potentials these terms do not contribute, but they
can contribute in some forms such as those used by Iwasaki \cite{iwa} and in
the Coulomb inspired form of Sucher \cite{sp}.

However, despite the agreement of many forms, the corrections which
we examine are often higher order in the relative velocity than
included in the Manohar-Stewart vNRQED exposition of \cite{sm}. Thus in the case
of the spin-orbit term, the one loop corrections which
we consider are order $v$ in the Manohar-Stewart expansion and are
therefore outside the quoted pieces of their potential. Likewise,
the spin-spin correlation corrections which we retain are order $v$
in vNRQED and again are not found in the Manohar-Stewart expressions of \cite{sm}.

\section{Spin-Dependent Scattering: Spin-Orbit Interaction}
\subsection{Spin-0 -- Spin-1/2}

Having determined the form of the scattering amplitude and the resulting
potential for the spinless scattering case we move on to the case of scattering
of particles carrying spin.  We begin with the scattering of a spinless
particle $a$ from a spin-1/2 particle $b$.  From the Dirac
Lagrangian density
\begin{equation}
{\cal L}=\bar{\psi}(x)(i\not\!\!{D}-m)\psi(x)
\end{equation}
we determine the one- and two-photon vertices for a spin-1/2 particle to be
\begin{eqnarray}
\tau^{(1)}_\mu(p_4,p_3)&=&-ie\gamma_\mu\nonumber\\
\tau^{(2)}_{\mu\nu}(p_4,p_3)&=&0\label{eq:vt}
\end{eqnarray}
and the resulting transition amplitude at tree level is found to be
\begin{equation}
{}^{1\over 2}{\cal M}^{(1)}(q)={4\pi\alpha\over q^2}{1\over m_a} \,
\bar{u}(p_4) \! \not\!{p}_1 u(p_3) \, \sqrt{m_a^2m_b^2\over
E_1E_2E_3E_4}
\end{equation}
where our spinors are normalized as $\bar u(p) \hspace*{1pt} u(p) =
1$. Defining the spin vector as
\begin{equation}
S_b^\mu={1\over 2}\bar{u}(p_4)\gamma_5\gamma^\mu u(p_3)
\end{equation}
where $\gamma_5=-i\gamma^0\gamma^1\gamma^2\gamma^3$, we find the
identity\footnote{Here $\epsilon^{0123}$ is taken to be +1.}
\begin{equation}
\bar{u}(p_4)\gamma_\mu u(p_3)=\left({1\over 1-{q^2\over
4m_b^2}}\right)\left[{(p_3+p_4)_\mu\over
2m_b}\bar{u}(p_4)u(p_3)-{i\over
m_b^2}\epsilon_{\mu\beta\gamma\delta}q^\beta p_3^\gamma
S_b^\delta\right]\label{eq:id}
\end{equation}
whereupon the nonanalytic part of the transition amplitude can be
written in the form
\begin{equation}
{}^{1\over 2}{\cal M}^{(1)}(q)= {4\pi\alpha\over
q^2}\left[\bar{u}(p_4)u(p_3)+{i\over
m_am_b^2}\epsilon_{\alpha\beta\gamma\delta}p_1^\alpha p_3^\beta
q^\gamma S_b^\delta \right].
\end{equation}
Now we again give the nonrelativistic amplitude in the symmetric
center of mass frame ($\vec p_1 = - \vec p_3 = \vec p + \vec q /2$)
where
\begin{equation}
S_b^\alpha\stackrel{NR}{\longrightarrow}(0,\vec{S}_b) \ \ \ \ \ {\rm
with} \ \ \ \ \ \vec{S}_b = {1\over
2}\chi_f^{b\dagger}\vec{\sigma}\chi_i^b,
\end{equation}
\begin{equation}
\bar{u}(p_4)u(p_3) \stackrel{NR}{\longrightarrow}
\chi_f^{b\dagger}\chi_i^b-{i\over 2m_b^2}\vec{S}_b\cdot \vec{p}
\times\vec{q} \label{eq:nr}
\end{equation}
and
\begin{equation} \label{eq:epsireduce}
\epsilon_{\alpha\beta\gamma\delta}p_1^\alpha p_3^\beta q^\gamma
S_b^\delta\stackrel{NR}{\longrightarrow}(m_a+m_b)\left(1+\frac{\vec
p^{\hspace*{1.4pt}2}}{2 m_a
m_b}\right)\vec{S}_b\cdot\vec{p}\times\vec{q},
\end{equation}
so that we find
\begin{equation}
{}^{1\over 2}{\cal M}^{(1)}(\vec{q})\simeq -{4\pi\alpha\over
\vec{q}^{\hspace{1.4pt} 2}}\left[\chi_f^{b\dagger}\chi_i^b+{i(m_a+2m_b)\over
2m_am_b^2}\vec{S}_b\cdot\vec{p}\times\vec{q} \, \right]
\label{eq:ampl10h}
\end{equation}
whereby the lowest order potential becomes
\begin{eqnarray}
{}^{1\over 2}V^{(1)}(\vec r) &=&-\int {d^3q\over (2\pi)^3} \,
\hspace*{1pt}
{}^{1\over 2}{\cal M}^{(1)}(\vec{q}) \, e^{-i\vec{q}\cdot\vec{r}} \nonumber\\
&\simeq&{\alpha\over r}\chi_f^{b\dagger}\chi_i^b-{m_a+2m_b \over
2m_am_b^2} \, \vec{S}_b\cdot\vec{p}\times\vec{\nabla}{\alpha\over
r}\nonumber\\
&\simeq&{\alpha\over r}\chi_f^{b\dagger}\chi_i^b-{\alpha\over
r^3}{m_a+2m_b\over 2m_am_b^2}\, \vec{L}\cdot\vec{S}_b\label{eq:oj}
\end{eqnarray}
where $\vec{L}=\vec{r}\times\vec{p}$ is the angular momentum and
$\vec r \equiv \vec r_a - \vec r_b$---the modification of the
leading spin-independent potential has a spin-orbit character.

When evaluating the one loop corrections we encounter an additional
complication: The calculation contains two independent kinematic
variables, the momentum transfer $q^2$ and $s - s_0$ which is to
leading order proportional to $p_0^2$ (where $p_0 \equiv
|\vec{p}_i|, \  i=1,2,3,4$) in the center of mass frame. 
We find that our results differ if we perform an
expansion first in $s - s_0$ and then in $q^2$ or vice versa. This
ordering issue only occurs for the box diagram, diagram (d) of 
Fig. \ref{fig_diags}, where it
stems from the reduction of vector and tensor box integrals. Their
reduction in terms of scalar integrals involves the inversion of a
matrix whose Gram determinant vanishes in the nonrelativistic threshold 
limit $q^2, s - s_0 \rightarrow 0$. More precisely, the denominators or 
the vector and tensor box integrals (see Appendix \ref{app_int}) involve a factor of 
$(4 p_0^2 - \vec q^{\hspace*{1.4pt} 2})$ when expanded in the 
nonrelativistic limit. Since $q^{\hspace*{1.4pt} 2} = 4 p_0^2 \sin^2 \frac{\theta}{2}$
with $\theta$ the scattering angle, we notice that $4 p_0^2 > \vec q^{\hspace*{1.4pt} 2}$ 
unless we consider backward scattering where $\theta = \pi$ and where the 
scattering amplitude diverges. And since $p_0^2$ originates from the relativistic 
structure $s - s_0$, we therefore have to first expand our vector
and tensor box integrals in $q^2$ and then in $s - s_0$.
Evaluating the diagrams (a)-(e) of Fig. \ref{fig_diags} and we find the results
\begin{eqnarray}
{}^{1\over 2}{\cal M}_{\ref{fig_diags}a}^{(2)}(q) \hspace*{-4pt} &= \hspace*{-4pt} &0\nonumber\\
{}^{1\over 2}{\cal M}_{\ref{fig_diags}b}^{(2)}(q) \hspace*{-4pt} &= \hspace*{-4pt} &0\nonumber\\
{}^{1\over 2}{\cal M}_{\ref{fig_diags}c}^{(2)}(q) \hspace*{-4pt} &= \hspace*{-4pt} &{\alpha^2\over m_am_b}\left[\bar{u}(p_4) u(p_3) \, S m_b\right]\nonumber\\
{}^{1\over 2}{\cal M}_{\ref{fig_diags}d}^{(2)}(q) \hspace*{-4pt} &= \hspace*{-4pt} &{\alpha^2\over
m_am_b}\left[\bar{u}(p_4)u(p_3)\left(-S{m_am_b(m_a+2m_b)\over s-s_0} \right.\right.\nonumber\\
&&\left. \left. \hspace*{100pt} - \hspace*{0.9pt} S m_b - L \, \frac{2m_a^2+3m_am_b-2m_b^2}{6m_am_b}\right)\right.\nonumber\\
&& \hspace*{33pt} + \left.{1\over m_a}\bar{u}(p_4) \! \not
\!{p}_1u(p_3)\left(4L{m_am_b\over q^2} +S{m_am_b(m_a+2m_b)\over
s-s_0}\right.\right.\nonumber\\
&& \hspace*{133pt} + \hspace*{-2.5pt} \left.\left.S(m_a \hspace*{-0.5pt} + \hspace*{-0.5pt} m_b) \hspace*{-0.6pt} + \hspace*{-0.6pt} L \, \frac{10m_a^2 \hspace*{-0.5pt} + \hspace*{-0.5pt} 11m_b^2}{12m_am_b} \hspace*{-0.2pt}\right) {}\hspace*{-3pt} {}\right]\nonumber\\
&&- \ i 4\pi \alpha^2 {L\over q^2}\sqrt{m_a m_b\over s-s_0}\ {1\over m_a}\bar{u}(p_4) \! \not \!{p}_1u(p_3)\nonumber\\
{}^{1\over 2}{\cal M}^{(2)}_{\ref{fig_diags}e}(q) \hspace*{-4pt} &= \hspace*{-4pt} &{\alpha^2\over
m_am_b}\left[\bar{u}(p_4)u(p_3)
\left(S \frac{m_a-2m_b}{4}+L{2m_a^2-3m_am_b -2m_b^2\over 6m_am_b}\right)\right.\nonumber\\
&&\hspace*{33pt}+ \left.{1\over m_a}\bar{u}(p_4) \! \not \!{p}_1u(p_3)\left(-4L{m_am_b\over q^2} - S \frac{3m_a+2m_b}{4} \right.\right.\nonumber\\
&&\hspace*{133pt} \left.\left.- L \, {10m_a^2+16m_am_b+11m_b^2\over
12m_am_b}\right) {}\hspace*{-3pt} {} \right]
\end{eqnarray}
Summing, we determine
\begin{eqnarray}
{}^{1\over 2}{\cal M}_{tot}^{(2)}(q) \hspace*{-4pt}
&= \hspace*{-4pt} &{\alpha^2\over m_am_b}\Bigg[L\left(-\bar{u}(p_4)u(p_3)-{4\over
3}{1\over m_a
}\bar{u}(p_4) \! \not\!{p}_1u(p_3)\right) \nonumber\\
&&\hspace*{32pt}+\hspace*{-2.3pt}\left.S\hspace*{-2pt}\left(\frac{m_a\hspace*{-1.4pt}-\hspace*{-1.1pt}2m_b}{4}
\, \bar{u}(p_4)u(p_3) \hspace*{-0.4pt} + \hspace*{-0.4pt} \frac{m_a
\hspace*{-1.4pt}+\hspace*{-1.1pt}2m_b}{4} {1\over m_a}
\bar{u}(p_4) \hspace*{-1.3pt} \! \not\!{p}_1u(p_3)\! \hspace*{-1.6pt} \right)\right.\nonumber\\
&&\hspace*{32pt}-
{(m_a\hspace*{-1pt}+\hspace*{-1pt}2m_b)m_am_bS\over
s-s_0}\hspace*{-1pt}\left(\bar{u}(p_4)u(p_3)-{1\over
m_a}\bar{u}(p_4) \! \not\!{p}_1u(p_3)\hspace*{-1pt}\right)\hspace*{-1pt}\Bigg]\nonumber\\
&&- \ i 4\pi \alpha^2 {L\over q^2}\sqrt{m_a m_b\over s-s_0}\ {1\over
m_a}\bar{u}(p_4) \! \not \!{p}_1u(p_3) . \label{eq:kf}
\end{eqnarray}
Using the identity Eq. (\ref{eq:id}) and
$$p_1\cdot(p_3+p_4)=2m_am_b+s-s_0+{q^2\over 2}$$
Eq. (\ref{eq:kf}) becomes
\begin{eqnarray}
{}^{1\over 2}{\cal M}_{tot}^{(2)}(q)\hspace*{-4pt}{}&= \hspace*{-4pt}&
{\alpha^2\over m_am_b}\left[L\left(-{7\over 3}\bar{u}(p_4)u(p_3) -
{4i\over 3m_am_b^2}\epsilon_{\alpha\beta\gamma\delta}
p_1^\alpha p_3^\beta q^\gamma S_b^\delta\right)\right.\nonumber\\
&& \hspace*{33pt}+\hspace*{-1.5pt}\left.S \! \left(\!\!(m_a
\hspace*{-1.3pt} + \hspace*{-1.3pt}
m_b)\bar{u}(p_4)u(p_3)\hspace*{-1pt}+\hspace*{-1pt}
{i(m_a\hspace*{-1.3pt}+\hspace*{-1.3pt}2m_b)\over
4m_am_b^2}\epsilon_{\alpha\beta\gamma\delta}
p_1^\alpha p_3^\beta q^\gamma S_b^\delta \!\right)\right.\nonumber\\
&& \hspace*{33pt}+\left.{iS(m_a+2m_b)\over
m_b(s-s_0)}\epsilon_{\alpha\beta\gamma\delta}
p_1^\alpha p_3^\beta q^\gamma S_b^\delta\right]\nonumber\\
&&-i4\pi \alpha^2 {L\over q^2}\sqrt{m_a m_b\over
s-s_0}\left(\bar{u}(p_4)u(p_3)+ {i\over
m_am_b^2}\epsilon_{\alpha\beta\gamma\delta} p_1^\alpha p_3^\beta
q^\gamma S_b^\delta\right) . \label{eq:oh}
\end{eqnarray}
Finally, working in the symmetric center of mass frame and taking
the nonrelativistic limit using Eqs. (\ref{eq:nr}),
(\ref{eq:epsireduce}) and
\begin{equation}
 \frac{1}{s-s_0} \stackrel{NR}{\longrightarrow} \frac{m_a m_b}{(m_a + m_b)^2} \frac{1}{p_0^2} + \frac{(m_a - m_b)^2}{4 m_a m_b (m_a + m_b)^2}
\end{equation}
we find
\begin{eqnarray}
{}^{1\over 2}{\cal M}_{tot}^{(2)}(\vec q)&\simeq&
\left[{\alpha^2\over m_am_b}\left((m_a+m_b)S-{7\over3}L\right) - i 4
\pi \alpha^2 \frac{L}{q^2} \frac{m_r}{p_0}\right]
\chi_f^{b\dagger}\chi_i^b \nonumber\\
&+&\Bigg[{\alpha^2 \over m_a m_b} \left(\frac{m_a^2 + 2 m_a m_b + 2
m_b^2}{2 m_a (m_a + m_b)} \, S
- \frac{m_a + 8 m_b}{6 m_a m_b} \, L\right)\nonumber\\
&&+ \frac{\alpha^2 (m_a + 2 m_b)}{(m_a + m_b)} \left(- i \frac{2 \pi
L}{p_0 q^2} + \frac{S}{p_0^2} \right)\Bigg] {i\over
m_b}\vec{S}_b\cdot\vec{p}\times\vec{q}\label{eq:fi}
\end{eqnarray}
We note from Eq. (\ref{eq:fi}) that the scattering amplitude
consists of two pieces---a spin-independent component proportional
to $\chi_f^{b\dagger}\chi_i^b$ whose functional form
\begin{equation}
{\alpha^2\over m_am_b}\left[(m_a+m_b)S-{7\over3}L\right] - i 4 \pi
\alpha^2 \frac{L}{q^2} \frac{m_r}{p_0}
\end{equation}
 is {\it identical}
to that of spinless scattering---together with a spin-orbit
component proportional to
$${i\over m_b}\vec{S}_b\cdot\vec{p}\times\vec{q}$$
whose functional form is
\begin{eqnarray}
&&{\alpha^2 \over m_a m_b} \left(\frac{m_a^2 + 2 m_a m_b + 2
m_b^2}{2 m_a (m_a + m_b)} \, S
- \frac{m_a + 8 m_b}{6 m_a m_b} \, L\right)\nonumber\\
&+& \frac{\alpha^2 (m_a + 2 m_b)}{(m_a + m_b)} \left(- i \frac{2 \pi
L}{p_0 q^2} + \frac{S}{p_0^2} \right)\label{eq:df}
\end{eqnarray}
We note in Eq. (\ref{eq:df}) the presence of an imaginary final
state rescattering term proportional to $i/p_0$ as before together
with a completely new type of kinematic form, proportional to
$1/p_0^2$ which diverges at threshold. Unlike the term proportional
to $i/p_0$, the spin-dependent piece proportional to $1/p_0^2$ or
proportional to $1/s-s_0$ is intriguing in that it is not imaginary
but real and therefore {\it does} contribute to observables.
(However, for bound state systems where the virial theorem
$\frac{1}{2} m v^2 \sim \frac{\alpha}{r}$ holds, this piece seems to
be of the same order as the leading order $\mathcal O(\alpha)$
spin-orbit piece.) Such a kinematic form has been seen before by
other researchers when looking at spin-dependent scattering.  It
appears in the form
$${1\over p_0^2 \cos^2{\theta\over 2}}={1\over p_0^2-{1\over 4}\vec{q}^{\hspace*{1.4pt} 2}}$$
and has been previously identified by Feinberg and Sucher in their
evaluation of spin-0 -- spin-1/2 scattering \cite{fs2} and by Manohar and Stewart \cite{sm}.
(Note that in our situation, since kinematics guarantees that $\vec{q}^{\hspace*{1.4pt} 2}\leq
4p_0^2$ we can expand via
$${1\over p_0^2-{1\over 4}\vec{q}^{\hspace*{1.4pt} 2}}={1\over p_0^2}\left(1+{\vec{q}^{\hspace*{1.4pt} 2}\over 4p_0^2}
+{\vec{q}^{\hspace*{1.4pt} 4}\over 16p_0^4}+\ldots\right)$$ and drop the terms higher
order in $\vec{q}^{\hspace*{1.4pt} 2}/4p_0^2$ since after Fourier-transforming, such
terms are higher order in $1/r^2$ and are therefore shorter distance
than the terms which we retain.) In any case, the presence of {\it
either} of these two forms proportional to $i/p_0$ and $1/p_0^2$
would prevent us from writing down a well defined second order
potential.

The solution to this problem is, as before, to properly subtract the
iterated first order potential
\begin{eqnarray}
{}^{1\over 2}{\rm Amp}­_C^{(2)}(\vec q) &=&- \int{d^3\ell\over
(2\pi)^3} \, \frac{\left<\vec p_f \left| {}^{\frac{1}{2}} \hat
V^{(1)}_C \right| \vec \ell \, \right> \left<\vec \ell \left|
{}^{\frac{1}{2}} \hat V^{(1)}_C \right| \vec p_i
\right>}{{p_0^2\over 2m_r} - \frac{\ell^2}{2 m_r} + i \epsilon}
\end{eqnarray}
where we now use the one-photon exchange potential ${}^{1\over
2}V_C^{(1)}(\vec r)$ given in Eq. (\ref{eq:oj}).  Splitting this
lowest order potential into spin-independent and spin-dependent
components---
\begin{equation}
\left<\vec p_f \left| {}^{\frac{1}{2}} \hat V^{(1)}_C \right| \vec
p_i \, \right> =
 \left<\vec p_f \left| {}^{\frac{1}{2}} \hat V^{(1)}_{S-I} \right| \vec p_i \, \right>
 + \left<\vec p_f \left| {}^{\frac{1}{2}} \hat V^{(1)}_{S-O} \right| \vec p_i \, \right>
\end{equation}
where
\begin{eqnarray}
 \left<\vec p_f \left| {}^{\frac{1}{2}} \hat V^{(1)}_{S-I} \right| \vec p_i \, \right>
 &=&{e^2\over \vec{q}^{\hspace*{1.4pt}2}} \, \chi_f^{b\dagger}\chi_i^b
  = \frac{e^2}{(\vec p_i - \vec p_f)^2} \, \chi_f^{b\dagger}\chi_i^b \nonumber\\
 \left<\vec p_f \left| {}^{\frac{1}{2}} \hat V^{(1)}_{S-O} \right| \vec p_i \, \right>
 &=&{e^2\over \vec{q}^{\hspace*{1.4pt}2}}{m_a+2m_b\over 2m_am_b}\, \frac{i}{m_b}\vec{S}_b\cdot\vec{p}\times\vec{q} \nonumber\\
 &=&{e^2\over (\vec p_i - \vec p_f)^2}{m_a+2m_b\over 2m_am_b}\, \frac{i}{m_b}\vec{S}_b\cdot \frac{1}{2} (\vec p_i + \vec p_f) \times (\vec p_i - \vec p_f) \nonumber\\
\end{eqnarray}
we find that the iterated amplitude splits also into
spin-independent and spin-dependent pieces.  The leading
spin-independent amplitude is
\begin{eqnarray}
{}^{1\over 2}{\rm Amp}^{(2)}_{S-I}(\vec q) \hspace*{-3pt}{} &=&-
\int{d^3\ell\over (2\pi)^3} \, \frac{\left<\vec p_f \left|
{}^{\frac{1}{2}} \hat V^{(1)}_{S-I} \right| \vec \ell \, \right>
\left<\vec \ell \left| {}^{\frac{1}{2}}
 \hat V^{(1)}_{S-I} \right| \vec p_i \right>}{\frac{p_0^2}{2 m_r} -
 \frac{\ell^2}{2 m_r} + i \epsilon} \nonumber\\
&=&i \sum_{s_\ell} \int {d^3\ell\over (2\pi)^3} {e^2
\chi_f^{b\dagger}\chi_{s_\ell}^b \over
|\vec{p}_f-\vec{\ell}|^2+\lambda^2}{i\over {p_0^2\over
2m_r}-{\ell^2\over 2m_r}+i\epsilon}{e^2
\chi_{s_\ell}^{b\dagger}\chi_i^b \over
|\vec{\ell}-\vec{p}_i|^2+\lambda^2}\nonumber\\
&\stackrel{\lambda\rightarrow 0}
{\longrightarrow}&\chi_f^{b\dagger}\chi_i^b H =-i4\pi
\alpha^2{L\over q^2}\frac{m_r}{p_0}\chi_f^{b\dagger}\chi_i^b
\label{eq:iteration0hSI}
\end{eqnarray}
and the leading spin-dependent term is
\begin{eqnarray}
{}^{1\over 2}{\rm Amp}^{(2)}_{S-O}(\vec q) \hspace*{-10pt} &= \hspace*{-10pt} &-
\int{d^3\ell\over (2\pi)^3} \, \frac{\left<\vec p_f \left|
{}^{\frac{1}{2}} \hat V^{(1)}_{S-I}
 \right| \vec \ell \, \right> \left<\vec \ell \left| {}^{\frac{1}{2}}
 \hat V^{(1)}_{S-O} \right| \vec p_i \right>}{\frac{p_0^2}{2 m_r} -
 \frac{\ell^2}{2 m_r} + i \epsilon} \nonumber\\
&\hspace*{-5pt}&- \int{d^3\ell\over (2\pi)^3} \, \frac{\left<\vec p_f \left|
{}^{\frac{1}{2}} \hat V^{(1)}_{S-O} \right| \vec \ell \, \right>
\left<\vec \ell \left| {}^{\frac{1}{2}} \hat V^{(1)}_{S-I} \right|
\vec p_i \right>}{\frac{p_0^2}{2 m_r} -
\frac{\ell^2}{2 m_r} + i \epsilon} \nonumber\\
&=\hspace*{-10pt}&{i(m_a+2m_b)\over 2 m_a m_b^2}
\vec{S}_b \cdot \nonumber\\
&\hspace*{-9pt}& \left(\hspace*{-2pt}i \hspace*{-3.2pt} \int \hspace*{-3.2pt}
{d^3\ell\over (2\pi)^3} {e^2\over |\vec{p}_f \hspace*{-1.1pt} -
\hspace*{-1.1pt} \vec{\ell} \hspace*{1pt}|^2 \hspace*{-1.1pt} +
\hspace*{-1.2pt} \lambda^2}{i \over {p_0^2\over 2m_r}
\hspace*{-1.1pt} - \hspace*{-1.1pt} {\ell^2\over 2m_r}
\hspace*{-1.1pt} + \hspace*{-1.1pt} i\epsilon}{e^2 \,
\frac{1}{2}(\vec p_i \hspace*{-1.1pt} + \hspace*{-1.1pt} \vec \ell)
\hspace*{-2.5pt} \times \hspace*{-2.5pt} (\vec p_i \hspace*{-1.1pt}
- \hspace*{-1.1pt} \vec \ell)\over |\vec{\ell} \hspace*{-1.1pt} -
\hspace*{-1.1pt} \vec{p}_i|^2 \hspace*{-1.1pt}
 + \hspace*{-1.2pt} \lambda^2}\right.\nonumber\\
&\hspace*{-9pt}&\left. \hspace*{-3.6pt}+ i \hspace*{-3.2pt} \int \hspace*{-3.2pt}
{d^3\ell\over (2\pi)^3} {e^2 \, \frac{1}{2}(\vec \ell
\hspace*{-1.3pt} + \hspace*{-1.4pt} \vec p_f) \hspace*{-2.7pt}
\times \hspace*{-2.7pt} (\vec \ell \hspace*{-1.3pt} -
\hspace*{-1.4pt} \vec p_f)\over |\vec{p}_f \hspace*{-1.1pt} -
\hspace*{-1.1pt} \vec{\ell} \hspace*{1pt}|^2 \hspace*{-1.1pt} +
\hspace*{-1.2pt} \lambda^2}{i \over {p_0^2\over 2m_r}
\hspace*{-1.1pt} - \hspace*{-1.1pt} {\ell^2\over 2m_r}
\hspace*{-1.1pt} + \hspace*{-1.1pt} i\epsilon}{e^2\over |\vec{\ell}
\hspace*{-1.1pt} - \hspace*{-1.1pt} \vec{p}_i|^2 \hspace*{-1.1pt}
 + \hspace*{-1.2pt}\lambda^2}\hspace*{-3pt}\right)\nonumber\\
&\stackrel{\lambda\rightarrow 0}{\longrightarrow}\hspace*{-8pt}&{}\, {i(m_a+2m_b)\over
2 m_a m_b^2} \, \vec{S}_b\cdot\vec{H}\times
\vec{q}\nonumber\\
&=\hspace*{-10pt}&\frac{\alpha^2 (m_a + 2 m_b)}{(m_a + m_b)} \left(- i \frac{2 \pi
L}{p_0 q^2} + \frac{S}{p_0^2} \right) {i\over
m_b}\vec{S}_b\cdot\vec{p}\times\vec{q} \label{eq:iteration0hSO}
\end{eqnarray}
which we have checked against Manohar and Stewart's expression for the iteration
amplitude to this order \cite{sm} setting $m_a = m_b = m$.
In principle we would also have to iterate the leading order
spin-orbit piece twice.  However this procedure yields only terms
higher order in $p^2$.  We observe that when the amplitudes Eqs.
(\ref{eq:iteration0hSO}) and (\ref{eq:iteration0hSI}) are subtracted
from the full one loop scattering amplitude Eq. (\ref{eq:df}) both the
terms involving $1/p_0^2$ and those proportional to $i/p_0$
disappear leaving behind a well-defined second order potential
\begin{eqnarray}
{}^{1\over 2}V^{(2)}_C(\vec r)&=&-\int{d^3q\over
(2\pi)^3}e^{-i\vec{q}\cdot\vec{r}}\left[{}^{1\over 2}{\cal
M}_{tot}^{(2)}(\vec{q})-{}^{1\over 2}{\rm
Amp}_C^{(2)}(\vec{q})\right]\nonumber\\
&=& - \frac{\alpha^2}{m_a m_b} \int{d^3q\over
(2\pi)^3}e^{-i\vec{q}\cdot\vec{r}}
\Bigg[\left((m_a+m_b)S-{7\over3}L\right)
\chi_f^{b\dagger}\chi_i^b\nonumber\\
&& {}\hspace*{15pt} + \bigg(\frac{m_a^2 + 2 m_a m_b + 2 m_b^2}{2 m_a
(m_a + m_b)} \, S - \frac{m_a + 8 m_b}{6 m_a m_b} \, L\bigg) {i\over
m_b}\vec{S}_b\cdot\vec{p}\times\vec{q}\Bigg]\nonumber\\
&=&\left[- {\alpha^2(m_a+m_b)\over 2m_am_br^2} -
{7\alpha^2\hbar\over 6\pi m_am_br^3}\right]
\chi_f^{b\dagger}\chi_i^b\nonumber\\
&+&{1\over m_b}\vec{S}_b\cdot \vec{p}\times\vec{\nabla}\left[
{\alpha^2(m_a^2+2m_a m_b+2m_b^2)\over 4m_a^2 m_b
(m_a+m_b)r^2}+{\alpha^2(m_a+8m_b)\hbar\over 12\pi
m_a^2m_b^2r^3}\right]\nonumber\\
&=&\left[- {\alpha^2(m_a+m_b)\over 2m_am_br^2} -
{7\alpha^2\hbar\over 6\pi m_am_br^3}\right]
\chi_f^{b\dagger}\chi_i^b\nonumber\\
&+&\left[{\alpha^2(m_a^2+2m_am_b+2m_b^2)\over
2m_a^2m_b^2(m_a+m_b)r^4}+{\alpha^2(m_a+8m_b)\hbar\over 4\pi
m_a^2m_b^3r^5}\right] \vec{L}\cdot\vec{S}_b \label{eq:mn}
\end{eqnarray}
We observe then that the second order potential for long range
Coulomb scattering of a spinless and a spin-1/2 particle has one
component which is independent of the spin of particle b and which
is identical to the potential found for the case of spinless
scattering, accompanied by a spin-orbit interaction with a new form
for its classical and quantum potentials.

Finally we would like to note that vertex corrections on the side of
the spin-1/2 particle, particle $b$, give corrections to the
g-factor of the spin-1/2 particle altering the tree level value
$g_b^{(0)} = 2$ to its $\mathcal O(\alpha)$ corrected value
$g_b^{(1)} = 2 + \frac{\alpha}{\pi}$. Since the g-factor is
implicitly a parameter of the spin-orbit coupling piece of our
leading order amplitude Eq. (\ref{eq:ampl10h}) and potential Eq.
(\ref{eq:oj}) the vertex correction which we have neglected will
yield a long range contribution of $\mathcal O(\alpha^2)$ with the
same distance dependence as the leading order contributions
proportional to $\vec L \cdot \vec S_b/r^3$. We will neglect these
contributions for now and include them later by writing our results
for particles with arbitrary charges and g-factors in Appendix \ref{app_general}.
Therefore by using the physical values of the mass, charge and
g-factor of the scattered particles it is sufficient to only
consider the two-photon exchange diagrams displayed in Fig.
\ref{fig_diags} when considering the leading long distance
corrections.

\subsection{Spin-0 -- Spin-1}

It is tempting to speculate that if we extend our considerations to
higher spin then this pattern continues---a spin-independent
component identical to the spin-0 -- spin-0 potential, accompanied
by a spin-orbit interaction which is the same for all spins, plus
additional terms which have no spin-0 or spin-1/2 analog.  In order
to test this hypothesis, we move to spin-0 -- spin-1 scattering, and
we take the spin-1 particle to be a $W^+$ boson. In order to
determine the correct interaction vertices we must recall that the
electroweak interaction is a non-abelian gauge theory \cite{gau}.
This means that the spin one Lagrangian which contains the charged-W
has the Proca form---
\begin{equation}
{\cal L}=-{1\over 4}(\vec{U}_{\mu\nu})^2+{m^2\over 2}\vec{U}_\mu^2
\end{equation}
but the SU(2) field tensor $\vec{U}_{\mu\nu}$ contains an additional
term on account of the required gauge invariance \cite{gau}
\begin{equation}
\vec{U}_{\mu\nu}=D_\mu\vec{U}_\nu-D_\nu\vec{U}_\mu-ic_{SU(2)}\vec{U}_\mu\times\vec{U}_\nu
\end{equation}
where $c_{SU(2)}$ is the SU(2) electroweak coupling constant.  This
additional term in the field tensor is responsible for the
interactions involving three and four W-bosons and for an ``extra''
interaction term which has the form of an anomalous magnetic moment
and, when added to the simple Proca moment, increases the predicted
gyromagnetic ratio from its naive value---$g_{W^\pm}^{\rm
naive}=1$---to its standard model value---$g_{W^\pm}^{\rm sm}=2$. As
discussed in \cite{Holstein:2006wi} there are various theoretical
reasons why elementary particles have a g-factor $g = 2$. The
resulting one- and two-photon vertices are then found to be
\begin{eqnarray}
\tau^{(1)}_{\mu, \beta \alpha}(p_4,p_3) &=& ie\left[\eta_{\alpha
\beta}(p_4+p_3)_\mu - \eta_{\mu \beta} (2 p_4 - p_3)_\alpha
 - \eta_{\mu \alpha}(2p_3 - p_4)_\beta\right] \nonumber \\
\tau^{(2)}_{\mu\nu, \beta \alpha}(p_4,p_3) &=& -ie^2 \left[2
\eta_{\mu\nu} \eta_{\alpha \beta} - \eta_{\mu \alpha} \eta_{\nu
\beta} - \eta_{\mu \beta} \eta_{\nu \alpha}\right]
\end{eqnarray}
for an incoming massive spin-1 particle with momentum $p_3$ and
Lorentz index $\alpha$ and the outgoing one with momentum $p_4$ and
Lorentz index $\beta$.

We assign particle $b$ to be the massive spin-1 particle with the
incoming polarization vector $\epsilon_i^b$ satisfying
$\epsilon_i^b\cdot p_3=0$ and the polarization vector $\epsilon_f^b$
satisfying $\epsilon_f^b\cdot p_4=0$. The lowest order scattering
amplitude then has the form
\begin{eqnarray}
{}^1{\cal M}^{(1)}(q)={8\pi\alpha\over \sqrt{2E_12E_22E_32E_4}} \!
\! \! \! \!
&\bigg[& \! \! \! \! \! \frac{s-m_a^2-m_b^2+\frac{1}{2}q^2}{q^2} \left(- \epsilon_f^{b*}\cdot\epsilon_i^b\right)\nonumber\\
&{}\hspace*{1pt} -& \! \! \! \!  \frac{2}{q^2}
\left(\epsilon_f^{b*}\cdot q \, \epsilon_i^b\cdot
p_1-\epsilon_f^{b*}\cdot p_1 \, \epsilon_i^b\cdot q\right)\bigg]
\end{eqnarray}
Now we rewrite this expression using the identity
\begin{equation}
 \epsilon_{f\mu}^{b*} \, \epsilon_i^b\cdot q - \epsilon_{i\mu}^b \, \epsilon_f^{b*}\cdot q
 =\frac{1}{1-\frac{q^2}{4 m_b^2}}\left[\frac{i}{m_b} \hspace*{1pt} \epsilon_{\mu\beta\gamma\delta}
 \hspace*{1pt} p_3^\beta q^\gamma S_b^\delta  - \frac{(p_3+p_4)_\mu}{2 m_b^2} \hspace*{1pt}
 \epsilon_f^{b*}\cdot q \, \epsilon_i^b \cdot q\right]\label{eq:kj}
\end{equation}
where we have defined the spin vector
\begin{equation}
S_{b\mu}={i\over 2m_b} \hspace*{1pt} \epsilon_{\mu\beta\gamma\delta}
\hspace*{1pt}
\epsilon_f^{b*\beta}\epsilon_i^{b\gamma}(p_3+p_4)^\delta
\end{equation}
The leading one-photon exchange amplitude can then be written as
\begin{equation}
{}^1{\cal M}^{(1)}(q) = {4\pi\alpha\over q^2} \left[-
\epsilon_f^{b*}\cdot\epsilon_i^b + {i\over m_am_b^2} \hspace*{1pt}
\epsilon_{\alpha\beta\gamma\delta} \hspace*{1pt} p_1^\alpha
p_3^\beta q^\gamma S_b^\delta-{1\over m_b^2} \hspace*{1pt}
\epsilon_f^{b*}\cdot q \, \epsilon_i^b\cdot q \right]
\end{equation}
Now in the nonrelativistic limit we have
\begin{equation}
\epsilon_i^{b0}\simeq {1\over m_b} \hspace*{1pt}
\hat{\epsilon}_i^b\cdot\vec{p_3},\quad\epsilon_f^{b0}\simeq {1\over
m_b} \hspace*{1pt} \hat{\epsilon}_f^b\cdot\vec{p_4}
\end{equation}
so that
\begin{eqnarray}
\epsilon_f^{b*}\cdot\epsilon_i^b&\simeq&
-\hat{\epsilon}_f^{b*}\cdot\hat{\epsilon}_i^b+{1\over
m_b^2} \hspace*{1pt} \hat{\epsilon}_f^{b*}\cdot\vec{p}_4 \hspace*{1pt} \hat{\epsilon}_i^b\cdot\vec{p}_3\nonumber\\
&\simeq&-\hat{\epsilon}_f^{b*}\cdot\hat{\epsilon}_i^b+{1\over
2m_b^2}\hat{\epsilon}_f^{b*}\times\hat{\epsilon}_i^b\cdot\vec{p}_4\times\vec{p}_3\nonumber\\
&&+{1\over
2m_b^2}\left(\hat{\epsilon}_f^{b*}\cdot\vec{p}_4\hat{\epsilon}_i^b\cdot\vec{p}_3
+\hat{\epsilon}_f^{b*}\cdot\vec{p}_3\hat{\epsilon}_i^b\cdot\vec{p}_4\right)
\label{eq:ss}
\end{eqnarray}
Since
\begin{equation}
-i\hat{\epsilon}_f^{b*}\times\hat{\epsilon}_i^b= \left<1,m_f \left|
\vec{S}_b \right| 1,m_i\right>,
\end{equation}
Eq. (\ref{eq:ss}) becomes
\begin{eqnarray}
\epsilon_f^{b*}\cdot\epsilon_i^b \hspace*{-0.2pt}{}& \simeq&
-\hat{\epsilon}_f^{b*}\cdot\hat{\epsilon}_i^b-{i\over 2m_b^2}
\hspace*{1pt} \vec{S}_b\cdot\vec{p}_3\times\vec{p}_4 +{1\over
2m_b^2} \! \hspace*{-0.2pt}
\left(\hat{\epsilon}_f^{b*}\cdot\vec{p}_4 \hspace*{1pt}
\hat{\epsilon}_i^b\cdot\vec{p}_3
+\hat{\epsilon}_f^{b*}\cdot\vec{p}_3 \hspace*{1pt} \hat{\epsilon}_i^b\cdot\vec{p}_4\right) \nonumber \\
& \simeq& -\hat{\epsilon}_f^{b*}\cdot\hat{\epsilon}_i^b +{1\over
m_b^2} \hspace*{1pt} \hat{\epsilon}_f^{b*}\cdot\vec{p} \
\hat{\epsilon}_i^b\cdot\vec{p} +{i \over 2m_b^2} \hspace*{1pt}
\vec{S}_b\cdot\vec{p} \times\vec{q} -{1\over 4 m_b^2} \hspace*{1pt}
\hat{\epsilon}_f^{b*}\cdot\vec{q} \ \hat{\epsilon}_i^b\cdot\vec{q}
\nonumber \\
\end{eqnarray}
in the symmetric center of mass frame and the transition amplitude assumes the form
\begin{eqnarray}
{}^1{\cal M}^{(1)}(\vec{q}) \simeq -{4\pi\alpha\over
\vec{q}^{\hspace*{1.4pt} 2}} \!\!\!\!\!&\bigg[& \!\!\!\!\!
\hat{\epsilon}_f^{b*}\cdot\hat{\epsilon}_i^b -{1\over m_b^2}
\hspace*{1pt} \hat{\epsilon}_f^{b*}\cdot\vec{p} \
\hat{\epsilon}_i^b\cdot\vec{p}
+{i(m_a+2m_b)\over 2m_am_b^2}\hspace*{1pt} \vec{S}_b\cdot\vec{p} \times\vec{q} \nonumber\\
&-& \!\!\!\! {3\over 4 m_b^2}\hspace*{1pt}
\hat{\epsilon}_f^{b*}\cdot \vec{q}\ \hat{\epsilon}_i^b\cdot
\vec{q}\bigg]
\end{eqnarray}
The spin-independent and spin-orbit terms here are identical in form
to those found in the spin-0 -- spin-1/2 case but now are
accompanied by new terms which are quadrupole in nature, as can be
seen from the identity
\begin{eqnarray}
T^b_{cd}&\equiv&{1\over 2}\left(\hat{\epsilon}_{fc}^{b*}
\hspace*{1pt}\hat{\epsilon}_{id}^b+ \hat{\epsilon}_{ic}^b
\hspace*{1pt}\hat{\epsilon}_{fd}^{b*}\right)-{1\over
3} \hspace*{1pt} \delta_{cd} \hspace*{1pt} \hat{\epsilon}_f^{b*}\cdot\hat{\epsilon}_i^b\nonumber\\
&=& - \left<1,m_f \left|{1\over 2} (S_cS_d+S_dS_c)-{2\over
3}\delta_{cd} \right|1,m_i \right>
\end{eqnarray}
The corresponding lowest order potential is then
\begin{eqnarray}
{}^1V^{(1)}_C(\vec r) &=&-\int {d^3q\over (2\pi)^3} \, \hspace*{1pt}
{}^1{\cal M}^{(1)}(\vec{q}) \, e^{-i\vec{q}\cdot\vec{r}} \nonumber\\
&\simeq &{\alpha\over
r}\left(\hat{\epsilon}_f^{b*}\cdot\hat{\epsilon}_i^b-{1\over
m_b^2}\hat{\epsilon}_f^{b*}\cdot\vec{p} \
\hat{\epsilon}_i^b\cdot\vec{p}\right)
- {m_a+2m_b\over 2m_am_b^2}\, \vec{S}_b\cdot\vec{p} \times\vec{\nabla}{\alpha\over r}\nonumber\\
&&+ \, {3 \over 4 m_b^2} \hspace*{1pt} \hat{\epsilon}_f^{b*}\cdot\vec{\nabla} \ \hat{\epsilon}_i^b\cdot\vec{\nabla} \, {\alpha\over r} \nonumber\\
&\simeq&{\alpha\over r}
\left(\hat{\epsilon}_f^{b*}\cdot\hat{\epsilon}_i^b-{1\over m_b^2}\
\vec p : T^b : \vec p\right)
 - {\alpha\over r^3}{m_a+2m_b\over 2m_am_b^2}\vec{L}\cdot\vec{S}_b \nonumber \\
&& + \, \frac{\alpha}{r^5} \frac{9}{4 m_b^2} \ \vec r : T^b : \vec r
\end{eqnarray}
where we have defined
$$\vec{w}:T^b:\vec{s}\equiv w_cT^b_{cd}s_d$$ and
which agrees precisely with its spin-1/2 analog---Eq.
(\ref{eq:oj})---up to tensor and quadrupole corrections.

The calculation of the one loop corrections proceeds as before, but
with increased complexity due to the unit spin.  We find
\begin{eqnarray}
{}^1{\cal M}_{\ref{fig_diags}a}^{(2)}(q) \hspace*{-5pt} & = \hspace*{-5pt} &{\alpha^2\over m_am_b} \hspace*{2pt} {3\over 2}L \, \epsilon_f^{b*}\cdot\epsilon_i^b\nonumber\\
{}^1{\cal M}_{\ref{fig_diags}b}^{(2)}(q) \hspace*{-5pt} & = \hspace*{-5pt} &{\alpha^2\over m_am_b} \left[(2L+m_aS)
\left(- \epsilon_f^{b*}\cdot\epsilon_i^b + {1\over m_a^2}
\epsilon_f^{b*}\cdot
p_1 \epsilon_i^b \cdot p_1 \right.\right.\nonumber\\
&& \hspace*{103pt} - \left.{1\over 2m_a^2}(\epsilon_f^{b*}\cdot
q\epsilon_i^b\cdot
p_1 + \epsilon_f^{b*}\cdot p_1 \epsilon_i^b\cdot q) \right) \nonumber\\
&& \hspace*{33pt} + \left. (16L+7m_aS){1\over
32m_a^2}\epsilon_f^{b*}\cdot q\epsilon_i^b\cdot
q\right]\nonumber\\
{}^1{\cal M}_{\ref{fig_diags}c}^{(2)}(q) \hspace*{-5pt} & = \hspace*{-5pt} & {\alpha^2\over m_am_b}\left[-
\frac{3L+2m_bS}{2} \hspace*{1pt} \epsilon_f^{b*}\cdot\epsilon_i^b
- m_bS{3\over 4m_b^2} \epsilon_f^{b*}\cdot q\epsilon_i^b\cdot q\right]\nonumber\\
{}^1{\cal M}_{\ref{fig_diags}d}^{(2)}(q) \hspace*{-5pt} & = \hspace*{-5pt} & {\alpha^2\over m_am_b}\left[
4{m_am_b\over q^2}L \! \left(\hspace*{-1pt}-\epsilon_f^{b*}
\hspace*{-2pt} \cdot\epsilon_i^b+{1\over m_am_b} \hspace*{-1pt}
\left(\epsilon_f^{b*} \hspace*{-1pt} \cdot \hspace*{-1pt}
p_1\epsilon_i^b \hspace*{-1pt} \cdot \hspace*{-1pt}
q-\epsilon_f^{b*} \hspace*{-1pt} \cdot \hspace*{-1pt} q\epsilon_i^b \hspace*{-1pt} \cdot \hspace*{-1pt} p_1 \hspace*{-1pt} \right) \hspace*{-3pt} \right)\right.\nonumber\\
&& \hspace*{33pt} +  {S\over s-s_0}
\Bigg(\hspace*{-3pt}\left(\epsilon_f^{b*}\cdot p_1\epsilon_i^b\cdot
q-\epsilon_f^{b*}\cdot q\epsilon_i^b\cdot p_1 \right)(m_a+2m_b) \nonumber\\
&& \hspace*{79pt} + \, \epsilon_f^{b*}\cdot q\epsilon_i^b\cdot q \, {m_a(m_a+3m_b)\over 2m_b}\Bigg)\nonumber\\
&& \hspace*{33pt} + S\Bigg(\hspace*{-1pt} -
\epsilon_f^{b*}\cdot\epsilon_i^b(m_a+m_b) - \epsilon_f^{b*}\cdot
p_1\epsilon_i^b\cdot p_1 \, {1\over 2m_a} \nonumber\\
&& \hspace*{51pt} - \hspace*{1pt} \epsilon_f^{b*} \hspace*{-1pt}
\cdot \hspace*{-1pt} q \epsilon_i^b \hspace*{-1pt} \cdot
\hspace*{-1pt} p_1 \hspace*{-2pt} \left({1\over m_b}
\hspace*{-1.2pt} + \hspace*{-1.2pt} {9\over 8m_a}\right)
\hspace*{-2pt}
+ \epsilon_f^{b*} \hspace*{-1pt} \cdot \hspace*{-1pt} p_1\epsilon_i^b \hspace*{-1pt} \cdot \hspace*{-1pt} q \hspace*{-2pt} \left({1\over m_b} \hspace*{-1.2pt} + \hspace*{-1.2pt} {13\over 8m_a} \right) \nonumber\\
&&\hspace*{51pt} - \hspace*{1pt} \epsilon_f^{b*}\cdot
q\epsilon_i^b\cdot q\left(-{1\over
2m_b}+{7\over 64m_a}\right) \hspace*{-2pt} \Bigg)\nonumber\\
&&\hspace*{33pt} +
L\Bigg(-\epsilon_f^{b*}\cdot\epsilon_i^b\left(-{1\over
2}+{5m_b\over 4m_a}+{m_a\over 4m_b}\right)\nonumber\\
&&\hspace*{51pt} - \hspace*{1pt} \epsilon_f^{b*}\cdot
p_1\epsilon_i^b\cdot p_1\left({1\over m_a^2}-{4\over
3m_am_b}\right) \nonumber\\
&&\hspace*{51pt}- \hspace*{1pt} \epsilon_f^{b*}\cdot
p_1\epsilon_i^b\cdot q\left(-{95\over 48m_a^2}+{2\over
3m_am_b}-{7\over
6m_b^2}\right) \nonumber\\
&&\hspace*{51pt}- \hspace*{1pt} \epsilon_f^{b*}\cdot
q\epsilon_i^b\cdot p_1\left({47\over 48m_a^2}+{2\over
3m_am_b}+{7\over
6m_b^2}\right)\nonumber\\
&&\hspace*{51pt}- \hspace*{1pt} \epsilon_f^{b*}\cdot
q\epsilon_i^b\cdot q\left({1\over 4m_a^2}-{1\over 3m_am_b}-{1\over
6m_b^2}-{7m_a\over
24m_b^3}\right) \hspace*{-2pt} \Bigg) \Bigg] \nonumber\\
&-\hspace*{-5pt} & i 4\pi \alpha^2 {L\over q^2}\sqrt{m_a m_b \over s-s_0}
\hspace*{-1pt} \left( \hspace*{-1pt} -\epsilon_f^{b*} \hspace*{-0.3pt}\cdot
\hspace*{-0.3pt} \epsilon_i^b  \hspace*{-0.3pt} -
\hspace*{-0.3pt}{1\over m_am_b} (\epsilon_f^{b*} \hspace*{-0.3pt}
\cdot \hspace*{-0.3pt} q \epsilon_i^b \hspace*{-0.3pt} \cdot
\hspace*{-0.3pt} p_1-
\epsilon_f^{b*} \hspace*{-0.3pt} \cdot \hspace*{-0.3pt} p_1\epsilon_i^b \hspace*{-0.3pt} \cdot \hspace*{-0.3pt} q) \hspace*{-1pt} \right) \nonumber\\
{}^1{\cal M}^{(2)}_{\ref{fig_diags}e}(q) \hspace*{-5pt} & = \hspace*{-5pt} & {\alpha^2\over m_am_b}\left[
4{m_am_b\over q^2}L \! \left(\hspace*{-1pt}\epsilon_f^{b*}
\hspace*{-2pt} \cdot\epsilon_i^b - {1\over m_am_b} \hspace*{-1pt}
\left(\epsilon_f^{b*} \hspace*{-1pt} \cdot \hspace*{-1pt}
p_1\epsilon_i^b \hspace*{-1pt} \cdot \hspace*{-1pt}
q-\epsilon_f^{b*} \hspace*{-1pt} \cdot \hspace*{-1pt} q\epsilon_i^b \hspace*{-1pt} \cdot \hspace*{-1pt} p_1 \hspace*{-1pt} \right) \hspace*{-3pt} \right)\right.\nonumber\\
&& \hspace*{33pt}+ S\Bigg(\epsilon_f^{b*}\cdot\epsilon_i^b(m_a+m_b)
- \epsilon_f^{b*}\cdot
p_1\epsilon_i^b\cdot p_1{1\over 2m_a}  \nonumber\\
&& \hspace*{51pt}- \hspace*{1pt} \epsilon_f^{b*} \hspace*{-2pt}
\cdot \hspace*{-1.4pt} p_1\epsilon_i^b \hspace*{-1.4pt} \cdot
\hspace*{-1.4pt} q \hspace*{-1pt} \left({5\over 8m_a}
\hspace*{-1.4pt} + \hspace*{-1.4pt} {3\over 4m_b}\right)
\hspace*{-2.5pt} + \hspace*{-1.5pt} \epsilon_f^{b*} \hspace*{-2pt}
\cdot \hspace*{-1.4pt} q\epsilon_i^b \hspace*{-1.4pt} \cdot
\hspace*{-1.4pt} p_1 \hspace*{-2pt} \left({3\over
4m_b} \hspace*{-1.4pt} + \hspace*{-1.4pt} {9\over 8m_a}\right) \nonumber\\
&& \hspace*{51pt} -\epsilon_f^{b*}\cdot q\epsilon_i^b\cdot
q\left({7\over 64m_a}-{1\over 8m_b}+{m_a\over
8m_b^2}\right)\Bigg)\nonumber\\
&& \hspace*{33pt}-L\Bigg(-\epsilon_f^{b*}\cdot\epsilon_i^b
\left({23\over
6}+{5m_b\over 4m_a}+{m_a\over 4m_b}\right)\nonumber\\
&& \hspace*{51pt} +\epsilon_f^{b*}\cdot p_1\epsilon_i^b\cdot
p_1\left({1\over m_a^2}+{4\over
3m_am_b}\right) \nonumber\\
&& \hspace*{51pt} - \epsilon_f^{b*}\cdot q\epsilon_i^b\cdot
p_1\left({95\over 48m_a^2}+{2\over 3m_am_b}+{7\over
6m_b^2}\right) \nonumber\\
&& \hspace*{51pt} + \epsilon_f^{b*}\cdot p_1\epsilon_i^b\cdot
q\left( {47\over 48m_a^2}+{2\over 3m_am_b}+{7\over
6m_b^2}\right)\nonumber\\
&& \hspace*{51pt} + \epsilon_f^{b*}\cdot q\epsilon_i^b\cdot q\left(
{1\over 4m_a^2} +{1\over 3m_am_b}-{1\over 6m_b^2}+{7m_a\over
24m_b^3}\right) \hspace*{-3pt}\Bigg) \Bigg]
\end{eqnarray}
Combining, we find the complete one loop amplitude
\begin{eqnarray}
{}^1{\cal M}_{tot}^{(2)}(q) \hspace*{-5pt} &= \hspace*{-5pt}&{\alpha^2\over m_am_b}\Bigg[{S\over
s-s_0}\Bigg(\! - \left(\epsilon_f^{b*}\cdot q\epsilon_i^b\cdot
p_1-\epsilon_f^{b*}\cdot p_1\epsilon_i^b\cdot q\right)
(m_a+2m_b) \nonumber\\
&&\hspace*{79pt} + \hspace*{1pt} \epsilon_f^{b*}\cdot
q\epsilon_i^b\cdot
q{m_a(m_a+3m_b)\over 2m_b}\Bigg)\nonumber\\
&& \hspace*{33pt} + S\Bigg(-
\epsilon_f^{b*}\cdot\epsilon_i^b(m_a+m_b) - \epsilon_f^{b*}\cdot
q\epsilon_i^b\cdot q{m_a+m_b\over 8m_b^2} \nonumber\\
&& \hspace*{51pt} - \left(\epsilon_f^{b*}\cdot q\epsilon_i^b\cdot
p_1-\epsilon_f^{b*}\cdot p_1\epsilon_i^b\cdot q\right){m_a+2m_b\over
4m_am_b}
\Bigg) \nonumber\\
&& \hspace*{33pt}+ L\Bigg(-\epsilon_f^{b*}\cdot\epsilon_i^b
\left(-{7\over 3} \right)+{1\over 3m_b^2} \epsilon_f^{b*}\cdot
q\epsilon_i^b\cdot q \nonumber\\
&& \hspace*{51pt} + {4\over 3m_am_b}\left(\epsilon_f^{b*}\cdot
q\epsilon_i^b\cdot p_1-\epsilon_f^{b*}\cdot p_1\epsilon_i^b\cdot q
\right)\Bigg)\Bigg] \nonumber\\
&-\hspace*{-5pt}& i 4\pi \alpha^2 {L\over q^2}\sqrt{m_a m_b \over s-s_0}
\hspace*{-1pt} \left( \hspace*{-1pt} -\epsilon_f^{b*} \hspace*{-0.3pt}\cdot
\hspace*{-0.3pt} \epsilon_i^b  \hspace*{-0.3pt} -
\hspace*{-0.3pt}{1\over m_am_b} (\epsilon_f^{b*} \hspace*{-0.3pt}
\cdot \hspace*{-0.3pt} q \epsilon_i^b \hspace*{-0.3pt} \cdot
\hspace*{-0.3pt} p_1-
\epsilon_f^{b*} \hspace*{-0.3pt} \cdot \hspace*{-0.3pt} p_1\epsilon_i^b \hspace*{-0.3pt} \cdot \hspace*{-0.3pt} q) \hspace*{-1pt} \right) \nonumber\\
\quad
\end{eqnarray}
Using the identity Eq. (\ref{eq:kj}) this becomes
\begin{eqnarray}
{}^1{\cal M}_{tot}^{(2)}(q)&=&{\alpha^2\over
m_am_b}\Bigg[-\epsilon_f^{b*}\cdot\epsilon_i^b\left(-{7\over
3}L+S(m_a+m_b)\right) \nonumber\\
&&\hspace*{33pt} + {i\over
m_am_b^2}\epsilon_{\alpha\beta\gamma\delta}p_1^\alpha p_3^\beta
q^\gamma S_b^\delta\left(-{4\over 3}L+{m_a+2m_b\over 4}S\right) \nonumber\\
&& \hspace*{33pt} + {1\over m_b^2}\epsilon_f^{b*}\cdot
q\epsilon_i^b\cdot q \, \left({5\over 3}L-S\left({7\over
8}m_a+{13\over
8}m_b\right)\right)\nonumber\\
&& \hspace*{33pt} + {iS(m_a+2m_b)\over
m_b(s-s_0)}\epsilon_{\alpha\beta\gamma\delta} p_1^\alpha p_3^\beta
q^\gamma S_b^\delta\nonumber\\
&& \hspace*{33pt}-{Sm_a(m_a+m_b)\over 2 m_b (s-s_0)} \epsilon_f^{b*}\cdot q\epsilon_i^b\cdot q \Bigg]\nonumber\\
&-& i 4\pi \alpha^2 {L\over q^2}\sqrt{m_a m_b \over s-s_0}
\hspace*{-1pt} \Bigg(-\epsilon_f^{b*} \hspace*{-0.3pt}\cdot
\hspace*{-0.3pt} \epsilon_i^b  \hspace*{-0.3pt} + {i\over
m_am_b^2}\epsilon_{\alpha\beta\gamma\delta}p_1^\alpha p_3^\beta
q^\gamma S_b^\delta \nonumber\\
&& \hspace*{95pt} - {1\over m_b^2}\epsilon_f^{b*}\cdot
q\epsilon_i^b\cdot q \Bigg) \label{eq:jh}
\end{eqnarray}
Notice here that without the $\epsilon_f^{b*}\cdot
q\epsilon_i^b\cdot q$ terms, Eq. (\ref{eq:jh}) has an identical
structure to that of the case of spin-0 -- spin-1/2 scattering---Eq.
(\ref{eq:kf})---provided we substitute
$\bar{u}(p_4)u(p_3)\longrightarrow -
\epsilon_f^{b*}\cdot\epsilon_i^b$.

Finally, taking the nonrelativistic limit we find
\begin{eqnarray}
{}^1{\cal M}_{tot}^{(2)}(\vec q) \hspace*{-5pt} & \simeq \hspace*{-5pt} &
\left[\hspace*{-1pt}{\alpha^2\over
m_am_b}\hspace*{-3pt}\left(\hspace*{-3pt}(m_a \hspace*{-2pt} +
\hspace*{-1pt} m_b)S \hspace*{-1.5pt} - \hspace*{-1.5pt} {7\over3}L
\hspace*{-1pt} \right) \hspace*{-3pt} - \hspace*{-1pt} i 4 \pi
\alpha^2 \frac{L}{q^2} \frac{m_r}{p_0}\right] \hspace*{-4.2pt}
\left(\hspace*{-1.5pt}\hat{\epsilon}_f^{b*} \hspace*{-1.5pt} \cdot
\hspace*{-1.5pt} \hat{\epsilon}_i^b \hspace*{-1.5pt} -
\hspace*{-1.5pt} {1\over
m_b^2}\, \vec p \hspace*{-1pt}: \hspace*{-1.5pt} T^b \hspace*{-3pt}: \hspace*{-1.5pt}\vec p\right) \nonumber\\
&+ \hspace*{-5pt} &\Bigg[{\alpha^2 \over m_a m_b} \left(\frac{m_a^2 + 2 m_a m_b + 2
m_b^2}{2 m_a (m_a + m_b)} \, S
- \frac{m_a + 8 m_b}{6 m_a m_b} \, L\right)\nonumber\\
&&+ \frac{\alpha^2 (m_a + 2 m_b)}{(m_a + m_b)} \left(-i \frac{2 \pi
L}{p_0 q^2} + \frac{S}{p_0^2} \right)\Bigg] {i\over
m_b} \hspace*{1pt} \vec{S}_b\cdot\vec{p}\times\vec{q} \nonumber \\
& + \hspace*{-5pt} &\Bigg[{\alpha^2 \over m_a m_b} \left(- \frac{3 m_a^2 + 7 m_a m_b + 6 m_b^2}{4 (m_a + m_b)} \, S + \frac{13}{12} \hspace*{1pt} L \right) \nonumber \\
&&+ \frac{\alpha^2 m_a m_b}{2(m_a + m_b)} \left(i \frac{6 \pi L}{p_0
q^2} - \frac{S}{p_0^2}\right) \Bigg] \frac{1}{m_b^2} \, \vec q : T^b
: \vec q
\end{eqnarray}
As found in the earlier calculations, there exist terms involving
both $i/p_0$ and $1/p_0^2$ which prevent the defining of a simple
second order potential.  The solution now is well
known---subtraction of the iterated first order potential.  Since
the form of the
spin-independent---$\hat{\epsilon}_B^*\cdot\hat{\epsilon}_A$---and
spin-orbit---$\vec{S}_b\cdot\vec{p} \times\vec{q}$---terms is
identical to that found for the case of spin-1/2, it is clear that
their subtraction goes through as before and that the corresponding
pieces of the second order potential have the same form as found for
spin-1/2.  However, there are now two new pieces of the amplitude,
the quadrupole structure $\vec{q}:T^b:\vec{q}$ which multiplies
terms involving both $i/p_0$ and $1/p_0^2$ and the tensor structure
$\vec{p}:T^b:\vec{p}$ multiplying only $i/p_0$. In order to remove
these we must iterate the full first order potential including these
quadrupole and tensor components. However, we find that our simple
nonrelativistic iteration fails to remove them! We suspect the
reason to be the presence of the tensor structure
$\vec{p}:T^b:\vec{p}$ in the lowest order potential which is in some
sense a relativistic correction but which when iterated yields also
quadrupole pieces $\vec{q}:T^b:\vec{q}$. A fully relativistic
iteration should thus be performed which we will not include in this
paper. It would be interesting to investigate if the requirement to
cancel all $i/p_0$ and $1/p_0^2$ forms in the quadrupole and tensor
pieces could clarify the ambiguity in the iteration of the leading
order potential as discussed for the spinless case.

Since we did not perform the proper iteration of the quadrupole and
tensor pieces we merely include the spin-independent and spin-orbit
pieces in the resulting second order potential
\begin{eqnarray}
{}^{1}V^{(2)}_C(\vec r)\hspace*{-3pt}& = \hspace*{-3pt} &-\int{d^3q\over
(2\pi)^3}e^{-i\vec{q}\cdot\vec{r}}\left[{}^{1}{\cal
M}_{tot}^{(2)}(\vec{q})-{}^{1}{\rm
Amp}_C^{(2)}(\vec{q})\right]\nonumber\\
&= \hspace*{-3pt} &\left[- {\alpha^2(m_a+m_b)\over 2m_am_br^2} -
{7\alpha^2\hbar\over 6\pi m_am_br^3}\right]
\hat{\epsilon}_f^{b*}\cdot\hat{\epsilon}_i^b \nonumber\\
&+ \hspace*{-3pt} &{1\over m_b}\vec{S}_b \hspace*{-2pt} \cdot
\hspace*{-2pt}\vec{p}\hspace*{-1.5pt}\times\hspace*{-2.5pt}\vec{\nabla}\hspace*{-2pt}\left[
\hspace*{-1pt}{\alpha^2(m_a^2\hspace*{-2pt}+\hspace*{-2pt}2m_a
m_b\hspace*{-2pt}+\hspace*{-2pt}2m_b^2)\over 4m_a^2 m_b
(m_a\hspace*{-2pt}+\hspace*{-2pt}m_b)r^2}\hspace*{-2pt}+\hspace*{-2pt}{\alpha^2(m_a\hspace*{-2pt}+\hspace*{-2pt}8m_b)\hbar\over
12\pi
m_a^2m_b^2r^3}\hspace*{-1pt}\right] \hspace*{-3pt}+ \hspace*{-1pt}{}^{1}V^{(2)}_T(\vec r)\nonumber\\
& = \hspace*{-3pt} &\left[- {\alpha^2(m_a+m_b)\over 2m_am_br^2} -
{7\alpha^2\hbar\over 6\pi m_am_br^3}\right]
\hat{\epsilon}_f^{b*}\cdot\hat{\epsilon}_i^b \nonumber\\
& + \hspace*{-3pt} &\left[{\alpha^2(m_a^2+2m_am_b+2m_b^2)\over
2m_a^2m_b^2(m_a+m_b)r^4}+{\alpha^2(m_a+8m_b)\hbar\over 4\pi
m_a^2m_b^3r^5}\right] \vec{L}\cdot\vec{S}_b + {}^{1}V^{(2)}_T(\vec r) \nonumber\\
\quad \label{eq:NLOpot01}
\end{eqnarray}
where ${}^{1}V^{(2)}_T(\vec r)$ denotes the tensor pieces not
explicitly shown. Comparison with the corresponding form of
${}^{1\over 2}V_C^{(2)}(\vec r)$ given in Eq. (\ref{eq:mn}) confirms
the universality which we have suggested---the spin-independent and
spin-orbit terms have identical forms. The next task is to see
whether this universality applies when both scattered particles
carry spin. For this purpose we consider the case of spin-1/2 --
spin-1/2 scattering.

\section{Spin-Dependent Scattering: Spin-Spin Interaction}

In order to check universality when both scattering particles carry
spin and to study possible spin-spin interactions, we now
consider the case where both particles $a$ and $b$ carry spin-1/2.
\subsection{Spin-1/2 -- Spin-1/2}
For this calculation the vertices have been given previously and the
calculation proceeds as before.  The one-photon exchange amplitude
is given by
\begin{equation}
{}^{{1\over 2}{1\over 2}}{\cal M}^{(1)}(q)= {4\pi\alpha\over q^2} \,
\bar{u}(p_2)\gamma_\alpha u(p_1) \, \bar{u}(p_4)\gamma^\alpha
u(p_3)\sqrt{m_a^2m_b^2\over E_1E_2E_3E_4}
\end{equation}
Using the spin-1/2 identity Eq. (\ref{eq:id}) and its pendant for
particle $a$
\begin{equation}
\bar{u}(p_2)\gamma_\mu u(p_1)=\left({1\over 1-{q^2\over
4m_a^2}}\right)\left[{(p_1+p_2)_\mu\over
2m_a}\bar{u}(p_2)u(p_1)+{i\over
m_a^2}\epsilon_{\mu\beta\gamma\delta}q^\beta p_1^\gamma
S_a^\delta\right]\label{eq:ida}
\end{equation}
where we have defined the spin vector
$$S_{a}^\mu={1\over 2}\bar{u}(p_2)\gamma_5\gamma^\mu u(p_1)$$ for
particle $a$, the one-photon exchange amplitude becomes
\begin{eqnarray}
{}^{{1\over 2}{1\over 2}}{\cal
M}^{(1)}(q)\hspace*{-2pt}&=&{4\pi\alpha\over q^2}
\Bigg[\bar{u}(p_2)u(p_1)\bar{u}(p_4)u(p_3)+{i\over
m_am_b^2}\bar{u}(p_2)u(p_1)\epsilon_{\alpha\beta\gamma\delta}p_1^\alpha
p_3^\beta q^\gamma S_b^\delta \nonumber\\
&&\hspace*{23pt}+\hspace*{1pt}{i\over
m_bm_a^2}\bar{u}(p_4)u(p_3)\epsilon_{\alpha\beta\gamma\delta}p_1^\alpha
p_3^\beta q^\gamma S_a^\delta \nonumber\\
&&\hspace*{23pt}+{1\over m_am_b}(S_a\cdot qS_b\cdot q-q^2S_b\cdot
S_a)\Bigg]
\end{eqnarray}
We observe that in addition to the spin-orbit pieces found
previously, a spin-spin interaction is also present. Taking the
nonrelativistic limit and working in the center of mass frame, we
find
\begin{eqnarray}
{}^{{1\over 2}{1\over 2}}{\cal
M}^{(1)}(\vec{q})&\simeq&-{4\pi\alpha\over \vec{q}^{\hspace*{1.4pt}
2}}\Bigg[\chi_f^{a\dagger} \chi_i^a \, \chi_f^{b\dagger}\chi_i^b
+{i(m_a+2m_b)\over
2m_am_b^2}\vec{S}_b\cdot\vec{p} \times\vec{q} \, \chi_f^{a\dagger}\chi_i^a \nonumber\\
&&\hspace*{31pt}+ {i(2m_a+m_b)\over
2m_a^2m_b}\vec{S}_a\cdot\vec{p} \times\vec{q} \, \chi_f^{b\dagger}\chi_i^b\nonumber\\
&&\hspace*{31pt}+{1\over m_am_b}\left(\vec{S}_a\cdot\vec{q} \,
\vec{S}_b\cdot\vec{q}-\vec{q}^{\hspace*{1.4pt} 2}
\vec{S}_a\cdot\vec{S}_b\right) \hspace*{-1pt}\Bigg]
\label{eq:LOMnonrelhh}
\end{eqnarray}
whereby the lowest order potential for spin-1/2 -- spin-1/2
scattering is of the form
\begin{eqnarray}
{}^{}\hspace*{-11pt} {}^{{1\over 2}{1\over 2}}V_C^{(1)}(\vec
r)\hspace*{-2pt}&=& -\int {d^3q\over (2\pi)^3} \, \hspace*{1pt}
{}^{{1\over 2}{1\over 2}}{\cal M}^{(1)}(\vec{q}) \, e^{-i\vec{q}\cdot\vec{r}} \nonumber\\
&\simeq&{\alpha\over
r}\chi_f^{a\dagger}\chi_i^a\chi_f^{b\dagger}\chi_i^b -
{(m_a+2m_b)\over 2m_am_b^2}\vec{S}_b\cdot\vec{p}
\times\vec{\nabla}{\alpha\over
r} \, \chi_f^{a\dagger}\chi_i^a\nonumber\\
&&-{(2m_a+m_b)\over 2m_a^2m_b}\vec{S}_a\cdot\vec{p}
\times\vec{\nabla}{\alpha\over r} \, \chi_f^{b\dagger}\chi_i^b -
{1\over
m_am_b}\vec{S}_a\cdot\vec{\nabla}\vec{S}_b\cdot\vec{\nabla}{\alpha\over r} \nonumber\\
&\simeq&{\alpha\over
r}\chi_f^{a\dagger}\chi_i^a\chi_f^{b\dagger}\chi_i^b -
\frac{\alpha}{r^3}{(m_a+2m_b)\over
2m_am_b^2} \vec L \cdot \vec S_b\, \chi_f^{a\dagger}\chi_i^a\nonumber\\
&&-\frac{\alpha}{r^3} {(2m_a+m_b)\over
2m_a^2m_b} \vec L \cdot \vec{S}_a \, \chi_f^{b\dagger}\chi_i^b \nonumber\\
&&- \frac{\alpha}{r^5}{1\over m_am_b}\left(3 \vec{S}_a\cdot\vec{r}
\, \vec{S}_b\cdot\vec{r} - r^2 \vec S_a \cdot \vec S_b \right)
\label{VLOhh}
\end{eqnarray}
Note that since the piece proportional to $\vec{S}_a\cdot\vec{S}_b$
in Eq. (\ref{eq:LOMnonrelhh}) is analytic in $\vec
q^{\hspace*{1.4pt} 2}$ it only gives a short distance contribution
which is omitted in the potential (\ref{VLOhh}).

In this case when we evaluate the loop diagrams (a)-(e) of Fig. \ref{fig_diags}, we
find that part of the spin-spin structure piece contains the form
$q^2 S_a \cdot S_b$ multiplying the nonanalytic structures $L$ and
$S$. Due to this extra factor of $q^2$ in this form, we must expand
all loop integrals to one order higher in $q^2$ than before in order
to be consistent. This has been done and does make a difference in
our results, but the very lengthy expressions for the vector and
tensor integrals prevent us from explicitly listing them to this
order in Appendix \ref{app_int}. The results for the diagrams (a)-(e) in Fig. \ref{fig_diags}
are then\footnote{The results found for the spin-1/2 -- spin-1/2
case are rather lengthy, and so we quote the results {\it after} the
identities Eqs. (\ref{eq:id}) and (\ref{eq:ida}) have been used.}
\begin{eqnarray}
{}^{{1\over 2}{1\over 2}}{\cal M}_{\ref{fig_diags}a}^{(2)}(q)\hspace*{-5pt}& = \hspace*{-5pt} & 0\nonumber\\
{}^{{1\over 2}{1\over 2}}{\cal M}_{\ref{fig_diags}b}^{(2)}(q)\hspace*{-5pt}& = \hspace*{-5pt} & 0\nonumber\\
{}^{{1\over 2}{1\over 2}}{\cal M}_{\ref{fig_diags}c}^{(2)}(q)\hspace*{-5pt}& = \hspace*{-5pt} & 0\nonumber\\
{}^{{1\over 2}{1\over 2}}{\cal M}_{\ref{fig_diags}d}^{(2)}(q)\hspace*{-5pt}& = \hspace*{-5pt} &
{\alpha^2\over m_am_b}\Bigg[{\mathcal U}_a {\mathcal U}_b
\hspace*{-2.5pt}\left(\hspace*{-2.5pt}L
\hspace*{-2pt}\left(\hspace*{-2.5pt} {4m_am_b\over
q^2}\hspace*{-2pt}+\hspace*{-2pt}\frac{3m_a^2\hspace*{-2pt}+\hspace*{-2pt}m_am_b\hspace*{-2pt}+\hspace*{-2pt}3m_b^2}{2m_am_b}\hspace*{-1.5pt}\right)\hspace*{-3pt}
+\hspace*{-1.5pt}S{3\over 2}(m_a\hspace*{-2.5pt}+\hspace*{-2pt}m_b)\hspace*{-3pt} \right) \nonumber\\
&&\hspace*{32pt}+  i \frac{{\mathcal E}_a {\mathcal U}_b}{m_a^2 m_b}\Bigg(L\left(\frac{4 m_a m_b}{q^2}+\frac{10m_a^2 + 11 m_b^2}{6 m_a m_b}\right) \nonumber\\
&& \hspace*{79pt} +S\left(\frac{m_a m_b(2m_a + m_b)}{s-s_0} + (m_a+m_b) \hspace*{-1pt}\right) \! \Bigg)\nonumber\\
&&\hspace*{32pt}+  i \frac{{\mathcal U}_a {\mathcal E}_b}{m_a m_b^2}\Bigg(L\left(\frac{4 m_a m_b}{q^2}+\frac{11m_a^2 + 10 m_b^2}{6 m_a m_b}\right) \nonumber\\
&& \hspace*{79pt} +S\left(\frac{m_a m_b(m_a + 2 m_b)}{s-s_0} + (m_a+m_b) \hspace*{-1pt}\right) \! \Bigg)\nonumber\\
&&\hspace*{32pt} + \frac{S_b\cdot qS_a\cdot q}{m_a m_b} \, L \left(\frac{4 m_a m_b}{q^2} + \frac{4 m_a^2 + 3 m_a m_b + 4 m_b^2}{3 m_a m_b} \right) \nonumber\\
&&\hspace*{32pt} - \frac{q^2 S_a \cdot S_b}{m_a m_b} \, L \left(\frac{2 m_a m_b}{q^2} + \frac{8 m_a^2 + 13 m_a m_b + 8 m_b^2}{6 m_a m_b}\right)\nonumber\\
&&\hspace*{32pt}+ \frac{S_a \cdot qS_b \cdot q - q^2 S_a \cdot
S_b}{m_a m_b}
\hspace*{1pt} S(m_a+m_b) \left({m_a m_b\over s-s_0} + 2\right)\nonumber\\
&&\hspace*{32pt} + \Big(2 S_a\cdot p_3 S_b\cdot p_1 + S_a\cdot q S_b\cdot p_1 - S_a\cdot p_3 S_b\cdot q\Big) \frac{7 L}{3 m_a m_b}\Bigg] \nonumber\\
&- \hspace*{-5pt} & i 4\pi \alpha^2 {L\over q^2}\sqrt{m_a m_b \over s-s_0} \,
\Bigg({\mathcal U}_a {\mathcal U}_b + i \frac{{\mathcal E}_a {\mathcal U}_b}{m_a^2 m_b} + i \frac{{\mathcal U}_a {\mathcal E}_b}{m_a m_b^2}\nonumber\\
&&\hspace*{89pt}  + \frac{S_a \cdot qS_b \cdot q - \frac{1}{2}q^2 S_a \cdot S_b}{m_a m_b}\Bigg) \nonumber\\
{}^{{1\over 2}{1\over 2}}{\cal M}_{\ref{fig_diags}e}^{(2)}(q)\hspace*{-5pt}& = \hspace*{-5pt} &
{\alpha^2\over m_am_b}\Bigg[{\mathcal U}_a {\mathcal U}_b \Bigg( L
\left(-{4m_am_b\over q^2}
- \frac{9m_a^2 + 17 m_am_b + 9m_b^2}{6m_am_b}\right)\nonumber\\
&& \hspace*{63pt}- S \, \frac{m_a + m_b}{2} \Bigg) \nonumber\\
&&\hspace*{32pt}+  i \frac{{\mathcal E}_a {\mathcal U}_b}{m_a^2 m_b}\Bigg(L\left(-\frac{4 m_a m_b}{q^2}-\frac{10m_a^2 + 8 m_a m_b + 11 m_b^2}{6 m_a m_b}\right) \nonumber\\
&& \hspace*{79pt} - S \, \frac{2 m_a + 3 m_b}{4} \Bigg)\nonumber\\
&&\hspace*{32pt}+  i \frac{{\mathcal U}_a {\mathcal E}_b}{m_a m_b^2}\Bigg(L\left(- \frac{4 m_a m_b}{q^2}- \frac{11m_a^2 + 8 m_a m_b + 10 m_b^2}{6 m_a m_b}\right) \nonumber\\
&& \hspace*{79pt} - S \, \frac{3 m_a + 2 m_b}{4} \Bigg)\nonumber\\
&&\hspace*{32pt} + \frac{S_b\cdot qS_a\cdot q}{m_a m_b} \, L \left(-\frac{4 m_a m_b}{q^2} - \frac{4 m_a^2 + 4 m_b^2}{3 m_a m_b} \right) \nonumber\\
&&\hspace*{32pt} - \frac{q^2 S_a \cdot S_b}{m_a m_b} \, L \left(- \frac{2 m_a m_b}{q^2} - \frac{8 m_a^2 + 9 m_a m_b + 8 m_b^2}{6 m_a m_b}\right)\nonumber\\
&&\hspace*{32pt}+ \frac{S_a \cdot qS_b \cdot q - q^2 S_a \cdot
S_b}{m_a m_b}
\hspace*{1pt} S(m_a+m_b) \left(- \frac{1}{4}\right)\nonumber\\
&&\hspace*{32pt} - \Big(2 S_a\cdot p_3 S_b\cdot p_1 + S_a\cdot q S_b\cdot p_1 - S_a\cdot p_3 S_b\cdot q\Big) \frac{7 L}{3 m_a m_b}\Bigg]\nonumber\\
\quad
\end{eqnarray}
where we have defined
\begin{equation}
{\mathcal U}_a=\bar{u}(p_2)u(p_1) \qquad \qquad {\mathcal
U}_b=\bar{u}(p_4)u(p_3)
\end{equation}
and
\begin{equation}
{\mathcal E}_i = \epsilon_{\alpha\beta\gamma\delta}p_1^\alpha
p_3^\beta q^\gamma S_i^\delta
\end{equation}
with $i = a, b$. The sum is found then to be
\begin{eqnarray}
{}^{{1\over 2}{1\over 2}}{\cal
M}_{tot}^{(2)}(q)\hspace*{-3pt}&=&{\alpha^2\over
m_am_b}\Bigg[{\mathcal U}_a {\mathcal U}_b\left((m_a+m_b)S-{7\over 3}L\right) \nonumber\\
&&\hspace*{32pt}+  i \frac{{\mathcal E}_a {\mathcal U}_b}{m_a^2 m_b}\Bigg(\frac{2 m_a + m_b}{4}S -\frac{4}{3}L + \frac{m_a m_b (2 m_a + m_b)}{s - s_0} S\Bigg)\nonumber\\
&&\hspace*{32pt}+  i \frac{{\mathcal U}_a {\mathcal E}_b}{m_a^2 m_b}\Bigg(\frac{m_a + 2 m_b}{4}S -\frac{4}{3}L + \frac{m_a m_b (m_a + 2 m_b)}{s - s_0} S\Bigg)\nonumber\\
&&\hspace*{32pt}+ S(m_a+m_b) \, \frac{S_a \cdot qS_b \cdot q - q^2
S_a \cdot S_b}{m_a m_b}
\left(\frac{7}{4} + {m_a m_b\over s-s_0}\right)\nonumber\\
&&\hspace*{32pt}+ L \, \frac{S_a \cdot qS_b \cdot q - \frac{2}{3}q^2 S_a \cdot S_b}{m_a m_b}\nonumber\\
&-& i 4\pi \alpha^2 {L\over q^2}\sqrt{m_a m_b \over s-s_0} \,
\Bigg({\mathcal U}_a {\mathcal U}_b + i \frac{{\mathcal E}_a {\mathcal U}_b}{m_a^2 m_b} + i \frac{{\mathcal U}_a {\mathcal E}_b}{m_a m_b^2}\nonumber\\
&&\hspace*{89pt}  + \frac{S_a \cdot qS_b \cdot q - \frac{1}{2}q^2
S_a \cdot S_b}{m_a m_b}\Bigg)
\end{eqnarray}
Comparison with the result Eq. (\ref{eq:oh}) reveals again the
universality which has been found in other cases---the forms for the
scalar density and antisymmetric tensor components is identical and
symmetric between particles $a$ and $b$.  However, there is also a
new component---a spin-spin interaction. Performing the
nonrelativistic reduction yields for the amplitude
\begin{eqnarray}
{}^{{1\over 2}{1\over 2}}{\cal M}_{tot}^{(2)}(\vec
q)\hspace*{-3pt}&\simeq& \left[{\alpha^2\over
m_am_b}\left((m_a+m_b)S-{7\over3}L\right) - i 4 \pi \alpha^2
\frac{L}{q^2} \frac{m_r}{p_0}\right]
\chi_f^{a\dagger}\chi_i^a \, \chi_f^{b\dagger}\chi_i^b \nonumber\\
&+&\Bigg[{\alpha^2 \over m_a m_b} \left(\frac{2m_a^2 + 2 m_a m_b +
m_b^2}{2 m_b (m_a + m_b)} \, S
- \frac{8 m_a + m_b}{6 m_a m_b} \, L\right)\nonumber\\
&&+ \frac{\alpha^2 (2 m_a + m_b)}{(m_a + m_b)} \left(- i \frac{2 \pi
L}{p_0 q^2} + \frac{S}{p_0^2} \right)\Bigg] {i\over
m_a}\vec{S}_a\cdot\vec{p}\times\vec{q} \ \chi_f^{b\dagger}\chi_i^b\nonumber\\
&+&\Bigg[{\alpha^2 \over m_a m_b} \left(\frac{m_a^2 + 2 m_a m_b + 2
m_b^2}{2 m_a (m_a + m_b)} \, S
- \frac{m_a + 8 m_b}{6 m_a m_b} \, L\right)\nonumber\\
&&+ \frac{\alpha^2 (m_a + 2 m_b)}{(m_a + m_b)} \left(- i \frac{2 \pi
L}{p_0 q^2} + \frac{S}{p_0^2} \right)\Bigg]
\chi_f^{a\dagger}\chi_i^a \ {i\over
m_b}\vec{S}_b\cdot\vec{p}\times\vec{q} \nonumber\\
&+& \frac{\alpha^2}{m_a m_b} \, \frac{2 m_a^2 + 3 m_a m_b + 2 m_b^2}{m_a + m_b}\, S \, \frac{\vec S_a \cdot \vec q \, \vec S_b \cdot \vec q - \vec q^{\hspace*{1.4pt}2} \vec S_a \cdot \vec S_b}{m_a m_b}\nonumber\\
&+& {\alpha^2\over m_am_b} \,  L \, \frac{\vec S_a \cdot \vec q \, \vec S_b \cdot \vec q - \frac{2}{3}\vec q^{\hspace*{1.4pt}2} \vec S_a \cdot \vec S_b}{m_a m_b}\nonumber\\
&+& \frac{\alpha^2 m_a m_b}{m_a + m_b} \, \frac{S}{p_0^2} \,  \frac{\vec S_a \cdot \vec q \, \vec S_b \cdot \vec q - \vec q^{\hspace*{1.4pt}2} \vec S_a \cdot \vec S_b}{m_a m_b}\nonumber\\
&+& \frac{\alpha^2 m_a m_b}{m_a + m_b} \left(-i \frac{4 \pi L}{p_0
q^2}\right) \frac{\vec S_a \cdot \vec q \, \vec S_b \cdot \vec q -
\frac{1}{2}\vec q^{\hspace*{1.4pt}2} \vec S_a \cdot \vec S_b}{m_a
m_b} \label{eq:NLOMhhNR}
\end{eqnarray}
Again we verify universality: the same spin-independent piece as in
the previous calculations and now two spin-orbit coupling pieces,
one for the spin of each particle, with again the same form as found
earlier. The novel spin-spin coupling piece consists of the last
four lines of Eq. (\ref{eq:NLOMhhNR}).

Note that unlike in the case of spin-0 -- spin-1 scattering where
there were relativistic forms $\hat{\epsilon}_f^{b*}\cdot\vec{p} \
\hat{\epsilon}_i^b\cdot\vec{p}$ along with the quadrupole forms
$\hat{\epsilon}_f^{b*}\cdot\vec{q} \ \hat{\epsilon}_i^b\cdot\vec{q}$
in the leading order potential, the leading order spin-1/2 --
spin-1/2 potential contains no analog relativistic term $\vec S_a
\cdot \vec p \, \vec S_b \cdot \vec p$ along with the spin-spin
terms $\vec S_a \cdot \vec q \, \vec S_b \cdot \vec q$. We will see
that now our nonrelativistic second Born iteration succeeds in
removing all terms involving $1/p_0^2$ and $i/p_0$ and we find the
spin-spin coupling piece of the $\mathcal O(\alpha^2)$ potential.

Due to the universalities we obtained, it is clear that the
iteration for the spin-independent piece and the spin-orbit pieces
proceeds as shown before in the spin-0 -- spin-1/2 case. As before,
the second Born amplitude is
\begin{eqnarray}
{}^{{1\over 2}{1\over 2}}{\rm Amp}_C^{(2)}(\vec q) &=&-
\int{d^3\ell\over (2\pi)^3} \, \frac{\left<\vec p_f \left|
{}^{{1\over 2}{1\over 2}} \hat V^{(1)}_C \right| \vec \ell \,
\right> \left<\vec \ell \left| {}^{{1\over 2}{1\over 2}} \hat
V^{(1)}_C \right| \vec p_i \right>}{\frac{p_0^2}{2 m_r} - \frac{\ell^2}{2 m_r} + i
\epsilon}
\end{eqnarray}
where we now use the one-photon exchange potential ${}^{{1\over
2}{1\over 2}} V_C^{(1)}(\vec r)$ given in Eq. (\ref{VLOhh}).
Splitting this lowest order potential into spin-independent,
spin-orbit and spin-spin components---
\begin{equation}
\left<\vec p_f \left| {}^{{1\over 2}{1\over 2}} \hat V^{(1)}_C
\right| \vec p_i \, \right> =
 \left<\vec p_f \left| {}^{{1\over 2}{1\over 2}} \hat V^{(1)}_{S-I} \right| \vec p_i \, \right>
 + \left<\vec p_f \left| {}^{{1\over 2}{1\over 2}} \hat V^{(1)}_{S-O} \right| \vec p_i \, \right>
 + \left<\vec p_f \left| {}^{{1\over 2}{1\over 2}} \hat V^{(1)}_{S-S} \right| \vec p_i \, \right>
\end{equation}
where
\begin{eqnarray}
 \left<\vec p_f \left| {}^{{1\over 2}{1\over 2}} \hat V^{(1)}_{S-I} \right| \vec p_i \, \right>
 &=&{e^2\over \vec{q}^{\hspace*{1.4pt}2}} \, \chi_f^{a\dagger}\chi_i^a \, \chi_f^{b\dagger}\chi_i^b \nonumber\\
 \left<\vec p_f \left| {}^{{1\over 2}{1\over 2}} \hat V^{(1)}_{S-O} \right| \vec p_i \, \right>
 &=&{e^2\over \vec{q}^{\hspace*{1.4pt}2}}{2 m_a+ m_b\over 2m_am_b}\, \frac{i}{m_a}\vec{S}_a\cdot\vec{p}\times\vec{q} \, \chi_f^{b\dagger}\chi_i^b\nonumber\\
 &+&{e^2\over \vec{q}^{\hspace*{1.4pt}2}}{m_a+2m_b\over 2m_am_b} \, \chi_f^{a\dagger}\chi_i^a \, \frac{i}{m_b}\vec{S}_b\cdot\vec{p}\times\vec{q} \nonumber\\
 \left<\vec p_f \left| {}^{{1\over 2}{1\over 2}} \hat V^{(1)}_{S-S} \right| \vec p_i \, \right>
 &=&{e^2\over \vec{q}^{\hspace*{1.4pt}2}} {1\over m_am_b} \, \vec{S}_a\cdot\vec{q} \, \vec{S}_b\cdot\vec{q}
\end{eqnarray}
we again find that the iterated amplitude splits also into
spin-independent, spin-orbit and spin-spin pieces. As mentioned
above the calculation of the leading spin-independent amplitude
${}^{{1\over 2}{1\over 2}} {\rm Amp}^{(2)}_{S-I}(\vec q)$ and the
leading spin-orbit amplitude ${}^{{1\over 2}{1\over 2}} {\rm
Amp}^{(2)}_{S-O}(\vec q)$ works out exactly as in the 0 -- 1/2 case,
cf. Eqs. (\ref{eq:iteration0hSI}) and (\ref{eq:iteration0hSO}), and
we will not repeat it here again.

The leading spin-spin term of the second Born iteration amplitude is
new and we compute
\begin{eqnarray}
{}^{{1\over 2}{1\over 2}} {\rm Amp}^{(2)}_{S-S}(\vec q)
\hspace*{-10pt} &= \hspace*{-10pt}&- \int{d^3\ell\over
(2\pi)^3} \, \frac{\left<\vec p_f \left| {}^{{1\over 2}{1\over 2}} \hat V^{(1)}_{S-I} \right| \vec \ell \, \right> \left<\vec \ell \left| {}^{{1\over 2}{1\over 2}} \hat V^{(1)}_{S-S} \right| \vec p_i \right>}{\frac{p_0^2}{2 m_r} - \frac{\ell^2}{2 m_r} + i \epsilon} \nonumber\\
&&- \int{d^3\ell\over
(2\pi)^3} \, \frac{\left<\vec p_f \left| {}^{{1\over 2}{1\over 2}} \hat V^{(1)}_{S-S} \right| \vec \ell \, \right> \left<\vec \ell \left| {}^{{1\over 2}{1\over 2}} \hat V^{(1)}_{S-I} \right| \vec p_i \right>}{\frac{p_0^2}{2 m_r} - \frac{\ell^2}{2 m_r} + i \epsilon} \nonumber\\
&=\hspace*{-10pt}&{1\over m_am_b} \, S_a^r \, S_b^s \nonumber\\
&& \left(\hspace*{-2pt}i \hspace*{-3.2pt} \int \hspace*{-3.2pt}
{d^3\ell\over (2\pi)^3} {e^2\over |\vec{p}_f \hspace*{-1.1pt} -
\hspace*{-1.1pt} \vec{\ell} \hspace*{1pt}|^2 \hspace*{-1.1pt} +
\hspace*{-1.2pt} \lambda^2}{i \over {p_0^2\over 2m_r}
\hspace*{-1.1pt} - \hspace*{-1.1pt} {\ell^2\over 2m_r}
\hspace*{-1.1pt} + \hspace*{-1.1pt} i\epsilon}{e^2 (p_i
\hspace*{-1.1pt} - \hspace*{-1.1pt} \ell)^r (p_i \hspace*{-1.1pt} -
\hspace*{-1.1pt} \ell)^s\over
|\vec{\ell} \hspace*{-1.1pt} - \hspace*{-1.1pt} \vec{p}_i|^2 \hspace*{-1.1pt} + \hspace*{-1.2pt} \lambda^2}\right.\nonumber\\
&&\left. \hspace*{-3.6pt}+ i \hspace*{-3.2pt} \int \hspace*{-3.2pt}
{d^3\ell\over (2\pi)^3} {e^2 (\ell \hspace*{-1.1pt} -
\hspace*{-1.2pt} p_f)^r (\ell \hspace*{-1.1pt} - \hspace*{-1.2pt}
p_f)^s \over |\vec{p}_f \hspace*{-1.1pt} - \hspace*{-1.1pt}
\vec{\ell} \hspace*{1pt}|^2 \hspace*{-1.1pt} + \hspace*{-1.2pt}
\lambda^2}{i \over {p_0^2\over 2m_r} \hspace*{-1.1pt} -
\hspace*{-1.1pt} {\ell^2\over 2m_r} \hspace*{-1.1pt} +
\hspace*{-1.1pt} i\epsilon}{e^2\over
|\vec{\ell} \hspace*{-1.1pt} - \hspace*{-1.1pt} \vec{p}_i|^2 \hspace*{-1.1pt} + \hspace*{-1.2pt}\lambda^2}\hspace*{-3pt}\right)\nonumber\\
&\stackrel{\lambda\rightarrow 0}{\longrightarrow}\hspace*{-8pt}& {}\, {1\over m_a m_b} \Bigg[\left(\vec{S}_a\cdot\vec{p_i} \, \vec{S}_b\cdot\vec{p_i} + \vec{S}_a\cdot\vec{p_f} \, \vec{S}_b\cdot\vec{p_f} \right) H\nonumber\\
&& \hspace*{34pt} - \vec S_a \cdot (\vec p_i + p_f) \vec S_b \cdot \vec H - \vec S_a \cdot \vec H \, \vec S_b \cdot (\vec p_i + p_f) \nonumber\\
&& \hspace*{34pt} + 2 \, S_a^r S_b^s \, H^{rs}\Bigg]\nonumber\\
&=\hspace*{-10pt}& \frac{\alpha^2 m_a m_b}{m_a + m_b} \, \frac{S}{p_0^2} \,  \frac{\vec S_a \cdot \vec q \, \vec S_b \cdot \vec q - \vec q^{\hspace*{1.4pt}2} \vec S_a \cdot \vec S_b}{m_a m_b}\nonumber\\
&+\hspace*{-10pt}& \frac{\alpha^2 m_a m_b}{m_a + m_b} \left(-i \frac{4 \pi L}{p_0
q^2}\right) \frac{\vec S_a \cdot \vec q \, \vec S_b \cdot \vec q -
\frac{1}{2}\vec q^{\hspace*{1.4pt}2} \vec S_a \cdot \vec S_b}{m_a
m_b}
\end{eqnarray}
in agreement with the corresponding terms in \cite{sm} in the equal mass limit $m_a = m_b = m$.
With that, the total second Born iteration amplitude becomes
\begin{eqnarray}
{}^{{1\over 2}{1\over 2}} {\rm Amp}_C^{(2)}(\vec q) \hspace*{-3pt}{}&=&{}^{{1\over 2}{1\over 2}} {\rm Amp}^{(2)}_{S-I}(\vec q) + {}^{{1\over 2}{1\over 2}} {\rm Amp}^{(2)}_{S-O}(\vec q) + {}^{{1\over 2}{1\over 2}} {\rm Amp}^{(2)}_{S-S}(\vec q) \nonumber \\
&=&- i 4 \pi \alpha^2
\frac{L}{q^2} \frac{m_r}{p_0} \,  \chi_f^{a\dagger}\chi_i^a \, \chi_f^{b\dagger}\chi_i^b \nonumber\\
&+& \frac{\alpha^2 (2 m_a + m_b)}{m_a + m_b} \left(- i \frac{2 \pi
L}{p_0 q^2} + \frac{S}{p_0^2} \right) \, {i\over
m_a}\vec{S}_a\cdot\vec{p}\times\vec{q} \ \chi_f^{b\dagger}\chi_i^b\nonumber\\
&+& \frac{\alpha^2 (m_a + 2 m_b)}{m_a + m_b} \left(- i \frac{2 \pi
L}{p_0 q^2} + \frac{S}{p_0^2} \right)
 \chi_f^{a\dagger}\chi_i^a \ {i\over
m_b}\vec{S}_b\cdot\vec{p}\times\vec{q} \nonumber\\
&+& \frac{\alpha^2 m_a m_b}{m_a + m_b} \, \frac{S}{p_0^2} \,  \frac{\vec S_a \cdot \vec q \, \vec S_b \cdot \vec q - \vec q^{\hspace*{1.4pt}2} \vec S_a \cdot \vec S_b}{m_a m_b}\nonumber\\
&+& \frac{\alpha^2 m_a m_b}{m_a + m_b} \left(-i \frac{4 \pi L}{p_0
q^2}\right) \frac{\vec S_a \cdot \vec q \, \vec S_b \cdot \vec q -
\frac{1}{2}\vec q^{\hspace*{1.4pt}2} \vec S_a \cdot \vec S_b}{m_a
m_b}
\end{eqnarray}
and we observe that when this amplitude is subtracted from the full
one loop scattering amplitude Eq. (\ref{eq:NLOMhhNR}), all terms
involving $1/p_0^2$ and $i/p_0$ disappear leaving behind a
well-defined second order potential
\begin{eqnarray}
{}^{{1\over 2}{1\over 2}} V^{(2)}_C(\vec r)&=&-\int{d^3q\over
(2\pi)^3}e^{-i\vec{q}\cdot\vec{r}}\left[{}^{{1\over 2}{1\over 2}}
{\cal M}_{tot}^{(2)}(\vec{q})- {}^{{1\over 2}{1\over 2}} {\rm
Amp}_C^{(2)}(\vec{q})\right]\nonumber\\
&=&\left[- {\alpha^2(m_a+m_b)\over 2m_am_br^2} -
{7\alpha^2\hbar\over 6\pi m_am_br^3}\right]
\chi_f^{a\dagger}\chi_i^a \, \chi_f^{b\dagger}\chi_i^b\nonumber\\
&+& {1\over m_a}\vec{S}_a \hspace*{-2pt}\cdot
\hspace*{-1.5pt}\vec{p}\hspace*{-1pt}\times\hspace*{-2.5pt}\vec{\nabla}\hspace*{-2.5pt}\left[
{\alpha^2(2m_a^2\hspace*{-1.3pt}+\hspace*{-1.3pt}2m_a
m_b\hspace*{-1.3pt}+\hspace*{-1.3pt}m_b^2)\over 4m_a m_b^2
(m_a\hspace*{-1.3pt}+\hspace*{-1.3pt}m_b)r^2}\hspace*{-1.3pt}+\hspace*{-1.3pt}{\alpha^2(8m_a\hspace*{-1.3pt}+\hspace*{-1.3pt}m_b)\hbar\over
12\pi
m_a^2m_b^2r^3}\right] \! \chi_f^{b\dagger}\chi_i^b\nonumber\\
&+&\chi_f^{a\dagger}\chi_i^a {1\over m_b}\vec{S}_b
\hspace*{-2pt}\cdot
\hspace*{-1.5pt}\vec{p}\hspace*{-1pt}\times\hspace*{-2.5pt}\vec{\nabla}\hspace*{-2.5pt}\left[
{\alpha^2(m_a^2\hspace*{-1.3pt}+\hspace*{-1.3pt}2m_a
m_b\hspace*{-1.3pt}+\hspace*{-1.3pt}2m_b^2)\over 4m_a^2 m_b
(m_a\hspace*{-1.3pt}+\hspace*{-1.3pt}m_b)r^2}\hspace*{-1.3pt}+\hspace*{-1.3pt}{\alpha^2(m_a\hspace*{-1.3pt}+\hspace*{-1.3pt}8m_b)\hbar\over
12\pi
m_a^2m_b^2r^3}\right]\nonumber\\
&+& \frac{\vec S_a \cdot \vec{\nabla} \vec S_b \cdot \vec{\nabla} -
\vec{\nabla}^2 \vec S_a \cdot \vec S_b}{m_a m_b}
\left[\frac{\alpha^2(2m_a^2 + 3 m_a m_b + 2 m_b^2)}{2 m_a m_b (m_a + m_b) r^2}\right]\nonumber\\
&+& \frac{\vec S_a \cdot \vec{\nabla} \vec S_b \cdot \vec{\nabla} -
\frac{2}{3} \vec{\nabla}^2 \vec S_a \cdot \vec S_b}{m_a m_b}
\left[-\frac{\alpha^2\hbar}{2 \pi m_a m_b r^3}\right]\nonumber\\
&=&\left[- {\alpha^2(m_a+m_b)\over 2m_am_br^2} -
{7\alpha^2\hbar\over 6\pi m_am_br^3}\right]
\chi_f^{a\dagger}\chi_i^a \, \chi_f^{b\dagger}\chi_i^b\nonumber\\
&+&\left[{\alpha^2(2m_a^2+2m_am_b+ m_b^2)\over
2m_a^2m_b^2(m_a+m_b)r^4}+{\alpha^2(8m_a+ m_b)\hbar\over 4\pi
m_a^3m_b^2r^5}\right] \vec{L}\cdot\vec{S}_a \, \chi_f^{b\dagger}\chi_i^b \nonumber\\
&+&\left[{\alpha^2(m_a^2+2m_am_b+2m_b^2)\over
2m_a^2m_b^2(m_a+m_b)r^4}+{\alpha^2(m_a+8m_b)\hbar\over 4\pi
m_a^2m_b^3r^5}\right] \chi_f^{a\dagger}\chi_i^a \, \vec{L}\cdot\vec{S}_b \nonumber\\
&+& \left[- \frac{2 \alpha^2 (2 m_a^2 + 3 m_a m_b + 2 m_b^2)}{m_a^2 m_b^2 (m_a + m_b) r^4} \right] \left(\vec S_a \cdot \vec S_b - 2 \vec S_a \cdot \vec r \, \vec S_b \cdot \vec r \, / r^2\right)\nonumber\\
&+& \left[\frac{\alpha^2 \hbar}{2 \pi m_a^2 m_b^2 r^5}\right]
\left(7 \vec S_a \cdot \vec S_b - 15 \vec S_a \cdot \vec r \, \vec
S_b \cdot \vec r \, / r^2\right)
\end{eqnarray}

\section{Conclusions}

Above we have analyzed the electromagnetic scattering of two charged
particles having nonzero mass.  In lowest order the interaction
arises from one-photon exchange and leads at threshold to the well
known Coulomb interaction $V(r)=\alpha/r$.  Inclusion of two-photon
exchange effects means adding the contribution from box, cross-box,
triangle, and bubble diagrams, which have a rather complex form. The
calculation can be simplified, however, by using ideas from
effective field theory.  The point is that if one is interested only
in the leading long-range behavior of the interaction, then one need
retain only the leading nonanalytic small momentum-transfer piece of
the scattering amplitude. Specifically, the terms which one retains
are those which are nonanalytic and behave as either
$\alpha^2/\sqrt{-q^2}$ or $\alpha^2\log -q^2$. When Fourier
transformed, the former leads to classical ($\hbar$-independent)
terms in the potential of order $\alpha^2/mr^2$ while the latter
generates quantum mechanical ($\hbar$-dependent) corrections of
order $\alpha^2\hbar/m^2r^3$. (Of course, there are also shorter
range nonanalytic contributions than these that are generated by
scattering terms of order $\alpha^2q^{2n}\sqrt{-q^2}$ or
$\alpha^2q^{2n}\log -q^2$.  However, these pieces are higher order
in momentum transfer and therefore lead to shorter distance effects
than those considered above and are therefore neglected in our
discussion.)

Specific calculations were done for particles with spin $0-0$,
$0-1/2$, $0-1$, and $1/2-1/2$ and various universalities were found.
In particular, we found that in each case there was a
spin-independent contribution of the form
\begin{eqnarray}
{}^{S_aS_b}{\cal M}^{(2)}_{tot}(q)&=& \left[{\alpha^2\over
m_am_b}\left((m_a+m_b)S-{7L\over 3}\right)
- i 4\pi \alpha {L\over q^2}\sqrt{m_a m_b \over s - s_0} {} \hspace*{3pt} \right]\nonumber\\
&& \times \left<S_a,m_{af}|S_a,m_{ai}\right> \left<S_b,m_{bf}|S_b,m_{bi}\right>
\end{eqnarray}
where $L=\log -q^2$ and $S=\pi^2/\sqrt{-q^2}$ and with $S_a$
the spin of particle $a$ and $S_b$ the spin of particle $b$ with
projections $m_a$ and $m_b$ on the quantization axis.
The imaginary component of the amplitude, which would not, when
Fourier-transformed lead to a real potential, is eliminated when the
iterated lowest order potential contribution is subtracted, leading
to a well defined spin-independent second order potential of universal
form
\begin{eqnarray}
{}^{S_aS_b}V_{S-I}^{(2)}(\vec r)&=&\left[- {\alpha^2(m_a+m_b)\over
2m_am_br^2}- {7\alpha^2\hbar\over 6\pi
m_am_br^3}\right]\nonumber\\
&& \times \left<S_a,m_{af}|S_a,m_{ai}\right> \left<S_b,m_{bf}|S_b,m_{bi}\right>
\end{eqnarray}
If either scattering particle carries spin then there is an
additional spin-orbit contribution, whose form is also universal
\begin{eqnarray}
{}^{S_aS_b}V_{S-O}^{(2)}(\vec r)&=&\left[{\alpha^2(2m_a^2+2m_am_b+ m_b^2)\over
2m_a^2m_b^2(m_a+m_b)r^4}+{\alpha^2(8m_a+ m_b)\hbar\over 4\pi
m_a^3m_b^2r^5}\right] \nonumber \\
&& \times \vec{L}\cdot\vec{S}_a \left<S_b,m_{bf}|S_b,m_{bi}\right> \nonumber\\
&+&\left[{\alpha^2(m_a^2+2m_am_b+2m_b^2)\over
2m_a^2m_b^2(m_a+m_b)r^4}+{\alpha^2(m_a+8m_b)\hbar\over 4\pi
m_a^2m_b^3r^5}\right] \nonumber \\
&& \times \left<S_a,m_{af}|S_a,m_{ai}\right> \vec{L}\cdot\vec{S}_b \nonumber\\
\quad
\end{eqnarray}
where we have defined
$$\vec{S}_a= \left<S_a,m_{af} \left|{}\, \vec{S} \, \right| S_a,m_{ai}\right>\quad{\rm and}
\quad \vec{S}_b= \left<S_b,m_{bf} \left|{}\, \vec{S} \, \right| S_b,m_{bi}\right>$$  In this case a
well defined second order potential required the subtraction of
infrared singular terms behaving as both $i/p_0$ {\it and}
$1/p_0^2$ which arise from the iterated lowest order potential.

In the calculation of spin-0 -- spin-1 scattering we encountered new
tensor structures including a quadrupole interaction. Unfortunately,
the subtraction of the $i/p_0$ and $1/p_0^2$ tensor pieces in the
two-photon exchange amplitude was not successful with our simple
nonrelativistic iteration of the leading order potential so that we
cannot at this time give the form of the quadrupole component of the
potential. Further work is needed to clarify this issue. The
corrections to the spin-spin interaction have only been calculated
in spin-1/2 -- spin-1/2 scattering where we found their
contributions to the scattering amplitude and to the potential.
Since we verified these forms only for a single spin configuration
we have not confirmed its universality which we, however, strongly
suspect. Of course, for higher spin configurations, there also exist
quadrupole-quadrupole interactions, spin-quadrupole interactions,
etc.  However, the calculation of such forms becomes increasingly
cumbersome as the spin increases, and the phenomenological
importance becomes smaller.  Thus we end our calculations here.

One point of view to interpret the universalities of the long distance
components of the scattering amplitudes and the resulting potentials
is that if we increase the spins of our scattered particles, all we do
is to add additional multipole moments. The spin-independent component
can then be viewed as a monopole-monopole interaction, the spin-orbit
piece as a dipole-monopole interaction etc. As long as we do not change
the quantum numbers that characterize the lower multipoles (such as the
charge for the monopole-monopole interaction or the g-factor for the
monopole-dipole interaction), an increase in spin of the scattered
particles merely adds new interactions that are less important at long
distances. In Appendix \ref{app_general} we show explicitly how this
multipole expansion structure arises.
While it is familiar from classical electrodynamics---i.e. at the one-photon exchange level---we
are not aware that this has been proven for two-photon exchange processes.

It is interesting that the same kinds of universalities of the long range
components of the scattering amplitudes are also found in gravitational 
scattering \cite{hrgr} and in mixed electromagnetic-gravitational 
scattering \cite{hrmix}.

\begin{center}
{\bf Acknowledgements}
\end{center}

We would like to thank John Donoghue for many clarifying discussions
and Walter Goldberger for important comments. This work was supported 
in part by the National Science Foundation under award PHY05-53304 
(BRH and AR) and by the the US Department of Energy under grant 
DE-FG-02-92ER40704 (AR).

\appendix

\section{One loop integration in EFT} \label{app_int}
In this appendix we sketch how our results were obtained.  The basic
idea is to calculate the Feynman diagrams shown in Fig. \ref{fig_diags}.
Since our calculations focus on long distance effects that stem from
nonanalytic contributions in the squared momentum transfer $q^2$, we only
evaluate these nonanalytic pieces of the one loop integrals neglecting
all short distance contributions which include the UV divergences.
In practice, that means that all one loop diagrams where $q$ does
not run through any part of the loop can be neglected. Furthermore,
only diagrams with at least two massless propagators yield nonanalytic
contributions, reducing the number of contributing diagrams and thus
integrals further. In the end, we need four different types of integrals
for our calculations: Bubble integrals with two massless propagators
and no massive propagator, triangle integrals with two massless propagators
and one massive propagator and box and cross-box integrals, each
with two massless propagators and two massive propagators.

For simplicity we shall assume spinless scattering. Thus for diagram (a)
of Fig. \ref{fig_diags}---the bubble diagram---we find
\begin{equation}
{\rm Amp}[\ref{fig_diags}a]={1\over 2!}\int{d^4k\over (2\pi)^4} \
{\tau_{\mu\nu}^{(2)}(p_2,p_1)\eta^{\mu\alpha}\eta^{\nu\beta}
\tau_{\alpha\beta}^{(2)}(p_4,p_3)\over k^2(k+q)^2} . \label{eqn:a}
\end{equation}
All vertex functions are listed in the main body of the paper, while for
the integrals, all that is needed is their nonanalytic behavior.
The exact expressions for the nonanalytic components of the bubble integrals read
\begin{eqnarray}
 I & = & \int \frac{d^4 k} {(2 \pi)^4} \ \frac {1} {k^2 (k+q)^2} = \frac {i} {16 \pi^2} \left(-L\right) \\
 I_\mu & = & \int \frac{d^4 k} {(2 \pi)^4} \ \frac {k_\mu} {k^2 (k+q)^2} = - \frac {1} {2} \, I \, q_\mu = \frac {i} {16 \pi^2} \left(\frac{1} {2} \, L\right) q_\mu\\
 I_{\mu \nu} & = &\int \frac{d^4 k} {(2 \pi)^4} \ \frac {k_\mu k_\nu} {k^2 (k+q)^2} = - \frac {1} {12} \, q^2 \, I \, \eta_{\mu \nu} + \frac {1} {3} \, I \, q_\mu q_\nu \nonumber \\
                                    & = &\frac {i} {16 \pi^2} \left(\frac{1} {12} \, q^2 \, L \ \eta_{\mu \nu} - \frac {1} {3} \, L \ q_\mu q_\nu\right)\\
 I_{\mu \nu \rho} & = &\int \frac{d^4 k} {(2 \pi)^4} \ \frac {k_\mu k_\nu k_\rho} {k^2 (k+q)^2} = \frac {1} {24} \, q^2 \, I \ 3 \, \eta_{(\mu \nu} q_{\rho)} - \frac {1} {4} \, I \, q_\mu q_\nu q_\rho \nonumber \\
                                    & = & \frac {i} {16 \pi^2} \left(- \frac{1} {24} \, q^2 \, L \ 3 \, \eta_{(\mu \nu} q_{\rho)} + \frac {1} {4} \, L \ q_\mu q_\nu q_\rho \right) \nonumber \\
 I_{\mu \nu \rho \sigma} & = & \int \frac{d^4 k} {(2 \pi)^4} \ \frac {k_\mu k_\nu k_\rho k_\sigma} {k^2 (k+q)^2} \nonumber \\
                                    & = & \frac {1} {240} \, q^4 \, I \ 3 \, \eta_{(\mu \nu} \eta_{\rho \sigma)} - \frac {1} {40} \, q^2 \, I \ 6 \, \eta_{(\mu \nu} q_\rho q_{\sigma)} + \frac{1}{5} \, I \, q_\mu q_\nu q_\rho q_\sigma \nonumber \\
                                    & = & \frac {i} {16 \pi^2} \left(- \frac {1} {240} \, q^4 \, L \ 3 \, \eta_{(\mu \nu} \eta_{\rho \sigma)} + \frac {1} {40} \, q^2 \, L \ 6 \, \eta_{(\mu \nu} q_\rho q_{\sigma)} - \frac{1}{5} \, L \, q_\mu q_\nu q_\rho q_\sigma\right) \nonumber \\ \quad
\end{eqnarray}
where our symmetrization convention is
$$A_{(\mu_1 \mu_2 \mu_3 \dots \mu_n)} = \frac {1} {n !} \left(A_{\mu_1 \mu_2 \mu_3 \dots \mu_n} + A_{\mu_2 \mu_1 \mu_3 \dots \mu_n} + \dots \right)$$
so that for example
$3 \, \eta_{(\mu \nu} q_{\rho)} = \eta_{\mu \nu} q_\rho + \eta_{\mu \rho} q_\nu + \eta_{\nu \rho} q_\mu$.

\begin{figure}
\begin{center}
\epsfig{file=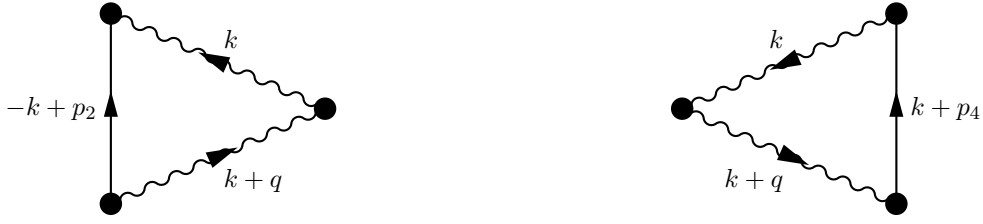,width=0.95\textwidth}
  \caption{Momentum labels for loops in triangle diagrams. On the left, the massive particle $a$ with mass $m_a$ runs through the loop, whereas on
           the right, particle $b$ with mass $m_b$ propagates in the loop.}
  \label{fig_triangles}
\end{center}
\end{figure}

There are two distinct triangle diagrams, (b) and (c) in Fig. \ref{fig_diags}, with two different masses that propagate inside the loop.
The momentum label conventions used in all triangle diagrams are shown in Fig. \ref{fig_triangles} for the two cases
so that the expression for the amplitude for diagram (b) for example reads
\begin{equation}
{\rm Amp}[\ref{fig_diags}b]=\int{d^4k\over (2\pi)^4} \ \frac{\tau^{(2)}_{\mu\nu}(p_4,p_3)\eta^{\mu\alpha}\eta^{\nu\beta}
\tau^{(1)}_\beta(p_2,p_2-k)\tau^{(1)}_\alpha(p_2-k,p_1)}{k^2(k+q)^2((k-p_2)^2-m_a^2)}. \label{eqn:b}
\end{equation}

In the evaluation of the integrals we use the on-shell relations
\begin{equation}
  p_2 \cdot q = - \frac{q^2}{2} \ \ \mbox{and} \ \ p_4 \cdot q = + \frac{q^2}{2}
\end{equation}
and the expressions for the triangle integrals read
\begin{eqnarray}
 J \hspace*{-7.5pt}& = \hspace*{-7pt}& \int \frac{d^4 k} {(2 \pi)^4} \ \frac {1} {k^2 (k+q)^2 ((k+p)^2 - m^2)} \nonumber \\
   & = \hspace*{-7pt}& \frac {i} {16 \pi^2} \, \frac {1} {m^2} \hspace*{-1pt} \left[\hspace*{-1pt} - \frac{1} {2} \hspace*{1pt} L \hspace*{-1pt} \left( \hspace*{-1.5pt} 1 \hspace*{-1pt} + \hspace*{-1pt} \frac{q^2}{6 m^2} \hspace*{-1pt} + \hspace*{-1pt} \mathcal O \hspace*{-2pt} \left[ \hspace*{-1pt} \left(\hspace*{-1pt} \frac{q^2}{m^2}\hspace*{-1pt} \right)^2\right] \hspace*{-1pt} \right) \right. \nonumber \\
    && \hspace*{48pt} \left. - \frac{m} {2} \hspace*{1pt} S \hspace*{-1pt} \left( \hspace*{-1pt} 1 \hspace*{-1pt} + \hspace*{-1pt} \frac{q^2}{8 m^2} \hspace*{-1pt} + \hspace*{-1pt} \mathcal O \hspace*{-2pt} \left[ \hspace*{-1pt} \left(\hspace*{-1pt} \frac{q^2}{m^2}\right)^2\right] \hspace*{-1pt} \right) \hspace*{-2pt} \right] \label{eq_ints_tria0} \\
 J_\mu \hspace*{-7.5pt} & = \hspace*{-5pt} & \int \frac{d^4 k} {(2 \pi)^4} \ \frac {k_\mu} {k^2 (k+q)^2 ((k+p)^2 - m^2)} \nonumber \\
                                    & = \hspace*{-7pt} & \mathcal F \left[\left( - \frac {1} {4 m^2} \, I - \frac {1} {2} \, J\right)  q_\mu + \left(\frac {1} {2 m^2} \, I  + \frac {1} {4} \frac {q^2} {m^2} \, J\right) p_\mu\right] \\
 J_{\mu \nu} \hspace*{-7.5pt} & = \hspace*{-7pt} & \int \frac{d^4 k} {(2 \pi)^4} \ \frac {k_\mu k_\nu} {k^2 (k+q)^2 ((k+p)^2 - m^2)} \nonumber \\
                                    & = \hspace*{-7pt} & \mathcal F \left(- \frac {q^2} {16 m^2} \, I - \frac{q^2} {8} \, J\right) \eta_{\mu \nu} \nonumber \\
                                    & + \hspace*{-7pt} & \mathcal F^2 \left[\frac {5} {16 m^2} \left(1 - \frac{1}{10} \, \frac{q^2}{m^2}\right) I + \frac{3} {8} \, J \right] q_\mu q_\nu \nonumber \\
                                    & + \hspace*{-7pt} & \mathcal F^2 \left[\frac {3 q^2} {16 m^4} \, I + \frac{q^2} {8 m^2} \left(1 + \frac{1}{2} \, \frac{q^2}{m^2}\right) J\right] p_\mu p_\nu \nonumber \\
                                    & + \hspace*{-7pt} & \mathcal F^2 \left[- \frac {1} {4 m^2} \left(1 + \frac{1}{8} \, \frac{q^2}{m^2}\right) I - \frac{3 q^2} {16 m^2} \, J\right] 2 \, p_{(\mu} q_{\nu)} \\
 J_{\mu \nu \rho} \hspace*{-7.5pt} & = \hspace*{-7pt} &\int \frac{d^4 k} {(2 \pi)^4} \ \frac {k_\mu k_\nu k_\rho} {k^2 (k+q)^2 ((k+p)^2 - m^2)} \nonumber \\
                                    & = \hspace*{-7pt} &\mathcal F^2 \left[\frac {5 q^2} {96 m^2} \left(1 - \frac{1}{10} \, \frac{q^2}{m^2}\right) I + \frac{q^2} {16} \, J\right] 3 \, \eta_{(\mu \nu} q_{\rho)} \nonumber \\
                                    & + \hspace*{-7pt} & \mathcal F^2 \left[- \frac {q^2} {24 m^2} \left(1 + \frac{1}{8} \, \frac{q^2}{m^2}\right) I - \frac{q^4} {32 m^2} \, J \right] 3 \, \eta_{(\mu \nu} p_{\rho)} \nonumber \\
                                    & + \hspace*{-7pt} & \mathcal F^3 \left[- \frac {11} {32 m^2} \left(1-\frac{13}{66} \, \frac{q^2}{m^2} + \frac{1}{66} \, \frac{q^4}{m^4}\right) I - \frac{5} {16} \, J\right] q_\mu q_\nu q_\rho \nonumber \\
                                    & + \hspace*{-7pt} & \mathcal F^3 \left[\frac {q^2} {12 m^4} \left(1 + \frac{11}{16} \, \frac{q^2}{m^2}\right) I + \frac{3 q^4} {32 m^4} \left(1 + \frac{1} {6} \, \frac{q^2}{m^2}\right) J\right] p_\mu p_\nu p_\rho \nonumber \\
                                    & + \hspace*{-7pt} & \mathcal F^3 \left[\frac {1} {6 m^2} \left(1 + \frac{9}{32} \, \frac{q^2}{m^2} - \frac{1} {64} \frac {q^4} {m^4}\right) I + \frac {5 q^2} {32 m^2} \, J\right] 3 \, q_{(\mu} q_\nu p_{\rho)} \nonumber \\
                                    & + \hspace*{-7pt} & \mathcal F^3 \left[- \frac {13 q^2} {96 m^4} \left(1 + \frac{1}{26} \, \frac{q^2}{m^2}\right) I - \frac{q^2} {16 m^2} \left(1 + \frac{q^2}{m^2}\right) J\right] 3 \, q_{(\mu} p_\nu p_{\rho)} \\
 J_{\mu \nu \rho \sigma} \hspace*{-7.5pt} & = \hspace*{-7pt} & \int \frac{d^4 k} {(2 \pi)^4} \ \frac {k_\mu k_\nu k_\rho k_\sigma} {k^2 (k+q)^2 ((k+p)^2 - m^2)} \nonumber \\
                                    & = \hspace*{-7pt} & \mathcal F^2 \left[\frac {5 q^4} {768 m^2} \left(1 - \frac{1}{10} \, \frac{q^2}{m^2}\right) I + \frac{q^4} {128} \, J\right] 3 \, \eta_{(\mu \nu} \eta_{\rho \sigma)} \nonumber \\
                                    & + \hspace*{-7pt} & \mathcal F^3 \left[- \frac {11 q^2} {256 m^2} \left(1-\frac{13}{66} \, \frac{q^2}{m^2} + \frac{1}{66} \, \frac{q^4}{m^4}\right) I - \frac{5 q^2} {128} \, J \right] 6 \, \eta_{(\mu \nu} q_{\rho} q_{\sigma)}\nonumber \\
                                    & + \hspace*{-7pt} & \mathcal F^3 \left[\frac {q^2} {48 m^2} \left(1 + \frac{9}{32} \, \frac{q^2}{m^2} - \frac{1}{64} \, \frac{q^4}{m^4}\right) I + \frac{5 q^4} {256 m^2} \, J\right] 12 \, \eta_{(\mu \nu} q_{\rho} p_{\sigma)} \nonumber \\
                                    & + \hspace*{-7pt} & \mathcal F^3 \left[- \frac {13 q^4} {768 m^4} \left(1+\frac{1}{26} \, \frac{q^2}{m^2}\right) I - \frac{q^4} {128 m^2} \left(1 + \frac{q^2}{m^2}\right) J \right] 6 \, \eta_{(\mu \nu} p_{\rho} p_{\sigma)}\nonumber \\
                                    & + \hspace*{-7pt} & \mathcal F^4 \left[\frac {93} {256 m^2} \hspace*{-1pt} \left(1 \hspace*{-1pt} - \hspace*{-0.5pt} \frac{163}{558} \, \frac{q^2}{m^2} \hspace*{-0.5pt} + \hspace*{-0.5pt} \frac{25}{558} \, \frac{q^4}{m^4} \hspace*{-0.5pt} - \hspace*{-0.5pt} \frac{1}{372} \, \frac{q^6}{m^6}\right) \hspace*{-2pt} I + \frac{35} {128} J\right] q_\mu q_\nu q_\rho q_\sigma \nonumber \\
                                    & + \hspace*{-7pt} & \mathcal F^4 \left[ \hspace*{-1pt} -\frac {1} {8 m^2} \hspace*{-2.5pt} \left(\hspace*{-2.4pt} 1 \hspace*{-2.4pt} + \hspace*{-1.4pt} \frac{29}{64}  \hspace*{1pt} \frac{q^2}{m^2} \hspace*{-1.4pt} - \hspace*{-1.4pt} \frac{19}{384} \hspace*{1pt} \frac{q^4}{m^4} \hspace*{-1pt} + \hspace*{-1pt} \frac{1}{384} \hspace*{1pt} \frac{q^6}{m^6} \hspace*{-2pt} \right) \hspace*{-2pt} I \hspace*{-1pt} - \hspace*{-0.5pt} \frac{35 q^2} {256 m^2} J\right] \hspace*{-1.5pt} 4 \hspace*{1pt} q_{(\mu} q_\nu q_\rho p_{\sigma)} \nonumber \\
                                    & + \hspace*{-7pt} & \mathcal F^4 \left[ \hspace*{-1pt} \frac {27 q^2} {256 m^4} \hspace*{-2.5pt} \left(\hspace*{-2.7pt} 1 \hspace*{-2.4pt} + \hspace*{-1.5pt} \frac{7}{81} \hspace*{0.5pt} \frac{q^2}{m^2} \hspace*{-1.4pt} - \hspace*{-1.4pt} \frac{1}{324} \hspace*{0.5pt} \frac{q^4}{m^4} \hspace*{-2.4pt}\right) \hspace*{-2.7pt} I \hspace*{-1.7pt} + \hspace*{-1.4pt} \frac{5 q^2} {128 m^2} \hspace*{-2.7pt} \left(\hspace*{-2.4pt}1 \hspace*{-2pt}+ \hspace*{-1pt} \frac{3}{2} \hspace*{0.5pt} \frac{q^2}{m^2} \hspace*{-2pt}\right) \hspace*{-2.7pt} J \hspace*{-1pt} \right] \hspace*{-2.5pt} 6 \hspace*{1pt} q_{(\mu} q_\nu p_\rho p_{\sigma)} \nonumber \\
                                    & + \hspace*{-7pt} & \mathcal F^4 \left[ \hspace*{-1.2pt} - \frac {q^2} {24 m^4} \hspace*{-2.5pt} \left(\hspace*{-2.4pt} 1 \hspace*{-2.4pt} + \hspace*{-1.5pt} \frac{83}{64} \hspace*{0.5pt} \frac{q^2}{m^2} \hspace*{-1.4pt} + \hspace*{-1.4pt} \frac{3}{128} \hspace*{0.5pt} \frac{q^4}{m^4} \hspace*{-2.4pt} \right) \hspace*{-2.7pt} I \hspace*{-1.7pt} - \hspace*{-1.4pt} \frac{15 q^4} {256 m^4} \hspace*{-2.7pt} \left(\hspace*{-2.4pt} 1 \hspace*{-2.2pt} + \hspace*{-1.4pt} \frac{1} {3} \hspace*{0.5pt}  \frac{q^2}{m^2} \hspace*{-2.2pt} \right) \hspace*{-2.7pt} J \hspace*{-1.2pt} \right] \hspace*{-3.2pt} 4 \hspace*{1pt}  q_{(\mu} p_\nu p_\rho p_{\sigma)} \nonumber \\
                                    & + \hspace*{-7pt} & \mathcal F^4 \left[\frac {55 q^4} {768 m^6} \hspace*{-1.5pt} \left(\hspace*{-1pt} 1 \hspace*{-1pt} + \hspace*{-0.5pt} \frac{5}{22} \hspace*{1pt} \frac{q^2}{m^2} \hspace*{-2pt}\right) \hspace*{-1pt} I + \frac{3 q^4} {128 m^4} \hspace*{-1pt} \left(\hspace*{-1pt} 1 \hspace*{-0.5pt} + \hspace*{-0.5pt} 2 \hspace*{1pt} \frac{q^2}{m^2} \hspace*{-0.5pt} + \hspace*{-0.5pt} \frac{1} {6} \hspace*{1pt} \frac{q^4}{m^4} \hspace*{-1pt}\right) \hspace*{-1pt} J\right] p_{\mu} p_\nu p_\rho p_{\sigma} \label{eq_ints_tria4} \nonumber \\ \quad
\end{eqnarray}
where we have defined
\begin{equation}
 \mathcal F \equiv \frac{1} {1- \frac{1}{4} \, \frac{q^2}{m^2}}
\end{equation}
in order to keep our notation more compact. Note that the scalar integral $J$ has been expanded in the limit
$q^2 \ll m^2$, however the expressions for the nonanalytic parts of the vector and tensor integrals
are exact to all orders in $q^2$ when expressed in terms of the scalar integrals $I$ and $J$.
The triangle integrals listed in Eqs. (\ref{eq_ints_tria0}-\ref{eq_ints_tria4}) must be used as
\begin{equation}
 \left[J, J_\mu, J_{\mu \nu}, J_{\mu \nu \rho}, J_{\mu \nu \rho \sigma}\right]\Bigg|_{p=-p_2, \, m=m_a}
\end{equation}
in diagrams where particle $a$ (incoming momentum $p_1$ and outgoing momentum $p_2$) propagates in the loop as sketched on the left
side of Fig. \ref{fig_triangles}, and as
\begin{equation}
 \left[J, J_\mu, J_{\mu \nu}, J_{\mu \nu \rho}, J_{\mu \nu \rho \sigma}\right]\Bigg|_{p=+p_4, \, m=m_b}
\end{equation}
when particle $b$ (incoming momentum $p_3$ and outgoing momentum $p_4$) propagates though the loop with momentum labels as seen on
the right hand side of Fig. \ref{fig_triangles}.

More challenging is the calculation of the box and cross-box
diagrams---diagrams (d) and (e) in Fig. \ref{fig_diags}. For the box diagram (d) we
have
\begin{eqnarray}
{\rm Amp}[\ref{fig_diags}d] \hspace*{-7pt}&=\hspace*{-7pt}&\int{d^4k\over (2\pi)^4} \ {1\over k^2(k+q)^2((k-p_2)^2-m_a^2)((k+p_4)^2-m_b^2)}\nonumber\\
&\times\hspace*{-7pt}&\tau^{(1)}_\nu(p_4, p_4 \hspace*{-1.5pt} + \hspace*{-1.5pt} k) \tau^{(1)}_\mu(p_4 \hspace*{-1.5pt} + \hspace*{-1.5pt} k, p_3) \, \eta^{\mu\alpha} \eta^{\nu\beta}
\tau^{(1)}_\beta(p_2, p_2 \hspace*{-1.5pt} - \hspace*{-1.5pt} k) \tau^{(1)}_\alpha(p_2 \hspace*{-1.5pt} - \hspace*{-1.5pt} k, p_1). \nonumber\\
\quad\label{eqn:d}
\end{eqnarray}
The evaluation of the box integrals has been performed earlier by others with Ref. \cite{VanNieuwenhuizen:1971yn} giving
a nice treatment with some of the calculational details. Unfortunately,
the exact expressions for the tensor integrals become extremely long so that we only give the form of the
vector box integral. The Passarino-Veltman reduction of the higher tensor integrals was performed with the
help of computer algebra, which is highly recommended. The expression for the scalar box integral is \cite{VanNieuwenhuizen:1971yn}
\begin{eqnarray}
 K \hspace*{-7pt} & = \hspace*{-7pt} & \int \frac{d^4 k} {(2 \pi)^4} \ \frac {1} {k^2 (k+q)^2 ((k-p_2)^2 - m_1^2) ((k+p_4)^2 - m_2^2)} \nonumber \\
   & = \hspace*{-7pt}& \frac {i} {16 \pi^2} \Bigg[-2 \hspace*{1pt} \frac{L}{q^2} \hspace*{3pt} \frac{1}{\sqrt{\Lambda}} \log \bigg|\frac{\sqrt{\Lambda} - (s-s_0)}{-\sqrt{\Lambda} - (s-s_0)} \bigg| \nonumber \\
   && {} \hspace*{25pt} - i \hspace*{1pt} 2 \pi \hspace*{1pt} \frac{L}{q^2} \hspace*{3pt} \frac{1}{\sqrt{\Lambda}} \ \theta(s-s_0) \hspace*{1pt} \Bigg] \nonumber \\
   & = \hspace*{-7pt}& \frac {i} {16 \pi^2} \Bigg[ \hspace*{-2pt} -2 \hspace*{1pt} \frac{L}{q^2} \left(- \frac{1}{2 m_a m_b}\right) \hspace*{-2pt} \left(1 - \frac{s-s_0}{6 m_a m_b} + \mathcal O\left((s-s_0)^2\right)\right) \nonumber \\
   && {} \hspace*{25pt} - i \hspace*{1pt} 2 \pi \hspace*{1pt} \frac{L}{q^2} \hspace*{1.3pt} \frac{1}{2 \sqrt{m_a m_b} \sqrt{s \hspace*{-1.5pt}- \hspace*{-1.5pt}s_0}} \hspace*{-1pt} \left(\hspace*{-1.8pt}1 \hspace*{-1.5pt}- \hspace*{-1.2pt}\frac{s \hspace*{-1.5pt} - \hspace*{-1.5pt} s_0}{8 m_a m_b} \hspace*{-1.2pt} + \hspace*{-1.3pt} \mathcal O \hspace*{-1.8pt}\left(\hspace*{-1.5pt}(s \hspace*{-1.5pt} - \hspace*{-1.5pt} s_0)^2\right) \hspace*{-2pt} \right) \theta(s-s_0) \hspace*{-0.3pt} \Bigg] \nonumber \\ \quad \label{eq_box_scalarint}
\end{eqnarray}
where
\begin{equation}
 \Lambda \equiv (s-s_0)(4 m_a m_b + s - s_0).
\end{equation}
Note that the expression is {\it exact} in $q^2$ and we only expand it in $s - s_0$ in our calculations.
The vector box integral is found to be
\begin{eqnarray}
 K^\mu \hspace*{-7pt} & = \hspace*{-8pt}& \int \frac{d^4 k} {(2 \pi)^4} \ \frac {k^\mu} {k^2 (k+q)^2 ((k-p_2)^2 - m_1^2) ((k+p_4)^2 - m_2^2)} \nonumber \\
       & = \hspace*{-8pt}& \frac{(2 m_a^2 \hspace*{-1.5pt} + \hspace*{-1.5pt}2 m_a m_b \hspace*{-1.5pt}+ \hspace*{-1.5pt}s \hspace*{-1.5pt}-\hspace*{-1.5pt}s_0) J_a \hspace*{-1pt} + \hspace*{-1pt}(2 m_b^2 \hspace*{-1.5pt} + \hspace*{-1.5pt} 2 m_a m_b \hspace*{-1.5pt} + s \hspace*{-1.5pt} - \hspace*{-1.5pt} s_0) J_b \hspace*{-1pt} - \hspace*{-1pt} \Lambda \, K} {2\left[((m_a + m_b)^2 + s - s_0) q^2 + \Lambda \right]} \, q^\mu \nonumber \\
       & + \hspace*{-8pt}& \frac{(4 m_a m_b \hspace*{-2.4pt} + \hspace*{-2.4pt} 2 (s \hspace*{-2.5pt} - \hspace*{-2.5pt}s_0) \hspace*{-2.5pt} + \hspace*{-2.4pt} q^2) \hspace*{-0.5pt} J_a \hspace*{-2pt} + \hspace*{-2pt}(4 m_b^2 \hspace*{-2.4pt} - \hspace*{-2.4pt} q^2) \hspace*{-0.5pt} J_b \hspace*{-2pt} + \hspace*{-2pt} q^2 \hspace*{-0.7pt}(2 m_b^2 \hspace*{-2.4pt} + \hspace*{-2.4pt}2 m_a m_b \hspace*{-2.4pt}+ \hspace*{-2.4pt} s \hspace*{-2.4pt} - \hspace*{-2.4pt} s_0) \hspace*{-0.5pt} K} {2\left[((m_a + m_b)^2 + s - s_0) q^2 + \Lambda \right]} \hspace*{1pt} p_2^\mu \nonumber \\
       & - \hspace*{-8pt}& \frac{(4 m_a^2 \hspace*{-2.4pt} - \hspace*{-2.4pt} q^2) \hspace*{-0.5pt} J_a \hspace*{-2pt} + \hspace*{-2pt} (4 m_a m_b \hspace*{-2.4pt} + \hspace*{-2.4pt} 2 (s \hspace*{-2.5pt} - \hspace*{-2.5pt}s_0) \hspace*{-2.5pt} + \hspace*{-2.4pt} q^2) \hspace*{-0.5pt} J_b \hspace*{-2pt} + \hspace*{-2pt} q^2 \hspace*{-0.7pt}(2 m_a^2 \hspace*{-2.4pt} + \hspace*{-2.4pt}2 m_a m_b \hspace*{-2.4pt}+ \hspace*{-2.4pt} s \hspace*{-2.4pt} - \hspace*{-2.4pt} s_0) \hspace*{-0.5pt} K} {2\left[((m_a + m_b)^2 + s - s_0) q^2 + \Lambda \right]} \hspace*{1pt} p_4^\mu \nonumber \\ \quad
\end{eqnarray}
with $J_i \equiv J\big|_{m=m_i}$, and we notice that its denominator vanishes in the limit $q^2, \, s - s_0 \rightarrow 0$.
More specifically, the denominator can be written as
\begin{eqnarray}
 D_{K^\mu} & = & 2\left[((m_a + m_b)^2 + s - s_0) q^2 + \Lambda \right] \nonumber \\
  & = & 2 \left[m_a^2 + m_b^2 + 2 \left(p_0^2 + \sqrt{m_a^2 + p_0^2}\sqrt{m_b^2 + p_0^2}\right)\right] \left(4 p_0^2 - \vec q^{\hspace*{1.4pt} 2} \right) \nonumber \\
  & = & 8 p_0^2 \left[m_a^2 + m_b^2 + 2 \left(p_0^2 + \sqrt{m_a^2 + p_0^2}\sqrt{m_b^2 + p_0^2}\right)\right] \left(1 - \sin^2 \frac{\theta}{2} \right) \nonumber \\
  & = & 8 p_0^2 \left[m_a^2 + m_b^2 + 2 \left(p_0^2 + \sqrt{m_a^2 + p_0^2}\sqrt{m_b^2 + p_0^2}\right)\right] \cos^2 \frac{\theta}{2} 
\end{eqnarray}
where we have used Eq. (\ref{eq_sminuss0_p0}) and $q^2 = - \vec q^{\hspace*{1.4pt} 2} = - 4 p_0^2 \sin^2 \frac{\theta}{2}$.
We see that the denominator vanishes for $p_0 \rightarrow 0$ and for backward scattering at $\theta = \pi$.
Unless we consider backward scattering where the denominator vanishes and thus the amplitude diverges, 
we have $4 p_0^2 > \vec q^{\hspace*{1.4pt} 2}$, and since $p_0^2$ originates from the relativistic structure 
$s - s_0$, we therefore expand our vector and tensor box integrals first in $q^2$ and then in $s - s_0$.
Denominators that vanish in the limit $q^2, \, s - s_0 \rightarrow 0$ are a common feature for all box 
vector and tensor integrals, and they are the source of the $1/(s-s_0) \sim 1/p_0^2$ components in our 
results for the scattering amplitude.

In the case of the cross-box diagram (e) the amplitude reads
\begin{eqnarray}
{\rm Amp}[\ref{fig_diags}e] \hspace*{-7pt}&=\hspace*{-7pt}&\int{d^4k\over (2\pi)^4} \ {1\over k^2(k+q)^2((k-p_2)^2-m_a^2)((k-p_3)^2-m_b^2)}\nonumber\\
&\times\hspace*{-7pt}&\tau^{(1)}_\nu(p_4, p_3 \hspace*{-1.5pt} - \hspace*{-1.5pt} k) \tau^{(1)}_\mu(p_3 \hspace*{-1.5pt} - \hspace*{-1.5pt} k, p_3) \, \eta^{\mu\alpha} \eta^{\nu\beta}
\tau^{(1)}_\beta(p_2, p_2 \hspace*{-1.5pt} - \hspace*{-1.5pt} k) \tau^{(1)}_\alpha(p_2 \hspace*{-1.5pt} - \hspace*{-1.5pt} k, p_1). \nonumber\\
\quad\label{eqn:e}
\end{eqnarray}
Now we need the cross-box scalar integral which can be deduced from the result for the box scalar integral by replacing the
set of Mandelstam variables $(s, t)$ by $(u, t)$ where $t = q^2$ and $s + t + u = 2 m_a^2 + 2 m_b^2$. Again, we only
give the scalar and vector integrals because the exact expressions for the higher tensor integrals become
very long. The resulting expressions are
\begin{eqnarray}
 K' \hspace*{-9pt} & = \hspace*{-8pt} & \int \frac{d^4 k} {(2 \pi)^4} \ \frac {1} {k^2 (k+q)^2 ((k-p_2)^2 - m_1^2) ((k-p_3)^2 - m_2^2)}\nonumber \\
    & = \hspace*{-8pt} & \frac {i} {16 \pi^2} \Bigg[-2 \hspace*{1pt} \frac{L}{q^2} \left(+ \frac{1}{2 m_a m_b}\right) \hspace*{-2pt} \Bigg\{\left(1 - \frac{s-s_0}{6 m_a m_b} + \mathcal O\left((s-s_0)^2\right)\right) \nonumber \\
    && \hspace*{133pt} - \, \frac{q^2}{6 m_a m_b} \left(1 - \frac{2(s-s_0)}{5 m_a m_b} + \mathcal O\left((s-s_0)^2\right)\right) \nonumber \\
    && \hspace*{133pt} + \, \frac{q^4}{30 m_a^2 m_b^2} \hspace*{-1pt} \left( \hspace*{-1pt} 1 \hspace*{-1pt} - \frac{9(s-s_0)}{14 m_a m_b} + \mathcal O\left((s-s_0)^2\right) \hspace*{-1pt} \right) \nonumber \\
    && \hspace*{133pt} + \, \mathcal O \left(q^6\right) \Big(1 + O\left(s-s_0\right)\Big) \Bigg\} \Bigg] \nonumber \\
 K'^\mu \hspace*{-9pt} & = \hspace*{-8pt} & \int \frac{d^4 k} {(2 \pi)^4} \ \frac {k^\mu} {k^2 (k+q)^2 ((k - p_2)^2 - m_1^2) ((k - p_3)^2 - m_2^2)} \nonumber \\
       & = \hspace*{-8pt} & \frac{(2 m_a^2 \hspace*{-2pt} - \hspace*{-2pt}2 m_a m_b \hspace*{-2pt} - \hspace*{-2.5pt}(s \hspace*{-2.5pt}-\hspace*{-2.5pt}s_0) \hspace*{-2.5pt} - \hspace*{-2.5pt} q^2) J_a \hspace*{-1.5pt} + \hspace*{-1.5pt}(2 m_b^2 \hspace*{-2.5pt} - \hspace*{-2.5pt} 2 m_a m_b \hspace*{-2.5pt} - \hspace*{-2.5pt}(s \hspace*{-2.5pt} - \hspace*{-2.5pt} s_0) \hspace*{-2.5pt} - \hspace*{-2.5pt} q^2) J_b \hspace*{-1.5pt} - \hspace*{-1.5pt} \tilde \Lambda \hspace*{1pt} K'} {2\left[((m_a + m_b)^2 + s - s_0) q^2 + \Lambda \right]} \, q^\mu \nonumber \\
       & - \hspace*{-8pt} & \frac{(4 m_a m_b \hspace*{-2.4pt} + \hspace*{-2.4pt} 2 (s \hspace*{-2.5pt} - \hspace*{-2.5pt}s_0) \hspace*{-2.5pt} + \hspace*{-2.4pt} q^2) J_a \hspace*{-2pt} - \hspace*{-2pt}(4 m_b^2 \hspace*{-2.4pt} - \hspace*{-2.4pt} q^2) J_b \hspace*{-2pt} - \hspace*{-2pt} q^2 \hspace*{-0.7pt}(2 m_b^2 \hspace*{-2.4pt} - \hspace*{-2.4pt}2 m_a m_b \hspace*{-2.4pt} - \hspace*{-2.4pt} (s \hspace*{-2.4pt} - \hspace*{-2.4pt} s_0) \hspace*{-2.5pt} - \hspace*{-2.5pt} q^2 \hspace*{-0.7pt}) K'} {2\left[((m_a + m_b)^2 + s - s_0) q^2 + \Lambda \right]} \hspace*{1pt} p_2^\mu \nonumber \\
       & + \hspace*{-8pt} & \frac{(4 m_a^2 \hspace*{-2.4pt} - \hspace*{-2.4pt} q^2) J_a \hspace*{-2pt} - \hspace*{-2pt} (4 m_a m_b \hspace*{-2.4pt} + \hspace*{-2.4pt} 2 (s \hspace*{-2.5pt} - \hspace*{-2.5pt}s_0) \hspace*{-2.5pt} + \hspace*{-2.4pt} q^2) J_b \hspace*{-2pt} + \hspace*{-2pt} q^2 \hspace*{-0.7pt}(2 m_a^2 \hspace*{-2.4pt} - \hspace*{-2.4pt}2 m_a m_b \hspace*{-2.4pt} - \hspace*{-2.4pt} (s \hspace*{-2.4pt} - \hspace*{-2.4pt} s_0) \hspace*{-2.4pt} - \hspace*{-2.4pt} q^2 \hspace*{-0.5pt}) K'} {2\left[((m_a + m_b)^2 + s - s_0) q^2 + \Lambda \right]} \hspace*{1pt} p_3^\mu \nonumber \\ \quad
\end{eqnarray}
where
\begin{equation}
 \tilde \Lambda \equiv (s-s_0 + q^2)(4 m_a m_b + s-s_0 + q^2).
\end{equation}
We point out that we did not include an imaginary component in the case of the cross-box scalar integral whereas
for the box integral in Eq. (\ref{eq_box_scalarint}) we included both an imaginary and a real part. The reason for
that is that the $\theta$-function multiplying the imaginary part in Eq. (\ref{eq_box_scalarint}) for the cross-box
integral becomes $\theta(u-s_0)$ and it vanishes in the kinematic region we are considering since
$$u = (m_a - m_b)^2 - (s - s_0) - q^2 < s_0 = (m_a + m_b)^2.$$

In this way all amplitudes quoted in the text can be generated.

\section{Fourier Transformations} \label{app_fts}

In this appendix we collect all Fourier transformation integrals needed to evaluate the potentials in coordinate space.
\begin{align}
 \int \frac{d^3q}{(2 \pi)^3} \ e^{-i \vec q \cdot \vec r} \, \frac{1}{\left|\vec q \hspace*{1pt}\right|^2} & = \frac{1} {4 \pi r} \notag \\
 \int \frac{d^3q}{(2 \pi)^3} \ e^{-i \vec q \cdot \vec r} \, \frac{q_i}{\left|\vec q \hspace*{1pt}\right|^2} & = - \frac{i \, r_i} {4 \pi r^3} \notag \\
 \int \frac{d^3q}{(2 \pi)^3} \ e^{-i \vec q \cdot \vec r} \, \frac{q_i q_j}{\left|\vec q \hspace*{1pt}\right|^2} & = - \frac{1} {4 \pi} \left(3 \frac{r_i r_j}{r^5} - \frac{\delta_{ij}}{r^3}\right) \notag \\
 \int \frac{d^3q}{(2 \pi)^3} \ e^{-i \vec q \cdot \vec r} \, \frac{1}{\left|\vec q \hspace*{1pt}\right|} & = \frac{1} {2 \pi^2 r^2} \notag \\
 \int \frac{d^3q}{(2 \pi)^3} \ e^{-i \vec q \cdot \vec r} \, \frac{q_i}{\left|\vec q \hspace*{1pt}\right|} & = - \frac{i \, r_i} {\pi^2 r^4} \notag \\
 \int \frac{d^3q}{(2 \pi)^3} \ e^{-i \vec q \cdot \vec r} \, \frac{q_i q_j}{\left|\vec q \hspace*{1pt}\right|} & = - \frac{1} {\pi^2} \left(4 \frac{r_i r_j}{r^6} - \frac{\delta_{ij}}{r^4}\right) \notag \\
 \int \frac{d^3q}{(2 \pi)^3} \ e^{-i \vec q \cdot \vec r} \, \log \left|\vec q \hspace*{1pt}\right|^2 & = - \frac{1} {2 \pi r^3} \notag \\
 \int \frac{d^3q}{(2 \pi)^3} \ e^{-i \vec q \cdot \vec r} \, q_i \log \left|\vec q \hspace*{1pt}\right|^2 & = \frac{3 i \, r_i} {2 \pi r^5} \notag \\
 \int \frac{d^3q}{(2 \pi)^3} \ e^{-i \vec q \cdot \vec r} \, q_i q_j \log \left|\vec q \hspace*{1pt}\right|^2 & = \frac{3} {2 \pi} \left(5 \frac{r_i r_j}{r^7} - \frac{\delta_{ij}}{r^5}\right) \notag \\
 \int \frac{d^3q}{(2 \pi)^3} \ e^{-i \vec q \cdot \vec r} \, \left|\vec q \hspace*{1pt}\right|^2 \log \left|\vec q \hspace*{1pt}\right|^2 & = \frac{3} {\pi r^5}
\end{align}

\section{Iteration Integrals} \label{app_iter}

In this appendix we evaluate the integrals
\begin{equation}
[H;H_r;H_{rs}]=i\int {d^3\ell\over (2\pi)^3} {e^2\over
|\vec{p}_f-\vec{\ell}|^2+\lambda^2}{i[1;\ell_r;\ell_r\ell_s]\over
{p_0^2\over 2m_r}-{\ell^2\over 2m_r}+i\epsilon}{e^2\over
|\vec{\ell}-\vec{p}_i|^2+\lambda^2}
\end{equation}
which are needed in order to perform the iteration of the lowest
order Coulomb potentials.  Note that we have introduced a
small photon mass $\lambda^2 \ll p_0^2$ in order to avoid
singularities, but since we are only interested in the long distance
effects we do not show the singularities in $\lambda$ in our expressions.
The evaluation of the integral $H$ has been given by
Dalitz as \cite{dal}
\begin{equation}
H = i4\pi\alpha^2 {m_r\over p_0} \frac{\log\vec{q}^{\hspace*{1.4pt}2}}{\vec{q}^{\hspace*{1.4pt}2}}
  = -i4\pi\alpha^2{m_r\over p_0}\frac{L}{q^2}.
\end{equation}
In order to determine the vector integral $H_r$ we define
\begin{equation}
H_r=A \, (p_i+p_f)_r
\end{equation}
and contracting with $(p_i+p_f)_r$, we find
\begin{equation}
2A (\vec{p}_i+\vec{p}_f)^2=(2\lambda^2+4p_0^2)H -4m_rY
-X(p_i)-X(p_f)
\end{equation}
where
\begin{equation}
Y =-\int{d^3\ell\over (2\pi)^3}{{e^2\over
|\vec{p}_f-\vec{\ell}|^2+\lambda^2}{e^2\over
|\vec{\ell}-\vec{p}_i|^2+\lambda^2}}=-{\pi^2\alpha^2\over
p_0\sin{\theta\over 2}}=-2\alpha^2S
\end{equation}
and
\begin{eqnarray}
X(p_i)=X(p_f)\hspace*{-5pt}&=\hspace*{-5pt}&ie^2\int {d^3\ell\over (2\pi)^3}{e^2\over
|\vec{\ell}-\vec{p}_i|^2+\lambda^2}{{i\over {p_0^2\over
2m_r}-{\ell^2\over 2m_r}+i\epsilon}} \nonumber \\
&= \hspace*{-5pt}&-i 4\pi \alpha^2  {m_r\over p_0}\log{i\lambda\over 2p_0+i\lambda}.
\end{eqnarray}
Note that the integrals X only depend on $p_i$ or $p_f$ and
therefore do not yield any terms nonanalytic in $q^2$. Thus we drop
the contributions of the X's and we have
\begin{equation}
A ={1\over 8p_0^2(1-\sin^2{\theta\over
2})}\left[4p_0^2H-4m_rY\right]\simeq \alpha^2 \! \left({m_r\over p_0^2} S - i \hspace*{0.5pt} 2\pi {m_r\over p_0} \frac{L}{q^2}\right)
\end{equation}
In the case of the tensor integral we define
\begin{equation}
H_{rs}=B \, \delta_{rs}+C \, (p_i+p_f)_r(p_i+p_f)_s + D \,
(p_i-p_f)_r(p_i-p_f)_s
\end{equation}
and we require three conditions in order to evaluate the
coefficients $B$, $C$ and $D$. Neglecting again the integrals X,
these are
\begin{itemize}
\item [i)]
$$\delta_{rs}H_{rs}:\,\,3B +(4p_0^2-\vec{q}^{\hspace*{1.4pt}2})C +\vec{q}^{\hspace*{1.4pt}2} D \simeq
p_0^2H -2m_rY$$
\item [ii)] $$(p_i+p_f)^rH_{rs}:\,\,
B +(4p_0^2-\vec{q}^{\hspace*{1.4pt}2})C \simeq {1\over
1-{\vec{q}^{\hspace*{1.4pt}2}\over 4p_0^2}}\left[p_0^2H -m_r
Y\right] - m_r Y$$
\item [iii)] $$(p_i-p_f)^rH_{rs}:\,\,B +\vec{q}^{\hspace*{1.4pt}2} D \simeq 0$$
\end{itemize}
Solving, we find
\begin{eqnarray}
B &\simeq& -{\vec{q}^{\hspace*{1.4pt}2}\over
4}\left(H -{m_r Y \over p_0^2}\right)\nonumber\\
C &\simeq&{1\over 4}\left(H -{2m_r Y \over
p_0^2} \right)\nonumber\\
D &\simeq&{1\over 4}\left(H -{m_r Y \over p_0^2} \right)
\end{eqnarray}
Keeping only the leading terms in $\vec{q}^{\hspace*{1.4pt}2}$ we
have then
\begin{eqnarray}
H &\simeq&-i 4\pi \alpha^2{m_r\over p_0} \frac{L}{q^2}\nonumber\\
H_r &\simeq&(p_i+p_f)_r\left(-i 2\pi \alpha^2 {m_r\over
p_0} \frac{L}{q^2} + \alpha^2 {m_r\over p_0^2} S +\ldots\right)\nonumber\\
H_{rs} &\simeq& \delta_{rs} \ \vec q^{\hspace*{1.4pt}2}
\left(i\pi\alpha^2{m_r\over
p_0} \frac{L}{q^2} - \frac{1}{2} \alpha^2 {m_r\over p_0^2} S + \ldots\right)\nonumber\\
&+&(p_i+p_f)_r(p_i+p_f)_s\left(- i\pi\alpha^2 {m_r\over p_0} \frac{L}{q^2}+
\alpha^2 {m_r\over p_0^2} S +\ldots\right)\nonumber\\
&+& (p_i-p_f)_r(p_i-p_f)_s\left(-i\pi\alpha^2{m_r\over p_0} \frac{L}{q^2}
+ \frac{1}{2} \alpha^2 {m_r\over p_0^2} S +\ldots\right)
\end{eqnarray}

\section{Generalized Results and Interpretation} \label{app_general}
How can we interpret the universalities we have found? As an example, let us
first consider two spinless charged particles of charge $e$ where the leading
order Coulomb interaction between these two charges is
\begin{equation}
V(\vec r) = \frac{\alpha}{r}.
\end{equation}
Now if we replace one of the two charges by a spin-1/2 particle of charge $-2e$
the spin-independent leading order Coulomb interaction becomes
\begin{equation}
V(\vec r) = \frac{-2 \alpha}{r}.
\end{equation}
We see that the universality of even the leading order Coulomb potential
depends on having equal charges. In this section, we extend our calculations
to arbitrary charges and g-factors, and our results lead us to the interpretation
that the universalities originate from a multipole expansion of the long range
scattering amplitudes and potentials.

\subsection{One-photon Exchange Potential}

It has long been known that a particle with spin $S$ has $2S+1$ multipole moments \cite{lal}.
The one-photon exchange potential thus exhibits a multipole expansion as we know it
from classical electrodynamics:

\subsubsection*{Spin-0 -- Spin-0}
The Lagrangian for a spin-0 particle with arbitrary charge $q = Z e$ reads
\begin{equation}
{\cal L}=(iD_\mu\phi)^\dagger iD^\mu\phi-m^2\phi^\dagger\phi
\end{equation}
with $D_\mu= \partial_\mu + ie Z A_\mu$ and the Feynman rules for the vertices become
\begin{eqnarray}
{}^0 \tau^{(1)}_\mu(p_2,p_1)&=&-iZe(p_2+p_1)_\mu\nonumber\\
{}^0 \tau^{(2)}_{\mu\nu}(p_2,p_1)&=&2i(Ze)^2\eta_{\mu\nu}.
\end{eqnarray}
The one-photon exchange potential for a spin-0 Particle $a$ with mass $m_a$, charge $q_a = Z_a e$ and
a spin-0 particle $b$ with mass $m_b$, charge $q_b = Z_b e$ is then
\begin{eqnarray}
{}^{0}V_C^{(1)}(\vec r)\hspace*{-2pt}&\simeq&
{Z_a Z_b \alpha\over r} \label{VLO00general}
\end{eqnarray}
and exhibits merely a monopole-monopole interaction, Coulomb's law, proportional to $1/r$.

\subsubsection*{Spin-0 -- Spin-1/2}
Now we introduce the Lagrangian for a spin-1/2 particle of arbitrary charge $q = Z e$ and
arbitrary g-factor $g$,
\begin{equation}
{\cal L}=\bar{\psi} (i\not\!\!{D}-m)\psi - \frac{Z e (g-2)}{8 m} F^{\mu \nu} \bar \psi \sigma_{\mu \nu} \psi
\end{equation}
with again $D_\mu= \partial_\mu + ie Z A_\mu$, which yields the Feynman vertex rules
\begin{eqnarray}
{}^{\frac{1}{2}}\tau^{(1)}_\mu(p_2,p_1)&=&-iZe\gamma_\mu + \frac{Z e(g-2)}{4 m} \sigma_{\mu \rho} (p_2 - p_1)^\rho \nonumber\\
{}^{\frac{1}{2}}\tau^{(2)}_{\mu\nu}(p_p,p_1)&=&0 .
\end{eqnarray}
The one-photon exchange potential for a spin-0 particle $a$ with mass $m_a$, charge $q_a = Z_a e$, g-factor $g_a$ and
a spin-1/2 particle $b$ with mass $m_b$, charge $q_b = Z_b e$, g-factor $g_b$ is
\begin{eqnarray}
{}^{{1\over 2}}V_C^{(1)}(\vec r)\hspace*{-2pt}&\simeq&
{Z_a Z_b \alpha\over r} \chi_f^{b\dagger}\chi_i^b - \frac{Z_a Z_b \alpha}{r^3}{(g_b-1)m_a+g_bm_b\over
2m_am_b^2} \vec L \cdot \vec S_b\  \label{VLO0hgeneral}
\end{eqnarray}
where $\vec L \cdot S_b = (\vec r \times \vec p) \cdot \vec S_b = \vec r \cdot (\vec p \times \vec S_b)$.
Thus, besides the leading monopole-monopole interaction -- Coulomb's law -- proportional to $1/r$,
we observe an additional monopole-dipole interaction, the spin-orbit coupling.
The coefficient of the monopole-monopole interaction does not depend on the g-factor whereas
the spin-orbit piece does.

Using the physical values for the masses, charges
and g-factors then includes all one-photon exchange effects to all orders in $\alpha$,
i.e. if we take $g = 2 + \frac{\alpha}{\pi}$  (for a particle of charge $\pm e$), we take
into account the $\mathcal O(\alpha^2)$ long-distance contribution from the one loop vertex correction diagram.

\subsubsection*{Spin-0 -- Spin-1}
For the spin-1 case we consider a particle of charge $Z e$ with g-factor $g$
in the Lagrangian
\begin{equation}
{\cal L}= - \frac{1}{2} U^\dagger_{\mu \nu} U^{\mu \nu} - m^2 \phi^\dagger_\mu \phi^\mu + i Z e (g-1) \phi^\dagger_\mu \phi_\nu F^{\mu \nu}
\end{equation}
where $U_{\mu \nu} = D_\mu \phi_\nu - D_\nu \phi_\mu$ with $D_\mu = \partial_\mu + ie Z A_\mu$.
Once the charge and the g-factor are determined, the quadrupole moment from this
Lagrangian is fixed. If one wants to include an arbitrary quadrupole moment
one has to add a dimension 6 operator \cite{couplings} which we will not do here since
it complicates the following two-photon exchange calculations considerably.
The resulting Feynman rules read
\begin{eqnarray}
{}^1 \tau^{(1)}_{\mu, \beta \alpha}(p_2,p_1) &=&
iZe\Big[\eta_{\alpha \beta}(p_2 + p_1)_\mu  \nonumber \\
&&\hspace*{17pt}- \eta_{\mu \beta} (g p_2 - (g-1) p_1)_\alpha
 - \eta_{\mu \alpha}(g p_1 - (g-1) p_2)_\beta\Big] \nonumber \\
{}^1 \tau^{(2)}_{\mu\nu, \beta \alpha}(p_2,p_1)
&=& -i(Ze)^2 \left[2 \eta_{\mu\nu} \eta_{\alpha \beta} - \eta_{\mu \alpha} \eta_{\nu \beta} - \eta_{\mu \beta} \eta_{\nu \alpha}\right]
\end{eqnarray}
and the one-photon exchange potential for a spin-0 particle $a$ with mass $m_a$, charge $q_a = Z_a e$ and
a spin-1 particle $b$ with mass $m_b$, charge $q_b = Z_b e$, g-factor $g_b$ is
\begin{eqnarray}
{}^1V^{(1)}_C(\vec r)
&\simeq&{Z_a Z_b \alpha\over r} \, \hat{\epsilon}_f^{b*}\cdot\hat{\epsilon}_i^b
 - {Z_a Z_b \alpha\over r^3}{(g_b-1)m_a+g_b m_b\over 2m_am_b^2}\vec{L}\cdot\vec{S}_b \nonumber \\
&+&  \frac{Z_a Z_b \alpha}{r^5} \frac{3(2g_b-1)}{4 m_b^2} \ \vec r : T^b : \vec r
\end{eqnarray}
where we neglected relativistic terms involving $\hat{\epsilon}_f^{b*}\cdot\vec{p} \ \hat{\epsilon}_i^b\cdot\vec{p}$.
Now besides the monopole-monopole and monopole-dipole pieces seen before, a new piece of monopole-quadrupole structure
constitutes the highest multipole in the expansion for the spin-1 particle.

\subsubsection*{Spin-1/2 -- Spin-1/2}
The one-photon exchange potential for a spin-1/2 particle $a$ with mass $m_a$, charge $q_a = Z_a e$, g-factor $g_a$ and
a spin-1/2 particle $b$ with mass $m_b$, charge $q_b = Z_b e$, g-factor $g_b$ is
\begin{eqnarray}
{}^{{1\over 2}{1\over 2}}V_C^{(1)}(\vec r)\hspace*{-2pt}&\simeq&
{Z_a Z_b \alpha\over r}\chi_f^{a\dagger}\chi_i^a\chi_f^{b\dagger}\chi_i^b \nonumber\\
&-& \frac{Z_a Z_b \alpha}{r^3}{(g_b-1)m_a+g_bm_b\over
2m_am_b^2} \vec L \cdot \vec S_b\, \chi_f^{a\dagger}\chi_i^a\nonumber\\
&-&\frac{Z_a Z_b \alpha}{r^3} {g_a m_a+ (g_a-1) m_b\over
2m_a^2m_b} \vec L \cdot \vec{S}_a \, \chi_f^{b\dagger}\chi_i^b \nonumber\\
&-& \frac{Z_a Z_b \alpha}{r^5}{g_a g_b\over 4 m_am_b}\left(3 \vec{S}_a\cdot\vec{r} \, \vec{S}_b\cdot\vec{r}
- r^2 \vec S_a \cdot \vec S_b \right) \label{VLOhhgeneral}
\end{eqnarray}
In this case, we observe a monopole-monopole piece, two monopole-dipole pieces aka spin-orbit pieces and a dipole-dipole piece, the
spin-spin interaction.

Clearly, the one-photon exchange potentials (and scattering amplitudes) for
two charged particles of various spins exhibit a multipole expansion,
and as one would expect for a multipole expansion, higher spins only
add higher multipole interactions while all lower multipole interactions are
universal, i.e. of identical form as for lower spins. That then implies that
the numerical coefficients in the multipole expansion do not depend on structures as
for example $\vec S^2$ or on coefficients that characterize higher multipoles, for example
Coulomb's law cannot depend on the g-factors but only on the charges and the spin-orbit
interaction does not depend on the quadrupole moment.

\subsection{Two-photon Exchange Potential}

At the two-photon exchange level the amplitudes and potentials we calculated
exhibit the same universalities as found in the one-photon exchange case
where they are explained in terms of a multipole expansion. For particles
with arbitrary charges and g-factors the results for the second order
potentials read
\begin{eqnarray}
{}^0V_C^{(2)}(\vec r)\hspace*{-4pt}
&\simeq \hspace*{-4pt} & -{(Z_a Z_b \alpha)^2(m_a+m_b)\over 2m_am_b r^2
}-{7(Z_a Z_b \alpha)^2\hbar\over 6\pi m_am_br^3}\nonumber\\
{}^{1\over 2}V^{(2)}_C(\vec r)\hspace*{-4pt}
&\simeq \hspace*{-4pt} &\Bigg[- {(Z_a Z_b \alpha)^2(m_a+m_b)\over 2m_am_br^2} -
{7(Z_a Z_b \alpha)^2\hbar\over 6\pi m_am_br^3}\Bigg]
\chi_f^{b\dagger}\chi_i^b\nonumber\\
&+ \hspace*{-4pt} &\Bigg[{(Z_a Z_b \alpha)^2\Big(\hspace*{-2pt}(g_b \hspace*{-1.8pt} - \hspace*{-1.8pt} 2)m_a^3 \hspace*{-1.8pt} + \hspace*{-1.8pt} (2g_b \hspace*{-1.8pt} - \hspace*{-1.8pt} 3) m_a^2 m_b \hspace*{-1.8pt} + \hspace*{-1.8pt} 2(g_b \hspace*{-1.8pt} - \hspace*{-1.8pt} 1 \hspace*{-0.4pt})m_a m_b^2 \hspace*{-1.4pt} + \hspace*{-1.4pt} g_b m_b^3\Big)\over
2m_a^2m_b^3(m_a+m_b)r^4}\nonumber\\
&& {}\hspace*{-7pt}+\!{(Z_a Z_b \alpha)^2 \hbar \Big(\hspace*{-2pt}(-3g_b^2 \hspace*{-1.8pt} + \hspace*{-2.4pt} 16g_b \hspace*{-1.8pt} - \hspace*{-2.4pt} 18)m_a \hspace*{-1.8pt} + \hspace*{-1.8pt} (-3g_b^2 \hspace*{-1.8pt} + \hspace*{-2.4pt} 16g_b \hspace*{-1.8pt} - \hspace*{-1.8pt} 4)m_b\Big)\over 8\pi
m_a^2m_b^3r^5}\Bigg] \vec{L}\hspace*{-1pt}\cdot \hspace*{-1pt} \vec{S}_b \nonumber\\
{}^{1}V^{(2)}_C(\vec r)\hspace*{-4pt}
&\simeq \hspace*{-4pt} &\Bigg[- {(Z_a Z_b \alpha)^2(m_a+m_b)\over 2m_am_br^2} -
{7(Z_a Z_b \alpha)^2\hbar\over 6\pi m_am_br^3}\Bigg]
\hat{\epsilon}_f^{b*}\cdot\hat{\epsilon}_i^b\nonumber\\
& + \hspace*{-4pt} &\Bigg[{(Z_a Z_b \alpha)^2\Big(\hspace*{-2pt}(g_b \hspace*{-1.8pt} - \hspace*{-1.8pt} 2)m_a^3 \hspace*{-1.8pt} + \hspace*{-1.8pt} (2g_b \hspace*{-1.8pt} - \hspace*{-1.8pt} 3) m_a^2 m_b \hspace*{-1.8pt} + \hspace*{-1.8pt} 2(g_b \hspace*{-1.8pt} - \hspace*{-1.8pt} 1 \hspace*{-0.4pt})m_a m_b^2 \hspace*{-1.4pt} + \hspace*{-1.4pt} g_b m_b^3\Big)\over
2m_a^2m_b^3(m_a+m_b)r^4}\nonumber\\
&& {}\hspace*{-7pt}+\!{(Z_a Z_b \alpha)^2 \hbar \Big(\hspace*{-2pt}(-3g_b^2 \hspace*{-1.8pt} + \hspace*{-2.4pt} 16g_b \hspace*{-1.8pt} - \hspace*{-2.4pt} 18)m_a \hspace*{-1.8pt} + \hspace*{-1.8pt} (-3g_b^2 \hspace*{-1.8pt} + \hspace*{-2.4pt} 16g_b \hspace*{-1.8pt} - \hspace*{-1.8pt} 4)m_b\Big)\over 8\pi
m_a^2m_b^3r^5}\Bigg] \vec{L}\hspace*{-1pt}\cdot \hspace*{-1pt} \vec{S}_b \nonumber\\
&+ \hspace*{-4pt} & {}^{1}V^{(2)}_T(\vec r) \nonumber\\
{}\hspace*{-5pt}{}^{{1\over 2}{1\over 2}}V^{(2)}_C(\vec r)\hspace*{-4pt}
&\simeq \hspace*{-4pt} &\Bigg[- {(Z_a Z_b \alpha)^2(m_a+m_b)\over 2m_am_br^2} -
{7(Z_a Z_b \alpha)^2\hbar\over 6\pi m_am_br^3}\Bigg]
\chi_f^{a\dagger}\chi_i^a \, \chi_f^{b\dagger}\chi_i^b\nonumber\\
& + \hspace*{-4pt} &\Bigg[{(Z_a Z_b \alpha)^2\Big(\hspace*{-2pt}g_a m_a^3 \hspace*{-1.8pt} + \hspace*{-1.8pt} 2(g_a \hspace*{-1.8pt} - \hspace*{-1.8pt} 1 \hspace*{-0.4pt})m_a^2 m_b \hspace*{-1.8pt} + \hspace*{-1.8pt} (2g_a \hspace*{-1.8pt} - \hspace*{-1.8pt} 3) m_a m_b^2 \hspace*{-1.8pt}+\hspace*{-1.8pt} (g_a \hspace*{-1.8pt} - \hspace*{-1.8pt} 2)m_b^3 \Big)\over
2m_a^3m_b^2(m_a+m_b)r^4}\nonumber\\
&& {}\hspace*{-7pt}+\!{(Z_a Z_b \alpha)^2 \hbar \Big(\hspace*{-2pt}(-3g_b^2 \hspace*{-1.8pt} + \hspace*{-2.4pt} 16g_b \hspace*{-1.8pt} - \hspace*{-1.8pt} 4)m_a \hspace*{-1.8pt} + \hspace*{-1.8pt} (-3g_b^2 \hspace*{-1.8pt} + \hspace*{-2.4pt} 16g_b \hspace*{-1.8pt} - \hspace*{-2.4pt} 18)m_b\Big)\over 8\pi
m_a^3m_b^2 r^5}\Bigg] SO_a \nonumber\\
& + \hspace*{-4pt}&\Bigg[{(Z_a Z_b \alpha)^2\Big(\hspace*{-2pt}(g_b \hspace*{-1.8pt} - \hspace*{-1.8pt} 2)m_a^3 \hspace*{-1.8pt} + \hspace*{-1.8pt} (2g_b \hspace*{-1.8pt} - \hspace*{-1.8pt} 3) m_a^2 m_b \hspace*{-1.8pt} + \hspace*{-1.8pt} 2(g_b \hspace*{-1.8pt} - \hspace*{-1.8pt} 1 \hspace*{-0.4pt})m_a m_b^2 \hspace*{-1.4pt} + \hspace*{-1.4pt} g_b m_b^3\Big)\over
2m_a^2m_b^3(m_a+m_b)r^4}\nonumber\\
&& {}\hspace*{-7pt}+\!{(Z_a Z_b \alpha)^2 \hbar \Big(\hspace*{-2pt}(-3g_b^2 \hspace*{-1.8pt} + \hspace*{-2.4pt} 16g_b \hspace*{-1.8pt} - \hspace*{-2.4pt} 18)m_a \hspace*{-1.8pt} + \hspace*{-1.8pt} (-3g_b^2 \hspace*{-1.8pt} + \hspace*{-2.4pt} 16g_b \hspace*{-1.8pt} - \hspace*{-1.8pt} 4)m_b\Big)\over 8\pi
m_a^2m_b^3r^5}\Bigg] SO_b \nonumber\\
& + \hspace*{-4pt}& \bigg[\! - \hspace*{-1pt} 8 g_a g_b m_a m_b \hspace*{-1pt} - \hspace*{-1pt} 5 g_a g_b (m_a^2 \hspace*{-1pt} + \hspace*{-1pt} m_b^2) \hspace*{-1pt} + \hspace*{-1pt} 2 (g_a m_a \hspace*{-1pt} + \hspace*{-1pt} g_b m_b) (m_a \hspace*{-1pt} + \hspace*{-1pt} m_b)\bigg] \nonumber \\
&& \hspace*{15pt} {} \times  \frac{(Z_a Z_b \alpha)^2 \vec S_a \cdot \vec S_b}{4 m_a^2 m_b^2 (m_a + m_b) r^4} \nonumber\\
& + \hspace*{-4pt}& \bigg[\hspace*{-1pt}(g_b^2 \hspace*{-1pt} + \hspace*{-1pt} 20 g_b \hspace*{-1pt} - \hspace*{-1pt} 12) g_a  m_a^2 \hspace*{-1pt} + \hspace*{-1pt} (g_a g_b (g_a \hspace*{-1pt} + \hspace*{-1pt} g_b \hspace*{-1pt} + \hspace*{-1pt} 32) \hspace*{-1pt} - \hspace*{-1pt} 12(g_a \hspace*{-1pt} + \hspace*{-1pt} g_b))m_a m_b  \nonumber \\
&& + (g_a^2 \hspace*{-1pt} + \hspace*{-1pt} 20 g_a \hspace*{-1pt} - \hspace*{-1pt} 12) g_b  m_b^2 \bigg]
 \times  \frac{(Z_a Z_b \alpha)^2 \vec S_a \cdot \vec r \, \vec S_b \cdot \vec r \, / r^2}{8 m_a^2 m_b^2 (m_a + m_b) r^4} \nonumber\\
& + \hspace*{-4pt}& \bigg[3 g_a^2 g_b^2 + 15 g_a g_b (g_a + g_b) - 92 g_a g_b + 36 (g_a + g_b) + 48 \bigg] \nonumber \\
&& \hspace*{15pt} {} \times  \frac{(Z_a Z_b \alpha)^2 \hbar \, \vec S_a \cdot \vec S_b}{32 \pi m_a^2 m_b^2 r^5} \nonumber\\
& + \hspace*{-4pt}& \bigg[g_a^2 g_b^2 \hspace*{-1pt} + \hspace*{-1pt} 14 g_a g_b (g_a \hspace*{-1pt} + \hspace*{-1pt} g_b) \hspace*{-1pt} - \hspace*{-1pt} 56 g_a g_b \hspace*{-1pt} + \hspace*{-1pt} 4 (g_a^2 \hspace*{-1pt} + \hspace*{-1pt} g_b^2) \hspace*{-1pt} + \hspace*{-1pt} 8 (g_a \hspace*{-1pt} + \hspace*{-1pt} g_b) \hspace*{-1pt} + \hspace*{-1pt} 16 \bigg] \nonumber \\
&& \hspace*{15pt} {} \times  \frac{5 (Z_a Z_b \alpha)^2 \hbar \, \vec S_a \cdot \vec r \, \vec S_b \cdot \vec r \, / r^2}{64 \pi m_a^2 m_b^2 r^5}
\end{eqnarray}
where we have introduced the short notations $SO_a \equiv \vec{L}\hspace*{-1pt}\cdot \hspace*{-1pt} \vec{S}_a \chi_f^{b\dagger}\chi_i^b $
and $SO_b \equiv \chi_f^{a\dagger}\chi_i^a \, \vec{L}\hspace*{-1pt}\cdot \hspace*{-1pt} \vec{S}_b$.

The two-photon exchange potential exhibits a similar structure as the
one-photon exchange potential, a ``generalized multipole expansion''. Its
{\it classical} part of $\mathcal O(\hbar^0)$ starts off with a monopole-monopole
piece proportional to $1/r^2$, then there is a monopole-dipole piece
proportional to $\vec L \cdot \vec S / r^4$ etc. Thus the ``generalized multipole
expansion'' of the {\it classical} part is similar to the multipole expansion of the
one-photon exchange potential, but it has one additional power of $r$ in the denominator.
The ``generalized multipole expansion'' of the {\it quantum} $\mathcal O(\hbar)$ part
of the two-photon exchange potential however is seen to start
with a monopole-monopole piece that falls off as $1/r^3$ followed
by monopole-dipole pieces that go as $\vec L \cdot \vec S / r^5$ etc.

That then suggests the interpretation of the universalities
we have found for the long-distance two-photon scattering potentials
and amplitudes as following from a ``generalized multipole expansion''
where ``generalized'' means that the multipole expansion of the potential
does not start with a monopole-monopole term proportional to $1/r$
as for a usual multipole expansion but proportional to $1/r^n$ with $n > 1$.

It would be interesting to see if one could prove this
multipole expansion scheme and thus the universalities found
using low-energy theorems for Compton scattering amplitudes
and combining two Compton scattering amplitudes to a two-photon
exchange scattering amplitude using dispersion relations.
Moreover, one could speculate that a three-photon exchange
potential would exhibit a similar structure with a
``generalized multipole expansion'' and universalities.

\end{document}